\documentclass[11pt,A4,twoside]{article} 
\usepackage[hmargin=1.4in,vmargin=1.2in]{geometry}
\usepackage{dsfont}
\newcommand\ind{\mathbf{1}}
\usepackage{xspace}
\usepackage{color,soul}
\usepackage[utf8]{inputenc}

\usepackage{afterpage}
\usepackage[nodisplayskipstretch]{setspace}

\usepackage{natbib}

\usepackage{hyperref}
\hypersetup{colorlinks=true,linkcolor=black, citecolor=black, urlcolor=black}
 
\usepackage{bbm}
\usepackage{multicol}
\usepackage{url}            
\usepackage{booktabs}       \usepackage{multirow}
\usepackage{adjustbox}
\usepackage{hhline}

\usepackage{pgfplots}
\usetikzlibrary{intersections}
\usepackage{amsthm,amstext,amsfonts,bm,amssymb}
\usepackage{amssymb,amsmath, amsfonts,amssymb, bbm}\usepackage{amsbsy}
\usepackage[linesnumbered,vlined,boxed,commentsnumbered]{algorithm2e}

\usepackage{tocbasic}
\usepackage{cancel}

\numberwithin{equation}{section}

\usepackage{float}

\renewcommand{\vec}{\mathbf}

\newcommand{\hex}{\widehat{\ex}}

\newcommand{\intc}{\dashint_{-\infty}^\infty}
\newcommand{\intcl}{\!\!\Phint{\ \,-}\limits_{-\infty}^\infty\!}

\newcommand\longmapsfrom{\mathrel{\reflectbox{\ensuremath{\longmapsto}}}}

\newcommand{\intcc}{\dashint}

\newcommand{\F}{\mathcal{F}}

\newcommand{\Fi}{\mathcal{F}^{-1}}

\renewcommand*\d{\mathop{}\!\mathrm{d}}
\newcommand{\norml}[1]{\left\Vert#1 \right\Vert_{1}}
\newcommand{\norm}[1]{\left|#1 \right|}
\newcommand{\Norm}[1]{\left\Vert#1 \right\Vert}
\newcommand{\inner}[1]{\left\langle#1 \right\rangle}

\newcommand{\eh}{\widehat{\ex}}

\renewcommand{\k}{\kappa}

\newcommand{\x}{\mathbf{x}}

\newcommand{\z}{\mathbf{z}}

\newcommand{\y}{\mathbf{y}}
\newcommand{\w}{\mathbf{w}}
\renewcommand{\sc}{\mathrm{sc}}

\newcommand{\ex}{\mathbbm{E}}

\newcommand{\R}{\mathbbm{R}}
\newcommand{\N}{\mathbbm{N}}

\newcommand{\Ra}{\mathcal{R}}

\renewcommand{\S}{\mathcal{S}}
\newcommand{\D}{\mathcal{D}}
\newcommand{\M}{\Sigma}

\newcommand{\C}{\mathcal{C}}

\newcommand{\nn}{\mathcal{NN}}

\newcommand{\intRd}{\int_{\R^d}}
\newcommand{\intRs}{\int_{\R }}
\newcommand{\intR}{\int_{\R}}

\DeclareMathOperator*{\argmin}{\arg\min}
\DeclareMathOperator*{\sgn}{sgn}

\def\Phint#1{\mathchoice
  {\XXint\displaystyle\textstyle{#1}}  {\XXint\textstyle\scriptstyle{#1}}  {\XXint\scriptstyle\scriptscriptstyle{#1}}  {\XXint\scriptscriptstyle\scriptscriptstyle{#1}}  \!\int}
\def\XXint#1#2#3{{\setbox0=\hbox{$#1{#2#3}{\int}$}
    \vcenter{\hbox{$#2#3$}}\kern-.5\wd0}}

\def\dashint{\Phint-}

\usepackage{placeins}

\usepackage{enumitem}
\usepackage{subcaption}

\makeatletter

\title{Spanning Multi-Asset Payoffs With ReLUs\footnote{Python code  
with documentation available at \url{https://github.com/hoangdungnguyen/Spanning_multi_asset_payoffs}. We thank Nisrine Madhar for her contributions to the numerical portion of this paper, as well as two anonymous reviewers for their helpful comments and suggestions.}}
  
\author{S\'ebastien Bossu\thanks{{\tt sbossu@charlotte.edu}. UNC Charlotte Department of Mathematics and Statistics.  The research of S. Bossu is partially supported by funds provided by the University of North Carolina at Charlotte.}, 
\href{https://perso.lpsm.paris/~crepey/}{St\'ephane Cr\'epey}\thanks{{\tt stephane.crepey@lpsm.paris}. Laboratoire de Probabilités, Statistique et Modélisation (LPSM), Universit\'e Paris-Cit\'e, 
France. The research of S.\ Crépey is supported by the Chairs ``Capital Markets Tomorrow: Modeling and Computational Issues'', a joint initiative of LPSM/Université Paris-Cité and Crédit Agricole CIB under the aegis of Institut Europlace de Finance, and ``Futures of Quantitative Finance'', a partnership between LPSM at Université Paris Cité, CERMICS at École des Ponts ParisTech, and BNP Paribas Global Markets.}, 
Hoang-Dung Nguyen\thanks{{\tt hdnguyen@lpsm.paris}. Laboratoire de Probabilités, Statistique et Modélisation (LPSM), Universit\'e Paris-Cit\'e, 
France. The research of H.D. Nguyen is funded by a CIFRE grant from Natixis.}
}

\date{\today}

\def\qqq{\;\;\;\;\;\;\;\;\;}
\def\qq{\;\;\;\;\;\;}

\def\bel{\begin{eqnarray*}\begin{aligned}}
\def\eel{\end{aligned}\end{eqnarray*}}
    \newcommand{\beqa}{\begin{eqnarray}}
    \newcommand{\eeqa}{\end{eqnarray}}
\def\bal{\begin{aligned}}
\def\eal{\end{aligned}}
\def\bll#1{
\beqa\label{#1}\bal
}\def\lel{\eal\eeqa}
\newcommand{\beql}[1]{\bll{#1}}
\newcommand{\eeql}{\lel}

\usepackage{amsthm}
\newtheorem{theorem}{Theorem}[section]
\newtheorem{proposition}[theorem]{Proposition}

\newtheorem{corollary}[theorem]{Corollary}
\newtheorem{lemma}[theorem]{Lemma}

\theoremstyle{definition}
\newtheorem{definition}{Definition}[section]
\newtheorem{problem}{Problem}[section]

\newtheorem{assumption}[definition]{Assumption}

\theoremstyle{remark}
\newtheorem{remark}{Remark}[section] 
\newtheorem{example}[remark]{Example}

\hypersetup{
 colorlinks  = true,  urlcolor   = black,  linkcolor  = black,  citecolor  = black }
\makeatletter
\renewcommand\@makefnmark{\hbox{\@textsuperscript{\normalfont\color{blue}\@thefnmark}}}
\renewcommand\@makefntext[1]{ \parindent 1em\noindent
      \hb@xt@1.8em{        \hss\@textsuperscript{\normalfont\@thefnmark}}#1}
\makeatother

\def\sp{,\;}
 
\def\AHl{{\rm AHa}\xspace}\def\AHl{{\rm (AH)}\xspace}

\def\z{\vec z}
\def\R{\mathbb{R}}

\usepackage{bm}

\def\varphie{\varphi_e}
\def\phie{\phi_e}
\def\finenv{\rule{4pt}{6pt}} \def\finenv{} 
\def\finproof {$\Box$ \vskip 5 pt }\def\finproof{\rule{4pt}{6pt}}\def\finproof{\ensuremath{\square}}
\def\proof{\noindent \emph{\textbf{Proof.} $\, $}}
\def\theh{\b{h}}\def\theh{h}
\def\thef{\b{f}}\def\thef{f}
\def\theS{D}
\def\cdots{\dots}
\pgfplotsset{compat=1.18}

\def\otimes{}
\def\bigotimes{\prod}
\def\then{l}\def\then{n}\def\them{n}\def\them{m}\def\thel{m}\def\thel{l}\def\thermi{{\rm i}}
\begin{document}
\allowdisplaybreaks
\maketitle

\setlength{\abovedisplayskip}{5pt}
\setlength{\belowdisplayskip}{5pt}
 
\begin{abstract} 
        We propose a distributional formulation of the spanning problem of 
    a multi-asset payoff by vanilla basket options. This problem is shown to have
   a unique solution  if and only if the payoff function is even and absolutely homogeneous, and we establish a Fourier-based formula to calculate the solution.      Financial payoffs are typically piecewise linear, resulting in a solution that may be derived explicitly,
        yet may also be hard to exploit numerically.
        One-hidden-layer feedforward neural networks instead provide a natural and efficient numerical alternative for discrete spanning.  We test this approach for a selection of archetypal payoffs and obtain
    better hedging results with vanilla basket options compared to industry-favored approaches based on single-asset vanilla hedges.
\end{abstract}
 {\let\thefootnote\relax\footnotetext{{{\it Corresponding author:}  {\tt stephane.crepey@lpsm.paris}.}} }

\def\keywordname{{\bfseries Keywords:}}
\def\keywords#1{\par\addvspace\baselineskip\noindent\keywordname\enspace
\ignorespaces#1}

\vspace{2mm}
\section{Introduction}

The popular \cite{carr-madan:1998} spanning formula shows that perfect
replication  may be achieved  
for single-asset options with twice differentiable payoffs,
using an infinite number of vanilla calls whose strikes span a one-dimensional continuum.  In honor of this formula, we refer to the practice of hedging a target payoff with a static portfolio of vanilla payoffs as ``spanning'' throughout this paper.
 \citet*{bossu-carr-papanicolaou:2022} and \cite{bossu2021static} spearheaded similar perfect spanning formulas for $\ell^2$ dispersion options and more general multi-asset absolutely homogeneous payoffs, using an infinite number of vanilla basket calls whose weights span a multidimensional continuum. Specifically, given a target European payoff function $F(x_1, \cdots, x_d,k )$ of $d$ underlying asset performances $x_j$ (terminal price ratios or returns) and moneyness or strike parameter $k$, one possible formulation of the continuum spanning problem in strong form is to find a combination of cash, underlying asset and vanilla basket options in respective quantities $\alpha, \mu_j, \nu(\cdots)$, such that
\beql{eq:repli-strong}
    & F(x_1, \cdots, x_d, k) = \alpha + \sum_{j=1}^d \mu_j x_j
    \\[-9pt]
    & \qqq + \idotsint \nu(w_1, \cdots, w_d) \left(\sum_{j=1}^d w_j x_j - k\right)^+ \d w_1 \cdots \d w_d,
    \qq \text{for all }x_1, \cdots, x_d, 
\eeql
wherein $w_1, \cdots, w_d$ are the asset weights of the basket call payoff $ \left(\sum_{j=1}^d w_j x_j - k\right)^+ $.  The primary aim of this paper is to deal with the delicate mathematical regularization aspects identified by the aforementioned authors.
To this end, we introduce a weak formulation of the continuum spanning problem to allow for distributional solutions in a rigorous setting.
 \begin{table}[!t]
\centering
\begin{tabular}{|l|l|c|c|}
\hline
\bf Target option\textsuperscript{*}              & \bf Notation & \bf Payoff $F(\vec x, k)$ & \bf AH variation\textsuperscript{\textdagger}   \\ \hline
Dispersion call/put      &   DC/DP        &      $ (\sum_j |x_j| - k)^\pm$ &  $  (\sum_j |x_j| - |k| )^\pm  $  \\ \hline
Best-of call/put         &   BOC/BOP     &    $(\max_j x_j -  k)^\pm$ & $(\max_j \norm{x_j} -  \norm k)^\pm $    \\ \hline
Worst-of call/put        &    WOC/WOP &   $  ( \min_j x_j -k )^\pm$  &  $  ( \min_j \norm{x_j} - \norm k )^\pm$   \\ \hline
Best-of-binary call/put  &  BOBC/BOBP       &       $\mathbf{1}_{\pm(\max_j x_j -k)> 0}$ & n/a  \\ \hline
Worst-of-binary call/put &  WOBC/WOBP        &        $\mathbf{1}_{\pm(\min_j x_j -k)> 0}$ & n/a \\
\hline
\end{tabular}
\\
\caption{Option payoffs covered in our numerics. $k$ is the strike price,  $\vec x= (x_1,x_2,\cdots,x_d)$ is the vector of asset performances (price ratios or returns), $z^\pm = \max(0,\pm z)$ is the positive or negative part of $z$.\\\footnotesize{Notes:~\textsuperscript{*}All options trade on over-the-counter financial markets, either directly or as building blocks for structured products \citep*{bossu:2014,schofield}.  \textsuperscript{\textdagger}Absolutely homogeneous (AH) variation of the payoff formula, provided in view of Theorem \ref{t:abs-homogeneous}.}}
\label{tab:paydes}
\end{table}

 In practice, only a finite number of basket calls and puts may be traded and spanning is imperfect. On top of contributing to the above theory, we study how a European multi-asset payoff may be partially hedged with a finite combination of cash, underlying assets, and vanilla basket calls and puts, corresponding to a discretization of the right-hand side of \eqref{eq:repli-strong}. For the selection of archetypal payoffs listed in Table \ref{tab:paydes},
we use neural network training techniques to empirically investigate how the discrete spanning error
\beql{eq:repli}
        F(x_1, \cdots, x_d, k) - \alpha - \sum_{j = 1}^d\mu_j x_j - \sum_{i=1}^{\then} \nu_i \left(w_1^{(i)} x_1 + \cdots + w_d^{(i)} x_d - k_i\right)^+   
\eeql
may be minimized with respect to some error metric such as MSE (mean squared error) for optimal quantities $\alpha$ of cash, $\mu_j$ 
of underlying assets $1\leq j\leq d,$ and $\nu_i$ of basket option payoffs $1\leq i\leq \then $ with associated basket weights $w_1^{(i)}, \cdots, w_d^{(i)}$ and strike parameter $k_i $.
The expressiveness power of the neural network family is well known in theory and in practice, and constitutes a particularly convincing paradigm for discrete spanning since \eqref{eq:repli} corresponds to the prediction error of a one-hidden-layer feedforward neural network.
Incidentally, the \citeauthor{carr-madan:1998} spanning formula can also be found in the neural network literature \citep*[see][Eqns\ (15)--(17)]{savarese2019infinite}.

 \FloatBarrier
\subsection{Background and Review}
 
The static hedging of a complex and illiquid payoff with more liquid instruments is  useful to derivatives practitioners for price discovery, trading, and risk management applications.  The topic of multi-asset payoffs has gained in popularity over the past fifteen years in the academic literature. \citet*{ilhan-jonsson-sircar:2009} study the problem of optimally hedging exotic derivatives positions with dynamic and static trading strategies when the performance is quantified by a convex risk measure. \citet*{hurd-zhou} derive a formula for spread option pricing based on Fourier analysis of the payoff function.  \citet*{carr-laurence} express the joint implied distribution of several underlying assets $x_1,\cdots, x_d$ at time $T$ as an inverse Radon transform of basket call prices with maturity $T$ and derive a multi-asset version of Dupire's formula. \citet*{alexander-venkatramanan:2012} derive analytic pricing approximations for European basket and rainbow payoffs using a decomposition as sum of compound payoffs. \citet*{Molchanov2014} show how the joint implied distribution may be recovered from best-of option prices, and investigate various symmetries of multi-asset derivatives. \citet*{cui-xu:2022} derive a multi-asset extension of the \citet*{carr-madan:1998} static replication formula in the form of multiple integral of products of call payoffs.  \citet*{chiu-cont:2023} propose a model-free approach to determine the superhedging cost of path-dependent payoffs, including Asian options.  \citet*{marinelli:2024} uses functional analysis and the theory of distributions to recover the implied distribution and price single-asset option payoffs with call prices. \citet*{leung-lorig-shirai:2024} use variational methods for utility maximization to find static hedges of basket options with vanilla options.

In the machine learning literature for finance, \citet*{lokeshwar2019neural} propose the use of neural network techniques for semi-static hedging of path-dependent exotics with short-term options. \citet*{lyons-nejad-perezarribas:2020} use ``signature payoffs'' to approximate path-dependent exotic derivatives payoffs. \citet*{antonov-piterbarg} discuss alternatives to neural networks for financial function approximation with an emphasis on linear regression concepts.

\subsection{Standing Notation and Organization of the Paper\label{ss:not}}

We write $ \ind_A \times(\cdot)$ for $(\cdot)$ if the Boolean expression $A$ is true, and 0 otherwise (even if $(\cdot)$ is not defined when $A$ is false, in accordance with the Iverson bracket definition). We use boldface for vectors. We denote by $\R_+^*, \R_+$, and $\R^*$ the sets of positive, nonnegative and nonzero real numbers, respectively. We write $\lim_{x\to 0+}$ for the limit when $x$ goes to 0 in $\R_+^*$. For any positive integer $q$, vector $\z \in \R^q$ and index $j \in  1\, .. \, q$, $\z_{\ne j}$ is the subvector excluding the $j$th coefficient of $\z$, $\z_{\leq j}\in\R^{j}$ is the subvector of the first $j$ coefficients of $\z$,  and $\z_{>j}\in\R^{q-j}$ is the subvector of the last $q-j$ coefficients of $\z$ (with $\z_{>q} = \O $);
$\Norm{\z}_1, \Norm{\z}_2 =|\z |,$ and $\Norm{\z}_\infty$ are respectively the $\ell^1$,  $\ell^2$ (Euclidean) and  $\ell^\infty $ (maximum) norms, and for any other vector $\z'\in\R^q, \z\cdot\z'$ is the dot product. We denote by $\F$ the Fourier transform operator, with $\thermi ^2 =-1$ (see Section \ref{a:fourier} for further notation and details).  Unless indicated otherwise, we work with real-valued functions, measures, and distributions, and their corresponding vector spaces and linear forms; in particular:
\begin{itemize}[leftmargin=*,topsep=2pt,itemsep=2pt,partopsep=2pt,parsep=0pt,listparindent=0pt]
\item $L^p(\R^q)$, for $p\in[1,+\infty)$, the space of Lebesgue-measurable functions on $\R^q$ such that $\Norm{f}_{L^p}:= \big(\int_{\R^q}|f(\x)|^p \d \x\big)^{1/p}<\infty$;
\item $\C^{\infty}(\R^q) $, the space of infinitely differentiable functions over $\R^q$;
\item $\S(\R^q)\subset \C^{\infty}(\R^q)$, the space of Schwartz functions over $\R^q$, i.e.\ $\varphi\in \C^{\infty}(\R^q) $ such that $\varphi(\z)$ and each of its derivatives vanish faster than any inverse power of $|\z|$ as $|\z| \to\infty$ \cite[page 395]{ramm1996radon}, \cite[page 480]{king_2009}; 
\item $\D(\R^q)\subset \S(\R^q)$, the Schwartz subspace of functions with compact support;
\item $\inner{T,\varphi}$, or $\inner{T_{\d\z},\varphi(\z)}_{\z}$ when there may be ambiguity about the variable, the action of a distribution $T$ over a Schwartz function $\varphi\in\S(\R^q)$ (generally, the action of a linear form on a test function in an appropriate functional space);
\item $\inner{\nu(\d\z) ,\varphi(\z)}_{\z} := \int \nu(\d\z)\varphi(\z)$, the integral of $\varphi\in\S(\R^q)$ with respect to a measure $\nu$ on $\R^q$, whenever this integral is well defined (e.g.\ for $\nu(\d\z)=f(\z)\d\z$ for $f$ locally integrable), and $\nu(\d\z)$, the corresponding linear form $\S(\R^q)\ni \varphi \mapsto \inner{\nu(\d\z) ,\varphi(\z)}_{\z}$;
\item $\inner{\delta_{\vec{a}}(\d\z), f(\z)}_\z := f(\vec{a})$, the Dirac mass at point $\vec {a}\in\R^q$ of any function $f : \R^q\to \mathbbm C$;
\item $\inner{\frac{\d z}{z-c} ,\varphi( z)}_{ z} :=\intc \frac{\d z}{z-c} \varphi( z)$, for $c\in\R$, the Cauchy integral of $\frac{\varphi( z)}{z-c}$, for $\varphi\in \S(\R)$, and $\frac{\d z}{z-c} $ the corresponding principal value distribution $\S(\R)\ni \varphi \mapsto \inner{\frac{\d z}{z-c} ,\varphi( z)}_{ z} \in \R$ (see Section~\ref{a:cauchy} for further notation and details).\\
\end{itemize}

The  paper is organized as follows. 
Section \ref{s:distreq} poses a weak, distributional formulation of the continuum spanning problem and examines its correspondence with strong formulations.  Section \ref{s:th} shows that
a unique solution exists if and only if the payoff function is even and absolutely homogeneous, 
and provides a Fourier-based formula \eqref{e:Tp1} for its calculation. 
Section \ref{ss:ex} derives explicit solutions of function-, measure-, and principal value-type for specific payoff examples. 
Section \ref{sec:benefits} showcases how feedforward neural networks, which have a natural interpretation for discrete payoff spanning, may constitute an effective numerical method compared to other restricted spanning strategies.
Section \ref{sec:conclusion} gives our conclusions and perspectives for future research. 
The most technical proof of the paper is built upon Sections \ref{a:cauchy} and \ref{a:fourier} and deferred to Section \ref{s:disp}.

The theoretical Sections \ref{s:distreq} to \ref{ss:ex} of this paper are dense and technical, hence the novice reader may wish to start with the numerical Section \ref{sec:benefits} which is more accessible, before proceeding with Sections \ref{s:distreq} to \ref{ss:ex} whose proofs may be skipped on first reading, with particular focus on:
\begin{itemize}[leftmargin=*]
    \item Problem \ref{def:span} page \pageref{def:span} is the weak, distributional formulation of the spanning problem;\
        \item Proposition \ref{t:S0} page \pageref{t:S0} shows that sufficiently regular solutions coincide with the strong problem;
        \item Theorem \ref{t:abs-homogeneous} page \pageref{t:abs-homogeneous} shows that the weak problem has a unique solution if and only the payoff is even and absolutely homogeneous (Definition \ref{d:ahl} page \pageref{d:ahl}), which can then be calculated by the Fourier-based formula \eqref{e:Tp1};
        \item Section \ref{ss:radon} page \pageref{ss:radon} shows that if the payoff is smooth enough, then the strong solution is the Radon transform of the second derivative with respect to the strike parameter $k$, in agreement with earlier research;
        \item Proposition \ref{ss:mh} page \pageref{ss:mh} derives the strong solution for a smooth payoff example;
        \item Proposition \ref{ex:1} page \pageref{ex:1} gives the weak solution (Definition \ref{d:dense} page \pageref{d:dense}) for the dispersion call payoff, while Proposition \ref{p:dc2} page \pageref{p:dc2} derives the corresponding strong, pointwise spanning formula in dimension $d=2$.
\end{itemize}

\section{The Continuum Spanning Problem}\label{s:distreq}

In dimension $d=1$, the \citet*{carr-madan:1998} spanning formula states that any twice differentiable European payoff function $F(x), x\in\R_+$, can be perfectly replicated by a ``continuous portfolio'' of vanilla calls in quantities $F''(K)\d K$ with strike prices spanning the continuum $K \in \R_+$, together with fixed amounts of cash and underlying asset positions, as 
   \begin{equation}\label{eq:carrmadan}
        F(x) = F(0) + F'(0) x + \int_0^\infty  ( x - K)^+ F''(K)\d K ,
        \qquad x \in \R_+ .
    \end{equation}
It is worth observing that the regularity condition on $F$ formally excludes most financial payoffs, such as a simple straddle payoff $F(x) = |x - 1| $ which is not differentiable at $x = 1$.  To reconcile the trivial replication identity ``long straddle = long 2 calls, short 1 forward contract'' with the Carr-Madan formula, we may write
\[
    |x - 1| =  -(x - 1) + 2(x-1)^+ = 1 - x + 2 \int_0^\infty  ( x - K)^+ \delta_1(\d K)
    \sp\  x\in\R_+,
\]
where $\delta_1$ is the Dirac mass at point $K=1$ introduced in Section \ref{ss:not}.  Here, a theoretical extension is needed to give definite meaning to such statements as ``$F''(K)\d K = 2 \delta_1(\d K)$'' to reconcile the above identity with \eqref{eq:carrmadan}.  This is 
the aim of Sections \ref{s:distreq} to \ref{ss:ex} of this paper, in general dimension $d\geq1$ for which solutions become nontrivial.

\citet*{bossu-carr-papanicolaou:2022} and \citet*{bossu2021static} explored replication identities for European multi-asset options with 
absolutely homogeneous payoffs
$F(\lambda\vec x, \lambda k) = \norm\lambda F(\x,k), \lambda\in\R^*$,
such as 
\begin{equation}\label{eq:cont-repli}
        F(\vec x, k) = \int_{\R^d} ( \vec w \cdot \vec x - k )^+  \nu(\d\vec w)  ,
        \qquad \vec x \in \R^d,
    \end{equation} 
where $k$ is a real parameter and $\nu(\d\vec w)$ is the quantity of vanilla basket calls with basket weights $\vec w$ spanning the continuum  $\vec w \in \R^d$. In this multidimensional version of the spanning problem, the cash and asset terms $\alpha+ \boldsymbol \mu\cdot\vec x$ are moved to the left-hand side and inside the arbitrary target payoff function $F(\vec x, k)$ for ease of writing. Section \ref{a:carr-madan} shows how the Carr-Madan formula \eqref{eq:carrmadan} may be rewritten as \eqref{eq:cont-repli} for $d=1$ under specific conditions.

\subsection{Technical Preliminaries}

Our continuum spanning results are established in general dimension $d\geq 1$ for a class of distributions which are not necessarily of measure-type.  This theoretical extension is necessary to give precise mathematical meaning to spanning solutions found for even simple financial payoffs, such as formula \eqref{N2} for the dispersion call in dimension 2.

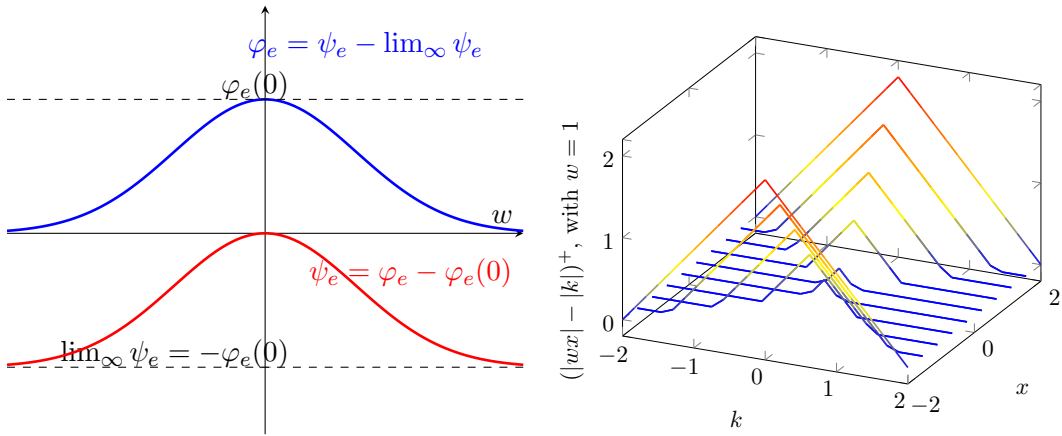
\begin{figure}[h]
\begin{subfigure}{.5\linewidth}
    \resizebox{.95\linewidth}{!}{\begin{tikzpicture}
    \begin{axis}[
        xlabel={$w$},
        ylabel={}, 
        axis lines=center,
        ymin=-1.5, ymax=1.7, 
        xtick=\empty, 
        ytick={-1, 1}, 
        yticklabels={$\lim_{  \infty} \psi_e = -\varphi_e(0) $, $\varphi_e(0)$}, 
        yticklabel style={yshift=5pt,xshift=15pt},
        legend style={at={(0.5,-0.15)}, anchor=north, legend columns=-1},
    ]

    \addplot[blue, domain=-2:2, samples=100, name path=A, line width=1pt] {exp(-x^2)} node[pos=0.95, anchor=south east, yshift=60pt] {$\varphi_e  = \psi_e  -\lim_  \infty \psi_e$};

    \addplot[red, domain=-2:2, samples=100, name path=B, line width=1pt] {exp(-x^2) - 1} node[pos=1, anchor=south east, yshift=25pt] {$\psi_e  = \varphi_e  -\varphi_e(0)$};
    \addplot[dashed, draw=black] coordinates {(-2,-1) (2,-1)};
\addplot[dashed, draw=black] coordinates {(-2, 1) (2, 1)};

    \addplot[draw=none] coordinates {(0,0)}; 


    \end{axis}
\end{tikzpicture}}
\end{subfigure}\begin{subfigure}{.5\linewidth}
     \resizebox{.95\linewidth}{!}{\begin{tikzpicture}
\begin{axis}[
    domain=-2:2,
    xlabel=$k$, ylabel=$x$, zlabel = {$(|wx| - |k|)^+$, with $w = 1$},
]
    \addplot3 [
        samples y=10,
        samples=25,
        mesh,
        patch type=line,
        thick,
    ] {max(abs(y)-abs(x),0)};
\end{axis}
\end{tikzpicture}}
\end{subfigure}
\caption{$d=1$. (\textit{Left}) One-to-one correspondence \eqref{e:bijdtc} between the example functions $\varphie (w) = e^{-w^2}$ and $\psi_e(w) = e^{-w^2} - 1$. (\textit{Right}) ``Basket call'' payoff $F(x,k) = (|wx|-|k|)^+$ in dimension $d=1$ used in the spanning formula \eqref{eq:distrib-repli} with $w = 1$. }\label{fig:testfunctions}
\end{figure}

In order to obtain the weak form \eqref{eq:distrib-repli} of the integral equation \eqref{eq:cont-repli}, we need to mollify both sides with rapidly decaying test functions $\varphi(\x,k)$, and replace $\nu(\d\w)$ by a distribution $N_{\d\w}$.
However, the $\nu(\d\w)$-integral presents a technical difficulty because a mollified basket call payoff $\int_{\R^{d+1}} (\w\cdot\x-k)^+ \varphi(\x,k) \d k\d\x$ may not be rapidly decaying as $\norm\w \to \infty$, which leads us to consider more general linear forms $N_{\d\w}$ defined on an appropriate functional space. To address these issues, we define the function set   
\begin{align*}
\S_0  &=\{\phi_0 \in \C^\infty(\R^{d+1}, {\mathbbm C}) ;\, \phi_0(\x,  k) = e^{-\thermi   r k} \theh (\x)  \mbox{ for some $r\in\R^* $ and $\theh$ in $\mathcal{S}(\R^d)$}\},
\end{align*}
and vector spaces
\begin{align*} 
 \S_e  &=\{\varphie  \mbox{ even in  } \mathcal{S}(\R^d) \} ,
\\
 \M_e &= \{\psi_e\in\C^\infty(\R^d)  ;\,   \psi_e  = \varphie -\varphie({\bf 0}) \mbox{ for some  }\varphie \in \S_e \} .
\end{align*}
Note that $\S_e$ and $\M_e$ are isomorphic via the maps
\beql{e:bijdtc}
   \S_e &  \longleftrightarrow  \M_e \\
    \varphie & \longmapsto  
     \varphie  -\varphie({\bf 0})  \\
 \psi_e -\lim_{\pmb{\infty}}\psi_e & \longmapsfrom  \psi_e,
\eeql
wherein $\lim_{\pmb{\infty}} \psi_e := \displaystyle \lim_{|\w|\to \infty} \psi_e(\w)$.  This one-to-one correspondence is illustrated in the left panel of Figure \ref{fig:testfunctions} for $d=1$.

The following lemma establishes that basket calls mollified by $\phi_0\in\S_0$ belong to the functional space $\Sigma_e$ of even Schwartz functions up to a constant.  Observe the introduction of absolute values to enforce compact support with respect to $k$ and  symmetry with respect to $(\x,k)$ (see also Remark \ref{rem:span}).
\begin{lemma}\label{l:phiO}
{\rm
   \textbf{(i)}} For any $\phi_0(\vec x, k)=e^{-\thermi  kr} \theh(\vec x) \in \S_0$ (with $r\neq 0$ and $h\in\S(\R^d)$), the function     \beql{eq:fou}
       &        \psi (\w)  :=
        - r^{2} \int_{\R^{d}} \d \x  \int_{\R}  \d k \big(|\w \cdot \x | - |k |\big)^+ \phi_0(\x, k)
        = 2\int_{\R^{d}} \d \x h(\x) \big(\cos (\w \cdot \x) - 1 \big)
    \eeql
    is in $\M_e.$
{\rm\hfill\break \textbf{(ii)}} The linear operator
\beql{e:ptop}\begin{array}{rcl}
     \S_0 & \longrightarrow & \M_e  \\
     \phi_0 & \longmapsto & \displaystyle \psi_e : \w \mapsto \int_{\R^{d+1}} \big(|\w \cdot \x|  - |k| \big)^+ \phi_0 (\x,k) \d k \d\x 
\end{array}
\eeql
is well defined.
\end{lemma}

\proof  
{\rm
   \textbf{(i)}}
For any $\phi_0(\vec x, k)=e^{-\thermi  kr} \theh(\vec x) \in \S_0$ with $r\neq 0$ and $h\in\S(\R^d)$, by definition, 
\bel
-\frac{1}{r^{2}}\psi(\w) =   \int_{\R^{d}} \d \x  h(\x) \int_{\R}  \d k e^{-\thermi rk} \big(|\w \cdot \x | - |k |\big)^+ =  \int_{\R^{d}} \d \x  h(\x)     \F_{k} \big[\big(|\w \cdot \x | - |k |\big)^+\big](r).
\eel
Substituting \eqref{e:c} for the Fourier transform of the basket call above,\bel
\psi(\w) & = 2\int_{\R^{d}} \d \x h(\x) \big(\cos (\w \cdot \x) - 1 \big)
\\
& = \int_{\R^{d}} e^{  \thermi  r \w \cdot \x}\theh(\x) \d \x   +\int_{\R^{d}} e^{  -\thermi   r \w \cdot \x}\theh(\x) \d \x - 2\int_{\R^d}\theh(\x)\d \x,
\eel
which establishes \eqref{eq:fou}. From $\lim_{|\w|\to\infty}\int_{\R^{d}} e^{  \pm \thermi  r \w \cdot \x}\theh(\x) \d \x =0$ (see Remark \ref{rem:fourl1}), we get $\lim_{\pmb{\infty}}\psi =- 2\int_{\R^d}\theh(\x)\d \x$. 
Let $\varphi = \psi - \lim_{\pmb{\infty}}\psi$.  Then $\varphi (\w)=\int_{\R^{d}} e^{  \thermi  r \w \cdot \x}\theh(\x) \d \x   +\int_{\R^{d}} e^{  -\thermi   r \w \cdot \x}\theh(\x) \d \x$ is even, and in $\mathcal{S}(\R^d)$ as a property of Fourier transforms \cite[Theorem 2 page 143]{kanwal}. Therefore, 
$\varphi$ belongs to $\S_e$, and
$\psi$ belongs to $\M_e$ in view of the bijection \eqref{e:bijdtc}.
{\rm\hfill\break \textbf{(ii)}} proceeds immediately from (i) since $r\neq 0$ and $\psi_e = -\psi/r^2$.\ \finproof\\

\subsection{Weak Formulation of the Continuum Spanning Problem}

Armed with the preceding definitions and lemma, we are ready to state the continuum spanning problem in weak form, beginning with the following payoff regularity condition that is assumed to hold throughout Sections 2 to 4.
\begin{assumption}\label{ass:pay} The payoff function $F(\x,k)$ is continuous in $\x$, in $k$, and such that
\beql{e:pay} \int_{\R^{d+1}} \d \x \d k |F(\x,k) \theh(\x)|   <\infty\sp\   \theh\in\mathcal{S}(\R^d).\eeql
\end{assumption}
\noindent
Observe that the above continuity assumption is less restrictive than the Carr-Madan differentiability condition in dimension $d=1$.  The following distributional equation is then a weak formulation of the continuum spanning problem.
\begin{problem}\label{def:span} Given a target payoff function $F(\x, k)$ satisfying Assumption \ref{ass:pay}, find a linear form $N$ defined on $\M_e$ such that
\beql{eq:distrib-repli}
  &  \intRd \d \x \intR  \d k F(\x,k)\phi_0 (\x,k)
  =  \\
    &\qqq\qqq\qqq   \inner{N_{\d \w}, \intRd \d\x \intR  \d k \big(|\w \cdot \x | - |k |\big)^+ \phi_0 (\x,k) }_{\w}\sp\  \phi_0 \in \S_0;
\eeql
or, equivalently by correspondence \eqref{e:bijdtc}, find a linear form $T$ on $\S_e$  such that
\beql{e:T}
    \inner{N , \psi_e   } := \inner{T , \varphi_e  } , \quad \mbox{where } \psi_e= \varphie  -   \varphie ({\bf 0}),
\eeql
solves \eqref{eq:distrib-repli}.
\end{problem} 
\begin{remark}  \label{rem:span}
    (i)~If such $T$ is continuous for the topology of $\S(\R^d)$, it defines a distribution on $\S_e$ \citep[Definitions page 22]{kanwal}. \\
    \noindent
    (ii)~By Lemma \ref{l:phiO}(ii), for any $\phi_0 (\x,k)\in\S_0$ the function $\w \mapsto \intRd \d\x \intR  \d k \big(|\w \cdot \x | - |k |\big)^+ \phi_0 (\x,k)$ is indeed in $\M_e$.  \\
    \noindent
    (iii)~The operator $N_{\d \w}$ represents the quantity of basket call payoffs $\big(|\w \cdot \x | - |k |\big)^+$ for each vector $\w$ of basket weights across the $\R^d$ continuum. Note that $\big(|\w \cdot \x | - |k |\big)^+ = \big(\w \cdot \x - |k |\big)^+ + \big(-\w \cdot \x - |k |\big)^+ ,$ which actually corresponds to two proper basket calls with opposite basket weights $\w$ and $-\w$ and identical nonnegative strike $|k|$ (see the right panel in Figure \ref{fig:testfunctions}).
\end{remark}

\subsection{Recovering Strong Solutions from Weak Solutions}

Proposition \ref{t:S0} below shows how the weak formulation \eqref{eq:distrib-repli} coincides with a strong formulation of the continuum spanning problem \eqref{eq:cont-repli} when the solution $N_{\d\w}$ is a measure-type distribution $\nu(\d\w)$  (see Remark \ref{rem:doubling}).  We begin with the following Dirac approximation lemma.

{\def\z{\x}  
\begin{lemma}\label{l:dd}
Let $\theh_n \in  \D(\R^d)$ be the
sequence of functions 
with support inside $\{\Norm{\z}_\infty < \frac{1}{n}\}$ given as
\beql{e:gtod-separable} 
&\theh_n(\z) = c_n \prod_{i=1}^d  
     \ind_{\norm{x_i}<\frac{1}{n}} \exp\Big(-\frac{1}{1-n^2 x_i ^2}\Big), 
\eeql
where $c_n$ is a  normalizing factor such that $\int_{\R^d}\theh_n(\z)\d \z = 1$. Then  $\int_{\R^d} \d\z\,  h_n(\z) g(\z)  \to g(\mathbf 0) $ as $n \to \infty$, for any function $g: \R^d \to \mathbb C$ which is continuous at the origin. 
\end{lemma}

\proof  $h_n(\x)$ is in $\D(\R^d)$ as normalized product of functions in $\D(\R)$ \citep[Lemma 2 page 181]{kanwal} $
    x_i\mapsto\ind_{\norm{x_i}<\frac{1}{n}} \exp\big(-\frac{1}{1-n^2 x_i ^2}\big),
$
and the support of $\theh_n  $ is inside $\{\Norm{\z}_\infty < \frac{1}{n}\}$ by \eqref{e:gtod-separable}.
By continuity of $g$ at the origin, for any $\epsilon>0,$  there exists $n_\epsilon$ such that
$| g(\z)- g({\bf 0})| \le \epsilon$  holds on $ \{\Norm{\z}_\infty < \frac{1}{n_\epsilon}\}$. Hence, for $n\ge n_\epsilon$, 
$\left|\int_{\R^d} \d\z\,  h_n(\z)( g(\z)-g({\bf 0})) \right|\le \int_{\{\Norm{\z}_\infty < \frac{1}{n_\epsilon}\}} \d\z\,  h_n(\z) | g(\z)-g({\bf 0}))| \le \epsilon $ since $h_n$ integrates to 1.
\finproof

\begin{remark}
Each sequence $\Big(\gamma_n \ind_{\norm{x }<\frac{1}{n}} \exp(-\frac{1}{1-n^2 x  ^2})\Big)_n$, where the constants $\gamma_n$ are such that the $L^1(\R)$ norm  is equal to 1, is a delta sequence in the sense of \citet{walter1979probability}---see also \citet[Section 1.2 page 4]{kanwal}.
Replacing $h_n$ in \eqref{e:gtod-separable}  by any other normalized product of delta sequences would do for our purposes.
\end{remark}

\begin{proposition}\label{t:S0} 
If a solution $N$ to the weak spanning problem \eqref{eq:distrib-repli} exists 
in the form of an integration operator on $\M_e$ against an even measure $\nu$ with finite second moment $\int_{\R^d} | \w|^2   \nu(\d\w) <\infty$, 
then the following strong, pointwise representation holds:
\beql{e:Fstrong}
    F(\x , k ) 
    =
   \int_{\R^d} (|\w\cdot\x|  - |k| )^+ \nu(\d\w), \qqq (\x, k ) \in\R^{d}\times \R.
\eeql 
\end{proposition}

\proof Let $  \phi_0(\x,k) = e^{-\thermi rk} \theh(\x) \in \S_0 $ with $r\neq 0$ and $h\in\S(\R^d)$.  Substituting $N_{\d \w} = \nu(\d\w)$ into \eqref{eq:distrib-repli},
\beql{eq:nu-repli}
 &   \intRd \d \x \int_\R \d k F(\x,k) \phi_0(\x, k)=
    \int_{\R^d}  \nu(\d\w)    \intRd \d \x \int_\R \d k \big(|\w \cdot \x | -|k |\big)^+ \phi_0(\x,k).
\eeql
The second moment condition on $\nu$ together with the Cauchy-Schwarz inequality implies
\bel
    &\int_{\R^d} \d\x| \theh(\x) |\int_{\R^{d+1}} (|\w\cdot\x|  - |k|)^+  \nu(\d\w) \d k  \leq    \int_{\R^d} \d\x| \theh(\x) |\int_{\R^d} (\w\cdot\x)^2   \nu(\d\w) \\ &\qqq  \leq \int_{\R^d} \d\x | \theh(\x) |  |\x|^2\int_{\R^d} |\w|^2   \nu(\d\w),
\eel 
which is finite for all  $\theh \in \mathcal{S}(\R^d)$ \cite[Theorem page 141]{kanwal}.  By dominated convergence left and right in \eqref{eq:nu-repli} (due respectively to \eqref{e:pay} and the above inequality), \eqref{eq:nu-repli}  also holds for $r$ formally set to 0 in $\phi_0$.  Replacing $\phi_0(\x,k)$ with $e^{-\thermi rk} \theh(\x)$  in \eqref{eq:nu-repli}, where $r\in \R $ may now be 0, and reordering integrals,
 \begin{align*}
     \int_{\R} \d k e^{-\thermi rk} \intRd \d \x  F(\x,k) \theh(\x)
    = 
    \int_{\R} \d k e^{-\thermi rk}  \int_{\R^d}  \nu(\d\w) \intRd \d \x  \big(|\w \cdot \x | - |k |\big)^+ \theh(\x).
 \end{align*}
Recognizing Fourier transforms over $k$,
 \[ \F_{k} \Big[ 
    \int_{\R^{d} } F(\x,k) \theh(\x) \d\x \Big] 
    =
  \F_{k} \Big[  \int_{\R^d} \d\x \theh(\x) \int_{\R^{d}} \nu(\d\w)\big(|\w \cdot \x | - |k |\big)^+   \Big] 
  .\]
Since the Fourier transform with respect to $k$ is injective on $L^1(\R)$  (see Remark \ref{rem:fourl1}), we recover for all $\theh\in\mathcal{S}(\R^d)$:
\beql{e:fn}  
    \int_{\R^{d} } F(\x,k) \theh(\x) \d\x 
    =
   \int_{\R^d} \d\x \theh(\x) \int_{\R^{d}} \nu(\d\w)\big(|\w \cdot \x | - |k |\big)^+  \eeql
 $\d k$-almost everywhere.  For each $k$ satisfying \eqref{e:fn}, by continuity in $\x$ of $F(\x,k) $ and $\int_{\R^{d}} \nu(\d\w)\big(|\w \cdot \x | - |k |\big)^+$, taking $h(\x) = h_n(\x-\x_0)$ and $n \to \infty$ by Lemma \ref{l:dd}, we recover $ F(\x_0 , k ) = \int_{\R^d} (|\w\cdot\x_0|  - |k| )^+ \nu(\d\w) $ for any $\x_0  \in\R^{d}$.  The validity of \eqref{e:Fstrong} then follows by continuity in $k$ applied to both sides.\ \finproof 

\begin{remark}\label{rem:doubling}
    For fixed $k\ge 0$, \eqref{e:Fstrong} can be seen as a variation of \eqref{eq:cont-repli} up to a doubling factor.  Indeed, substituting $ (|\w\cdot\x|  - |k| )^+ = (\w\cdot\x  - k )^+ + (-\w\cdot\x  - k )^+ $ and splitting the integrand,
\[
F(\x , k ) 
    =
   \int_{\R^d} (\w\cdot\x  - k )^+ \nu(\d\w) + \int_{\R^d} (-\w\cdot\x  - k )^+ \nu(\d\w).
\] 
Substituting $\w \mapsto -\w$ then $\nu(-\d\w) = \nu(\d\w)$ into the second integral above,
\[
    F(\x , k ) 
    =
   2\int_{\R^d} (\w\cdot\x  - k )^+ \nu(\d\w), \qqq \x\in\R^d, k\in\R_+.
\] 
\end{remark}

If $N$ is not induced by a measure, then integration over $k$ and the action of $N$ do not  necessarily commute and the proof of Proposition \ref{t:S0}  no longer works. Hence, in general, strong representations of the form
``$F(\x,k) 
    =   \big\langle N_{\d \w},   \big(|\w \cdot \x | - |k |\big)^+  \big\rangle_{\w}, (\x, k ) \in\R^{d+1}$'' cannot be readily obtained from  
\eqref{eq:distrib-repli}. Nevertheless, we have the following regularized representation.

\begin{proposition} \label{prop:strong-always}
    For any linear form $N$ on $\Sigma_e$ solving \eqref{eq:distrib-repli}, 
    if $ k \mapsto F(\x,k)$ and its Fourier transform $ r \mapsto \F_k\big[ F(\x,k)\big](r)$ are both in $L^1(\R)$ for each $\x\in\R^d $, and $ \x \mapsto \F_k\big[ F(\x,k)\big](r)$ is continuous on $\R^d$ for each $r\in\R^*$,
    then the following strong, pointwise representation holds $d k$-almost everywhere, for each $\x\in\R^d$:
    \begin{align*}
        F(\x,k) & = \Fi_r\left[\lim_{n\to\infty
        } \mathbf{1}_{r\neq 0} \inner{N_{\d \w}, \intRd \d\y \theh_n(\y-\x)
        \F_\kappa\left[ \big(|\w \cdot \y | - |\kappa |\big)^+\right](r) }_{\w}\right](k),
    \end{align*}    
where $h_n$ is the sequence of Dirac approximation functions in \eqref{e:gtod-separable}.
\end{proposition}
\proof Let $  \phi_0(\x,k) = e^{-\thermi rk} \theh(\x) \in \S_0 $ with $r\neq 0$ and $h\in\S(\R^d)$.
Substituting into \eqref{eq:distrib-repli},
\[
    \intRd \d \x \theh(\x) \intR  \d k F(\x,k) e^{-\thermi rk} 
    =\inner{N_{\d \w}, \intRd \d\x \theh(\x) \intR  \d k \big(|\w \cdot \x | - |k |\big)^+ e^{-\thermi rk}  }_{\w}.  
\]
Recognizing Fourier transforms, 
replacing $\x$ with $\y$ on both sides and $k$ with $\kappa$ on the right-hand side,
\[
    \intRd \d \y \theh(\y) \F_k\big[ F(\y,k)\big](r)
    = 
    \inner{N_{\d \w}, \intRd \d\y \theh(\y) \F_\kappa\left[ \big(|\w \cdot \y | - |\kappa|\big)^+ \right](r) }_{\w}.
\]
Let $\x\in\R^d$. Substituting $\theh(\y)\equiv \theh_n(\y-\x)$,
\begin{align*}
    \intRd \d \y \theh_n(\y-\x)
    & \F_k\big[ F(\y,k)\big](r)
    = 
    \inner{N_{\d \w}, \intRd \d\y \theh_n(\y-\x)
    \F_\kappa\left[ \big(|\w \cdot \y | - |\kappa|\big)^+ \right](r) }_{\w}.
\end{align*}
Taking $ n\to+\infty$, by Lemma \ref{l:dd} the limit of the left-hand side exists as $\F_k\big[F(\x,k)\big](r)$, and thus on both sides as
\[
  \F_k \big[F(\x,k)\big](r) = \lim_{n\to+\infty
  } \inner{N_{\d \w}, \intRd \d\y  \theh_n(\y-\x)
  \F_\kappa\left[ \big(|\w \cdot \y | - |\kappa|\big)^+ \right](r) }_{\w}
  \sp r\neq 0.
\]
As such, the functions $\R \ni r \mapsto \F_k \big[F(\x,k)\big](r)$
and
\bel    r \mapsto \ind_{r\neq 0} \lim_{n\to+\infty
  }
  \inner{N_{\d \w}, \intRd \d\y  \theh_n(\y-\x)
  \F_\kappa\left[ \big(|\w \cdot \y | - |\kappa|\big)^+ \right](r) }_{\w} 
    \eel 
coincide on $\R^*$. The inverse Fourier transform of  $ r \mapsto \F_k \big[F(\x,k)\big](r)$ exists as $k \mapsto F(\x,k)$ (see Remark \ref{rem:fourl1}}), hence both functions admit inverse Fourier transforms, which coincide as functions in $L^1(\R)$ and therefore $\d k$-almost everywhere.\ \finproof
 
\section{Solution to the Continuum Spanning Problem}\label{s:th}

In this section we establish our main theoretical result that the weak spanning equation \eqref{eq:distrib-repli} has a unique solution if and only if the payoff is of class \AHl per Definition \ref{d:ahl} below.  The following technical lemma and corollary show that if a solution exists it is unique.

\begin{lemma} \label{l:pf}
The following identity holds between any two bijectively related test functions $\varphie$ and $\theh_e={-\frac{1}{2}}\F^{-1}\varphie$ that are both in $\S_e$: \beql{e:df}
&   \psi_e(\w) := \varphie(\w) - \varphie(\mathbf 0) = 
 \intRd \d\x  \theh_e(\x) \int_\R  \d k  \,e^{-\thermi k}  \big(|\w\cdot\x| -| k|\big)^+   .
\eeql 
\end{lemma} 
\proof Let $\varphie \in\S_e$.  By the Fourier inversion theorem (see Remark \ref{rem:fourcont}), $\displaystyle
\varphie (\w) = \F_\x [\F^{-1} \varphie(\x)](\w) =-2 \int_{\R^d} e^{-\thermi   \x \cdot\w} \theh_e(\x) \d \x
$.  Because $h_e$ is even,
\beql{e:lm1}
\varphie (\w) &= - \int_{\R^d} e^{-\thermi   \x \cdot\w} \theh_e(\x) \d \x -  \int_{\R^d} e^{\thermi  \x \cdot\w} \theh_e(\x) \d \x
 =- 2\int_{\R^d} \cos( \x \cdot\w) \theh_e(\x) \d \x\\
& =  \intRd \d \x  \int_{\R}  \d k e^{\thermi k} \theh_e(\x) \big(|\w \cdot \x | - |k |\big)^+  - 2 
\intRd \d\x    \theh_e( \x) ,\eeql
where we used \eqref{eq:fou} with $r =- 1$ in the last step.  Substituting $k \mapsto -k$ and 
\[
   -2 \intRd \d\x    \theh_e( \x) =\intRd \d\x   (  \F^{-1}\varphie)( \x)= \F\F^{-1}\varphie (\mathbf 0)= \varphie (\mathbf 0),
\]
then rearranging terms yields the required result. \finproof

\begin{corollary}\label{c:uniq} The function space
 $\{\w\mapsto \intRd \d\x \intR  \d k \big(|\w \cdot \x | - |k |\big)^+ \phi_0 (\x,k)\ ; \ \,   \phi_0\in\S_0\}$ coincides with $\Sigma_e$ and
Problem \ref{def:span} 
 admits at most one solution.
\end{corollary}
\proof This space is a subset of $\Sigma_e$ by Lemma \ref{l:phiO}(ii), and a superset by Lemma \ref{l:pf}. Hence any two solutions $N$ and $N'$ to Problem \ref{def:span} coincide on their entire domain $\Sigma_e$, i.e.\ $N=N'$.\ \finproof

\begin{definition}\label{d:ahl}
The payoff function $F(\vec x, k)$  is said to be of class \AHl if it is absolutely homogeneous in $(\x, k )$ and even in both  $\x$ and $k$, i.e.
\bel& F(\lambda\x,\lambda k) = \norm{\lambda} F(\x, k), \lambda\in \R^*\sp\mbox{\ and\ \ }F(\x, k ) = F(\x, -k ) = F(-\x, k )\sp (\x, k )\in\R^{d+1}.\ \finenv\eel
\end{definition}
\begin{remark}
    Examples of absolutely homogeneous payoffs can be found in the last column in Table \ref{tab:paydes}, page \pageref{tab:paydes}.  By continuity of $F$  in $\x$ or $k$ postulated in Assumption \ref{ass:pay}, the payoff function  must vanish at the origin, since e.g.\ $F(\vec 0, 0) = \lim_{n \to \infty }F(\vec 0, 1/n)=\lim_{n \to \infty } 1/n  \, F(\vec 0, 1)=0$.
    \end{remark}
 
\subsection{General Weak Solution}

The following lemma defines two linear forms $N^F, T^F$ associated with any payoff $F$ that will be shown in Theorem \ref{t:abs-homogeneous} below to uniquely solve Problem \ref{def:span} if and only if $F$ is of class \AHl.

\begin{lemma}[and definition]     \label{l:abs-homogeneous}
Given a payoff $F(\x,k)$, the associated linear forms $N^F, T^F$ below are well defined:
\beql{e:Tp1}
   \inner{N^F_{\d \w},\psi_e(\w)}_{\w}
   & =
    \inner{T^F_{\d \w}, \varphie (\w) }_{\w}:=  -\frac{1}{2}\intRd  \d \x\Fi \varphie (\x) \F_k[ F(\x,k) ](1)
    \\
     &\qqq := -\frac{1}{2(2\pi)^d}\intRd  \d \x \intRd \d\w e^{\thermi w\cdot\x} \varphie (\w)  \intRs \d k e^{-\thermi k}  F(\x,  k),
\eeql
for any $\varphie  \in \S_e , \psi_e=  \varphie-  \varphie(\mathbf{0}) \in\Sigma_e$.  If, in addition, 
$$ \d\x\,\F_k[F(\x,k)](1) = \d\x \left(\intR \d k F(\x,k)e^{-\thermi k}\right) $$ is a function-type distribution on $ \S(\R^d)$,
then $T^F$ is a distribution on $\S_e$. 
\end{lemma} 
 \proof 
 Observe that $\Fi\varphie\in\S_e$ for any $\varphie\in\S_e$ \citep[Theorem 3.3 and 4.1]{milton1974fourier}.  By Assumption \ref{ass:pay}, $\intRd  \d \x\Fi \varphie (\x) \F_k[ F(\x,k) ](1)$ is therefore well defined for any $\varphie\in\S_e$, and it is linear in $\varphie$. 
In addition, if $ (\int \d k  F(\x,  k) e^{-\thermi k})\d\x$ is a distribution on $\S(\R^d)$ and thus on $\S_e$ \citep[Theorem 4.4]{milton1974fourier}, because $\Fi$ is a continuous linear operator on $\S(\R^d) \supset \S_e$ (see~Remark \ref{rem:fourcont}), the right-hand side of \eqref{e:Tp1} is  continuous with respect to $\varphie\in\S_e$, hence~$T^F$ given by \eqref{e:Tp1} defines a distribution  on $\S_e$.\ \finproof 
 
\begin{theorem}     \label{t:abs-homogeneous}
Problem \ref{def:span} admits a solution $N$, or equivalently $T$ by correspondence \eqref{e:T}, if and only if the payoff $ F(\x, k)$ is of class \AHl, in which case the solution is unique and given by 
\eqref{e:Tp1}. 
\end{theorem}

\proof The uniqueness of a solution to Problem \ref{def:span} was established in Corollary \ref{c:uniq}.  

\noindent($\Rightarrow$)~Suppose that a solution $T$ to \eqref{eq:distrib-repli}-\eqref{e:T} exists, and let $\phi_0 \in \S_0$ and $\lambda\in\R^*$. The function $(\x,k) \mapsto \phi_0(\lambda^{-1}\x,\lambda^{-1}k)$ is again in $\S_0$ and may be substituted into \eqref{eq:distrib-repli} to obtain
\beql{e:pahl0}
         &\intRd \d \x \intR \d k  F(\x,  k) \phi_0(\lambda^{-1}\x, \lambda^{-1}k) =
     \\
        &\qqq\qqq\qqq  \inner{N_{\d \w}, \intRd \d \x \intR \d k \big(|\w \cdot \x | - |k |\big)^+ \phi_0 (\lambda^{-1}\x, \lambda^{-1}k) }_{\w}. 
\eeql
By linear change of variable $(\x, k) \mapsto (\lambda \x, \lambda k)$ on both sides, and absolute homogeneity of $ \big(|\w\cdot\x| -| k| \big)^+ $ in $(\x, k)$,
\begin{align*}
    \intRd \d \x \intR \d k & |\lambda|^{-d-1} F( \lambda \x, \lambda k) \phi_0 ( \x, k)
    \\ =&\ 
      \inner{N_{\d \w}, \intRd \d \x \intR \d k |\lambda|^{-d-1} \big(|\w\cdot \lambda \x| -| \lambda k| \big)^+ \phi_0 (\x, k) }_{\w}
     \\ =&\ 
     |\lambda|^{-d} \inner{N_{\d \w}, \intRd \d \x \intR \d k \big(|\w\cdot  \x| - | k| \big)^+ \phi_0 (\x, k) }_{\w}
     \\ =&\ 
     |\lambda|^{-d}\intRd \d \x \intR \d k  F(\x,  k) \phi_0 ( \x, k),
\end{align*}
where we substituted \eqref{eq:distrib-repli} in the last step.  Multiplying both sides of the above by $\norm\lambda^{d+1}$,
    \begin{align}\label{e:ahproof}
        \intRd \d \x \intR \d k 
        F( \lambda \x \lambda k) \phi_0 ( \x, k) = |\lambda|\intRd \d \x \intR \d k  F(\x,  k) \phi_0 ( \x, k),
    \end{align}
    which holds for any $\phi_0\in\S_0$, and also for $r$ formally set to 0 in $\phi_0$ as a result of dominated convergence left and right in \eqref{e:ahproof}. By the same reasoning that concluded the proof of Proposition \ref{t:S0}, we recover $ F( \lambda \x, \lambda k)  = |\lambda|F(\x,  k)$ for all $(\x , k)\in\R^{d+1}$. In addition, $(\x,k) \mapsto \phi_0(\x,-k)$ and $(\x,k) \mapsto \phi_0(-\x,k)$ are also in $\S_0$, and we similarly obtain $F(\x,k) =F(\x,-k) =F(-\x,k)$ for $(\x , k)\in\R^{d+1}$. Therefore, $F$ is of class \AHl as required.
\vspace{1.5mm}

\noindent ($\Leftarrow$)~Suppose that $F$ is of class \AHl, and
let $\psi_e\in \M_e$, $\varphie = \psi_e - \lim_{\pmb{\infty}} \psi$, and $r \in \R^*$. By absolute homogeneity of $F(\vec x, k)$, we may substitute $F(\vec x, k) = \norm r F(r^{-1}\x, r^{-1}k)$ in \eqref{e:Tp1} to get
\[
     \inner{N^F_{\d \w},\psi_e (\w) }_{\w}
    = -\frac{1}{2}\intRd \d \x \intR \d k |r| F(r^{-1}\x, r^{-1} k) e^{-\thermi k} \Fi \varphie(\x).
\]
By linear change of variables $(\y,\kappa)=(r^{-1}\x,r^{-1} k)$,
\begin{align}   \label{e:repl0}
     \inner{N^F_{\d \w},\psi_e (\w) }_{\w}
    = 
    -\frac{1}{2}\intRd \d \y \intR \d \kappa |r|^{d+2}F(\y,\kappa) e^{-\thermi  r\kappa} \Fi \varphie(r\y).
\end{align} 
By Lemma \ref{l:pf} with $ \theh_e =-\frac{1}{2} \Fi \varphie$, the left-hand side satisfies
\beql{e:repl}
&  \inner{N^F_{\d \w},\psi_e (\w)  }_{\w}=
  \inner{N^F_{\d \w}, \intRd \d\x \int_\R  \d k  \,  e^{-\thermi k} \theh_e(\x)  \big(|\w\cdot\x| -| k|\big)^+    }_{\w}\\
  &\quad=  \inner{N^F_{\d \w},|r|^{d+2} \intRd \d\y \int_\R  \d \kappa  \,  e^{-\thermi  r\kappa} \theh_e(r\y)  \big(|\w\cdot\y| -| \kappa|\big)^+    }_{\w},
\eeql  
where we applied the change of variable $(\x,k)\mapsto (\y,\kappa)=(r^{-1}\x,r^{-1} k)$ in the last step.  Connecting with \eqref{e:repl0} and dividing both sides by $\norm r^{d+2}$,
\beql{e:ss}
\intRd \d \y \intR \d \kappa F(\y,\kappa) e^{-\thermi  r\k} \theh_e(r\y) = \inner{N^F_{\d \w},\intRd \d \y \intR \d \k e^{-\thermi  r\k} \theh_e(r\y)  \big(|\w\cdot\y| -| \k|\big)^+  }_{\w},
\eeql
which holds for any $\theh_e(r\y) = -\frac{1}{2} \Fi \varphie( r\y ) \in\S_e$. In addition, for any $\theh_o$ odd in $\mathcal{S}(\R^d)$, the even-odd product $F(\y,\kappa)h_o(r\y)$ is odd in $\y$, so that
\[
    \intRd \d \y \intR \d \k F(\y,\k) e^{-\thermi  r\k} \theh_o(r\y)   
    = \intR \d \k e^{-\thermi  r\k} \intRd \d \y F(\y,\k)  \theh_o(r\y)   
    = 0
\]
as the $\d\y$ integral over $\R^d$ of an odd function vanishes.
Consequently, replacing $\theh_e$ by any odd function $\theh_o$ to the left-hand side of \eqref{e:ss} yields 0, and the equation thus holds for any $\theh \in \S (\R^d)$, so that $N^F$ solves \eqref{eq:distrib-repli} as required.\ \finproof 

\subsection{Connection With Radon Transform}\label{ss:radon}

In general, the integrals in \eqref{e:Tp1} may not commute, so that the solution $T$ (or $N$) is not necessarily ``of function-type''.

The following proposition connects our Theorem \ref{t:abs-homogeneous} with earlier research based on Radon transforms and the Fourier slice inversion formula \eqref{eq:sol-L1-der2} in the  Lebesgue-integrable case.  In particular, Proposition \ref{prop:sol-L1}(iii) is tantamount to saying that $f(\w)$ is the inverse Radon transform of $ \frac12\partial^2_{k^2} F(\x,k)$, in agreement with the solution derived in \citet[Section 3]{bossu2021static} for the related strong spanning problem \eqref{eq:cont-repli} when $\nu(\d\w)=f(\w)\d\w$.    
\begin{proposition} \label{prop:sol-L1}
    Let $F$ be a payoff function of class \AHl which is $L^1(\R)$ in $k$ and such that $\F_k[ F(\x,k)](1)$ is $L^1(\R^d)$ in $\x$.  Then:

\noindent{\emph{\textbf{(i)}}}
        The corresponding solution \eqref{e:Tp1} to Problem \ref{def:span} is of function-type  $T^F_{\d \w} = \thef(\w)\d\w$, where
    \begin{equation}    \label{eq:sol-L1}
        \thef(\w) = -\frac{1}{2}\F^{-1}_\x\big[\F_k[ F(\x,k)](1)\big] (\w);
    \end{equation}
{\emph{\textbf{(ii)}}}
If $F(\x,k)$ is $n$-times continuously differentiable against $k$ and all the derivatives are $L^1(\R)$ in $k$, then we may replace $F$ by $(-\thermi  )^n \partial^n_{k^n} F$ in \eqref{eq:sol-L1}, e.g., for $n=2$,
    \begin{equation}    \label{eq:sol-L1-der2}
        \thef(\w) = \frac{1}{2}\F^{-1}_\x\big[\F_k[ \partial^2_{k^2} F(\x,k)](1)\big] (\w);
    \end{equation}
{\emph{\textbf{(iii)}}}
    If $\partial_k F, \partial^2_{k^2} F$ exist and are $L^1(\R)$ in $k$, and $\frac12\partial^2_{k^2} F(\x,k) $ is the Radon transform $\Ra g(\x,k)$ defined below of some function $g\in L^1(\R^d)$, then $f=g$ almost everywhere:
    \beql{e:defrad}
      \Ra g(\x,k) := \frac{\ind_{\x\ne\mathbf0}}{|x_j|} \int_{\R^{d-1}}  g\left(\y_{<j},\frac{k- \x_{\ne j}  \cdot \y_{\ne j}}{x_j},\y_{>j}\right)  \d \y_{\ne j},
    \eeql
    which is independent from the choice of $j$ such that $x_j \neq 0$, for any $\x\ne\mathbf 0$ \citep[page 130]{rubin}.
\end{proposition}

\proof
\textbf{(i)}~All integrands being now Lebesgue-integrable, for any $\psi_e\in \M_e$ and $\varphie $ such that  $\psi_e = \varphie   -  \varphie (\bf 0) $, we may reorder integrals in \eqref{e:Tp1} to get
\begin{align*}
    \inner{T^F_{\d \w}, \varphie (\w) }_{\w}    & = -\frac{1}{2}\intRd \d\w \varphie (\w) \frac{1}{(2\pi)^d}\intRd \d \x e^{\thermi w\cdot\x}  \intRs \d k e^{-\thermi k}  F(\x,  k) 
    \\
    & = -\frac{1}{2}\intRd \d \w \varphie(\w) \F^{-1}_\x\big[\F_k[ F (\x,k)](1)\big] (\w),
    \qq\text{as required.}
\end{align*}

\noindent
\textbf{(ii)} follows from the derivative rule for Fourier transforms applied to $\F_k[ F(\x,k)](1) $ \cite[Theorem 1.8 page 12]{iosevich2014decay}.

\noindent
\textbf{(iii)} 
Substituting $\frac12\partial^2_{k^2} F(\x,k) = \Ra g(\vec x,k)$ and \eqref{e:defrad} into the right-hand side of \eqref{eq:sol-L1-der2},
\bel
&f(\w )=\frac{1}{2}\F^{-1}_\x\big[\F_k[ \partial^2_{k^2} F(\x,k)](1)\big] (\w) = \frac1{(2\pi)^d}\intRd \d\x e^{\thermi \w\cdot\x}\int_\R \d k e^{-\thermi k} \intRd \delta_{\x\cdot\y = k}(\d\y) g(\vec y)\\
&=\frac1{(2\pi)^d}\intRd \d\x e^{\thermi \w\cdot\x}\int_\R \d k e^{-\thermi k} \frac{\ind_{\x\ne 0}}{|x_{j_\x}|} \int_{\R^{d-1}}  g\left(\y_{<j_\x}, \frac{k- \x_{\ne j_\x} \cdot \y_{\ne j_\x}}{x_{j_\x}}, \y_{>j_\x}\right)  \d \y_{\ne {j_\x}},
\eel
where $j_\x$ measurably picks a nonzero coordinate of $\x$ (i.e.\ $x_{j_\x}\ne 0$), for every $\x\ne 0.$
By change of variable $k \mapsto y_{j_\x} = \frac{k- \x_{\ne {j_\x}} \cdot \y_{\ne {j_\x}}}{x_{j_\x}}$ for each $\x\ne 0$,
\begin{align*}
 f(\w )&= \frac1{(2\pi)^d}\intRd \d\x e^{\thermi \w\cdot\x}\int_\R \d y_{j_\x}  \int_{\R^{d-1}}\d \y_{\ne {j_\x} } e^{-\thermi \x\cdot \y} g(\y)\\
 &=\frac1{(2\pi)^d}\intRd \d\x e^{\thermi \w\cdot\x}\int_\R \d \y e^{-\thermi \x\cdot \y} g(\y) = \F^{-1}\F g(\w) = g (\w)  \text{ a.e.}.\ \finproof  
 \end{align*}

\section{Explicit Examples}\label{ss:ex}

In this section, we provide explicit solutions to Problem \ref{def:span}, encompassing the function, measure and principal value types that may be encountered depending on the regularity of the target payoff $F$.

\subsection{Smooth Payoff Function\label{ss:mh}}
Let $G_d(\x,k) = \ind_{\x\ne\mathbf0} |\x| e^{-{\frac{k^2}{|\x|^2} }}$
be a smooth payoff function.  This payoff does not trade but provides a valuable example where the solution is of function-type and available in closed form, as shown below.

\begin{proposition}\label{e:Mdkk}
For $F=G_d$, Problem \ref{def:span} admits the unique solution $T^{G_d}_{\d \w}=
g_d(\w) \d \w$ on $\S_e$ and  $
N^{G_d}_{\d \w}= g_d(\w) \d \w$ on $\M_e$, where $g_d(\w)=-\frac{\sqrt{\pi}}{2}\Fi_\x\big[|\x|^2e^{-\frac{|\x|^2}{4}}\big](\w)$. In particular,
\begin{equation}\label{e:mex}
 \begin{gathered}
    g_1 (w) = e^{-{w^2}}(2w^2-1) \sp  g_2 (w_1,w_2) = \frac{2}{\sqrt{\pi}}e^{-|\w|^2}(|\w|^2-1),\\
    g_3 (w_1,w_2,w_3) = \frac{1}{\pi}e^{-|\w|^2}(2|\w|^2-3).   
\end{gathered}
\end{equation}
 \end{proposition}

\proof The function $G_d$ is of joint class $\C^\infty(\R^{d+1})$, $\AHl$, and $L^1(\R)$ in $k$. By \citet[Example page 139]{kammler2008},
$\F_{k}[G_d(\x,k)](1) = \sqrt{\pi}|\x|^2e^{-\frac{|\x|^2}{4}}$ is $L^1(\R^d)$ in $\x$. Hence $F=G_d$ satisfies the assumptions of Proposition \ref{prop:sol-L1}(i), and the corresponding solution $T^{G_d}$ is of function-type  $T^{G_d}_{\d\w}= g_d (\w)\d\w$ with
\bel
 g_d (\w) = -\frac{1}{2}\F^{-1}_\x\big[\F_k[ G_d(\x,k)](1)\big] (\w) = -\frac{\sqrt{\pi}}{2}\Fi_\x\left[|\x|^2e^{-\frac{|\x|^2}{4}}\right](\w).
\eel
The expressions for $g_1,g_2$ and $g_3$ given in \eqref{e:mex} can be obtained by Hankel transform computations \citep[Theorem 4.1 page 93]{iosevich2014decay}. In addition,
\beql{eq:me3}
&\intRd g_d(\w) \d \w = \F g_d(\mathbf 0) =  -\frac{\sqrt{\pi}}{2} |\x|^2e^{-\frac{|\x|^2}{4}}\Big|_{\x = \mathbf{0}} = 0 .
\eeql
Consequently, for any pair $(\varphie, \psi_e)$ in correspondence \eqref{e:bijdtc}, we have $ \int_{\R} \varphie (\w) g_d(\w)\d \w =  \int_{\R} \psi_e(\w) g_d(\w)\d \w$, so that 
$N^{G_d}_{\d\w}=    g_d(\w)\d \w$ holds on $\M_e$ as required.\ \finproof  

\begin{corollary}\label{ex:Mex2}
The following strong spanning formula holds, for the solution $g_d$ given in Proposition \ref{e:Mdkk}:
\begin{align}\label{e:spanmh}
    G_d(\x, k) = \int_{\R^d} (|\x \cdot \w| - |k|)^+ g_d (\w) \d \w \sp (\x,k) \in \R^{d+1}  .
\end{align}
\end{corollary}
\proof  Since $|\x|^2e^{-\frac{|\x|^2}{4}}\in \S(\R^d)$, $g_d(\w)$ is also in $\S(\R^d)$ and thus $ \intRd |\w|^2 | g_d(\w) |\d\w<\infty$ \cite[Theorem page 141]{kanwal}, yielding that the even measure $ N_{\d \w} = g_d(\w)\d\w $ has finite second moment.  \eqref{e:spanmh}  then follows by an application of Propositions \ref{e:Mdkk} and \ref{t:S0}.\ \finproof

\begin{remark}[sanity check]
    For $d=1$, we may verify by the following direct computation that, for $(x,k) \in  \R^{*}\times \R $,
\begin{align*}
     &\int_{\R} (|xw| -|k|)^+ g_1(w) \d w  =\int_{\frac{|k|}{|x|}}^\infty (|x|w -|k|) g_1(w) \d w + \int^{-\frac{|k|}{|x|}}_{-\infty} (-|x|w -|k|) g_1(w) \d w \\
     &\qqq= 2\int_{\frac{|k|}{|x|}}^\infty (|x|w -|k|) g_1(w) \d w
     = \int_{\frac{|k|}{|x|}}^\infty (|x|w -|k|) e^{-{w^2}}(2w^2-1) \d w \\
    &\qqq= \left.-e^{-w^2}(2w^2|x|+|x|-2kw)\right|^\infty_{\frac{|k|}{|x|}} = |x|e^{-\frac{k^2}{x^2}} = G_1(x,k),
\end{align*}
while $\int_{\R} (|xw| -|k|)^+ g_1(w) \d w = G_1(x, k) = 0$ for $x=0$.
In dimension $d=2$ and $d=3$, \eqref{e:spanmh} may be verified by change of variables to polar coordinates.\ \finenv
\end{remark}

\subsection{Dispersion Call\label{ss:dp}}

For $d \geq 1$,
let 
\beql{e:dcd}
C_d (\vec x, k)= \left( \sum_{j=1}^d \norm{x_j} - \norm k \right)^+
= \big(\Norm\x_1 - \norm k\big)^+\sp\  (\x,k) \in \R^{d+1 }
\eeql 
denote the absolutely homogeneous variation of the $\ell^1$ dispersion call payoff listed in Table \ref{tab:paydes}, page \pageref{tab:paydes}.  This type of payoff trades on exotic derivatives markets \citep*[Figure 1]{bossu-carr-papanicolaou:2022}.
 
\begin{definition}\label{d:dense}
{\def\jmath{i}
Let $S_e \subset \S_e$ be the Schwartz subspace generated by functions of the form $\varphi(\w)+\varphi(-\w)$ where $\varphi\in\S(\R^d)$ is multiplicatively separable, i.e.\ $\varphi(\w)= \prod_{i=1}^d \varphi_i( w_i)$ for $\varphi_i \in \S(\R), i=1,\dots,d$. (For $d=1, S_e = \S_e $.)
{\rm\hfill\break \textbf{(i)}}
By Definition \ref{d:ppv}(iii) and Lemma \ref{prop:pvcom}, we define the following maps, where $c$ is a constant and $\varphi_i \in \S(\R)$ for every $i, j \in  1\, ..\,d$:
\beql{e:defcomppv}
&  \inner{ 
 \delta_c(\d w_{j}) \otimes    \bigotimes_{\jmath\, \in\, 1\,..\,d\, \setminus\,  \{j\}} \frac{\d w_\jmath}{w_{\jmath}\pm w_j}   ,\prod_{i=1}^d \varphi_i( w_i)}_{\w}   := 
       \varphi_j( c) \prod_{\jmath\, \in\, 1\,..\,d\, \setminus\,  \{j\}}\dashint_{-\infty}^\infty \frac{\d w_{\jmath}}{w_{\jmath} \pm  c }\varphi_{\jmath}( w_{\jmath}), \\
    &
    \inner{      \frac{\d w_j}{w_j - c}  \otimes     \bigotimes_{\jmath\, \in\, 1\,..\,d\, \setminus\,  \{j\}}  \frac{\d w_{\jmath}}{w_{\jmath} \pm w_j }    , \prod_{i=1}^d \varphi_i( w_i)}_{\w}
    := \\
    &\qqq\qqq\qqq\qqq  \dashint_{-\infty}^\infty \frac{\d w_j}{w_j - c}\Big(\varphi_j( w_j)\prod_{\jmath\, \in\, 1\,..\,d\, \setminus\,  \{j\}}\dashint_{-\infty}^\infty \frac{\d w_{\jmath}}{w_{\jmath} \pm  w_j}\varphi_{\jmath}( w_{\jmath})\Big)
\eeql
and, in line with the distributivity principle,
\beql{e:defcomppv2}
    &
\inner{ \Big(\delta_1(\d w_{j}) +  \delta_{-1}(\d w_{j}) \Big)\bigotimes_{\jmath\, \in\, 1\,..\,d\, \setminus\,  \{j\}} \Big(\frac{\d w_{\jmath}}{w_{\jmath}+w_j  }-\frac{\d w_{\jmath}}{w_{\jmath}-   w_j }\Big) ,  \phi_e( \w)}_{\w}
    := \\
    &  
\qqq\sum_{\substack{c\,\in\{-1, 1\}\\(\varepsilon_\jmath )_{\jmath\in \, 1\,..\,d\, \setminus\,  \{j\}}\,\in\{-1, 1\}^{d-1 }}}  \varepsilon_\jmath \inner{ \delta_c (\d w_{j})   \bigotimes_{\jmath\, \in\, 1\,..\,d\, \setminus\,  \{j\}}  \frac{\d w_{\jmath}}{w_{\jmath} +\varepsilon_\jmath w_j} ,  \phi_e( \w)}_{\w} ,\\
    &
\inner{ \Big( \frac{\d w_j}{w_j -1}+\frac{\d w_j}{w_j + 1} \Big)\bigotimes_{\jmath\, \in\, 1\,..\,d\, \setminus\,  \{j\}} \Big(\frac{\d w_{\jmath}}{w_{\jmath}+w_j  }-\frac{\d w_{\jmath}}{w_{\jmath}-   w_j }\Big) ,  \phi_e( \w)}_{\w}
    := \\
    &  
\qqq\sum_{\substack{c\,\in\{-1, 1\}\\(\varepsilon_\jmath )_{\jmath\in \, 1\,..\,d\, \setminus\,  \{j\}}\,\in\{-1, 1\}^{d-1 }}}  \varepsilon_\jmath\inner{ \frac{\d w_j}{w_j +c} \bigotimes_{\jmath\, \in\, 1\,..\,d\, \setminus\,  \{j\}}  \frac{\d w_{\jmath}}{w_{\jmath} +\varepsilon_\jmath w_j} ,  \phi_e( \w)}_{\w}
\sp \phie\in S_e ,
\eeql 
which we both extend by linearity as linear forms on $S_e$.
{\rm\hfill\break \textbf{(ii)}}
We denote by $T^d$ the following linear form on $S_e$:   
\begin{itemize}[leftmargin=30pt,topsep=1pt,itemsep=0pt,partopsep=0pt,parsep=0pt,listparindent=0pt]
    \item  $T^1_{\d w} = \frac12\delta_1(\d w) + \frac12\delta_{-1}(\d w)$ , 
    \beql{N2}
  T^{2}_{\d\w}& = \frac{1}{4\pi^2} \Big(
  \frac{\d w_1}{w_1+1}+\frac{\d w_1}{w_1-1}\Big) \Big(\frac{\d w_2}{w_2+w_1} - \frac{\d w_2}{w_2-w_1} \Big)
        \\ 
    &\ + \frac{1}{4\pi^2} \Big(\frac{\d w_2}{w_2+1}+\frac{\d w_2}{w_2-1}\Big)\Big(\frac{\d w_1}{w_1+w_2} - \frac{\d w_1}{w_1-w_2} \Big);
\eeql
    \item more generally, for $d=2l-1$ odd or $2l$ even (with $l$ positive integer),
  \beql{e:tdoe}
    &  \hspace*{-1cm}T^{2l-1}_{\d \w} = \frac{(-1)^{l-1}\pi}{(2\pi)^{d}     }
     \sum_{j = 1}^{d}
   \Bigg( 
   \Big(\delta_1(\d w_{j}) +  \delta_{-1}(\d w_{j})\Big)\otimes    \bigotimes_{\jmath\, \in\, 1 \,..\,d\, \setminus\,  \{j\}} \left(\frac{\d w_\jmath}{w_\jmath+w_{j}} - \frac{\d w_\jmath}{w_\jmath-w_{j}} \right) \Bigg) \\
            &
    \hspace*{-1cm}T^{2l}_{\d \w} = \frac{(-1)^{l-1}}{(2\pi)^{d}} \sum_{j = 1}^{d}  \Bigg(\Big(\frac{\d w_\jmath}{w_j-1} + \frac{\d w_\jmath}{w_j+1}\Big)\otimes     \bigotimes_{\jmath\, \in\, 1\,..\,d\, \setminus\,  \{j\}} \Big(\frac{\d w_\jmath}{w_\jmath+w_{j}} - \frac{\d w_\jmath}{w_\jmath-w_{j}} \Big) \Bigg).
  \eeql
    \end{itemize}
    }\end{definition}
\begin{remark}\label{r:dc-sol-combin}
The underlying combinatorial structure of $T^d$ is apparent in \eqref{e:defcomppv2}.  In particular, $T^d$ embeds a sum of $d \times 2^d $ products and/or iterations of Cauchy integrals. For large $d$ (e.g. $d>20$), the corresponding quadrature would be numerically infeasible. 
\end{remark}

\begin{proposition}\label{ex:1} For $F=C_d$, Problem \ref{def:span} admits a unique solution
$ T^{C_d}$, which is a distribution on $\S_e$, namely the continuous extension of $ T^d  - \delta_{\mathbf{0}}$ to $\S_e$. In particular,
\beql{e:TNt}
   & 
\inner{T^{C_d}_{\d \w},\phi_e(\w)}_{\w}=  \inner{T^d_{\d \w},\phi_e(\w)}_{\w}-\phi_e(\mathbf{0})\sp\  \phi_e \in S_e . \eeql

\end{proposition} 
\proof The function $C_d$ is of class \AHl and, by equivalence of finite-dimensional norms, there exist positive constants $B,M$ such that
\beql{}
\Big|\int_\R \d k  C_d(\x,  k) e^{-\thermi k}\Big|\leq \int_\R \d k(\norml{\x}-|k|)^+ = \norml{\x}^2 \leq B(1+|\x|)^M\sp \x\in \R^d.
\eeql
Consequently, the function  $ \x\mapsto \int \d k  C_d(\x,  k) e^{-\thermi k}$ induces a distribution on $\S(\R^d)$ \citep[Eqn.\ (8.3.2) page 97]{friedlander1998introduction}. Hence by Theorem \ref{t:abs-homogeneous} and Lemma \ref{l:abs-homogeneous} the corresponding unique solution $T^{C_d}$ defines a distribution on $\S_e$.
The subspace of $\D(\R^d)$ generated by functions of the form $\prod_{i=1}^d\varphi_i(x_i),\, \varphi_i \in \D(\R), i=1,\dots,d$, is dense in $\D(\R^d)$ \citep[Lemma 2 page 181]{kanwal}, which itself is dense in $\S(\R^d)$ \citep[Remark page 140]{kanwal}.  As a result, $S_e$ is dense in $\S_e$. 

Turning our attention to \eqref{e:TNt}, for $\psi_e\in \M_e$ and $\varphie $ such that  $\psi_e = \varphie  -  \varphie (\mathbf 0) $, \eqref{e:Tp1} yields
\beql{e:dc_span}
    \inner{N^{C_d}_{\d \w}, \psi_e(\w)}_{\w}
    & = -\frac{1}{2}\intRd \d\x\intRs \d k C_d (\x,k) e^{-\thermi k} \Fi \varphie (\x)\\
    &= -\frac{1}{2}\intRd \d\x\Fi \varphie (\x)\F_{k}\big[ (\norml\x - \norm k)^+\big](1).
\eeql
Substituting \eqref{e:c} with $c=\norml\x$ and $r=1$,
\beql{e:dcc3}
    & \inner{N^{C_d}_{\d \w}, \psi_e(\w)}_{\w}
    = \intRd \d\x\Fi \varphie (\x) (\cos\norml\x - 1)=
    \\
     &\qqq -\int_{\R^d} \d \x\Fi \varphie (\x) +  \frac{1}{2}  \intRd \d\x  e^{-\thermi  \norml{\x}} \Fi \varphie (\x)
    +  \frac{1}{2}  \intRd \d\x  e^{ \thermi  \norml{\x}} \Fi \varphie (\x).
\eeql
As shown in Section \ref{s:disp}, for $\varphie=\phi_e \in S_e,$
\beql{e:TNbis}
&\frac{1}{2}  \intRd \d\x  e^{-\thermi  \norml{\x}} \Fi \phi_e (\x)+  \frac{1}{2}  \intRd \d\x  e^{ \thermi  \norml{\x}} \Fi \phi_e (\x)=
  \inner{T^d_{\d \w},\phi_e (\w)}_{\w}.
\eeql 
Plugging $\int_{\R^d} \d \x\Fi \varphie (\x) =\varphie (\mathbf{0}) $ and
  \eqref{e:TNbis}
into \eqref{e:dcc3},
\beql{e:TN}
 \inner{T^d_{\d \w},\phi_e(\w)}_{\w}-\phi_e(\mathbf{0}) =\inner{N^{C_d}_{\d \w}, \psi_e(\w)}_{\w}
  =  \inner{T^{C_d}_{\d \w},\varphi_e(\w)}_{\w}=\inner{T^{C_d}_{\d \w},\phi_e(\w)}_{\w},
\eeql
which proves \eqref{e:TNt}.\ \finproof

\begin{remark}[sanity check]\label{ex:expldisp}
    For $d=1$ and $T^1 = \frac12(\delta_1 + \delta_{-1})$, \eqref{e:TN} yields
    \bel
   & \inner{N^{C_1}_{\d w}, \psi_e(w)}_{w}  
   =   \inner{T^1_{\d w},\phi_e (w)}_{w} -\phi_e (0)   = \inner{\frac12\big(\delta_1(\d w) + \delta_{-1}(\d w)\big),\phi_e  }_{w} -\phi_e (0)\\&\qqq
   =\inner{\frac12\big(\delta_1(\d w) + \delta_{-1}(\d w)\big),\phi_e -\phi_e (0) }_{w}= \frac12\big(\psi_e(1) + \psi_e(-1)\big) , \eel 
   hence $N^{C_1}_{\d w}=\frac12\big(\delta_1(\d w) + \delta_{-1}(\d w)\big) $
    on $\M_e$ (recalling that $S_e=\S_e$ for $d=1$).  
    Since $N^{C_1}$ is a measure-type distribution with finite second moment $ \frac12\intR w^2 \big(\delta_1(\d w) + \delta_{-1}(\d w)\big) = 1$, we recover via Proposition \ref{t:S0} the trivial spanning identity, for $(x,k) \in \R^2$,
 \beql{EQ:PROP31}
    (|x|-|k|)^+  = \inner{N^{C_1}_{\d w}, (|wx|-|k|)^+}_{w}
    = \frac{1}{2} \big((|x|-|k|)^+ + (|-x|-|k|)^+ \big).
\eeql 
\end{remark}

In higher dimension $d\ge 2$, the distribution
$T^{C_d}$  is not of measure-type and there is no strong spanning representation readily following from \eqref{eq:distrib-repli}. However, by \eqref{e:c},
$\F_\kappa\left[ \big(|\w \cdot \y | - |\kappa |\big)^+\right](r)      =
     \frac{2-2\cos(r \w\cdot\y)}{r^2} ,$ which is continuous in $\y$. The strong representation from Proposition \ref{prop:strong-always} is thus applicable and 
can be turned into the following more explicit representation, derived for ease of writing in dimension $d=2$ only.
\begin{lemma}\label{l:dc2}
For any $r\in\R^*$ and multiplicatively separable $h\in\S(\R^2)$, letting $\psi_e(\w) = \int_{\R^2} \d\y h(\y) \F_\kappa\left[ (|\w \cdot \y | - |\kappa |)^+\right](r)$ and $\phie :=\psi_e-\lim_{\boldsymbol\infty}\psi_e$, then $\phie \in S_e$ and
    \beql{e:t2p}
    & \inner{T^2, \phie} - \phie(\boldsymbol{0})    =  \frac2{r^2}\int_{\R^2} \d\y h(\y)
    \\
    &\qq - \frac1{4\pi^2}\intcl \left(\frac{\d w_1}{ w_1+1}+\frac{\d w_1}{ w_1-1}\right)\intcl\!\left(\frac{\d w_2}{ w_2+w_1}-\frac{\d w_2}{ w_2-w_1}\right)\!\frac2{r^2}\int_{\R^2} \d\y h(\y) \cos(r \w\cdot\y)
    \\
    &\qq - \frac1{4\pi^2} \intcl \left(\frac{\d w_2}{ w_2+1}+\frac{\d w_2}{ w_2-1}\right)\intcl\!\left(\frac{\d w_1}{ w_1+w_2}-\frac{\d w_1}{ w_1-w_2}\right)\!\frac2{r^2} \int_{\R^2} \d\y h(\y) \cos(r \w\cdot\y).\eeql
\end{lemma}
\proof  By Lemma \ref{l:phiO}, $\psi_e(\w) $ is in $\Sigma_e.$ Substituting \eqref{e:c} into the definition of $\psi_e $ yields $\psi_e(\w) = \frac2{r^2}\int_{\R^2} \d\y h(\y) (1-\cos(r \w\cdot\y)) $. By Riemann-Lebesgue's lemma (see Remark~\ref{rem:fourl1}), $\lim_{\pmb{\infty}}\psi_e = \frac2{r^2} \int_{\R^2} \d\y h(\y)$, so that \beql{e:phiek}\phie(\w) = -\frac{2}{r^2}\int_{\R^2} \d\y h(\y) \cos(r \w\cdot\y) ,\eeql 
which is in $\S(\R^2)$ as Fourier transform of an $\S(\R^2)$ function.  As $h(\y)
$ is multiplicatively separable, expanding $\cos(r \w\cdot\y) = \cos(rw_1y_1)\cos(rw_2y_2) - \sin(rw_1y_1)\sin(rw_2y_2)$ and separating integrals 
proves that $\phie \in S_e$.  Substituting \eqref{e:phiek} into the left-hand side of \eqref{e:t2p},
\bel
    &   \inner{T^2, \phie} - \phie(\boldsymbol{0}) 
    =& -\inner{T_{\d \w}^2, \frac{2}{r^2}\int_{\R^2} \d\y h(\y) \cos(r \w\cdot\y)}_{\w}+ \frac2{r^2}\int_{\R^2} \d\y h(\y).
\eel
Substituting \eqref{N2} into the right-hand side yields the required result.\ \finproof

\begin{proposition}\label{p:dc2}
    In dimension $d=2$ the dispersion call payoff admits the strong representation   
    \beql{e:dc2}
     &\left(|x_1| + |x_2| - |k|\right)^+ \\
     =& \Fi_r\Bigg[\lim_{n\to\infty}\mathbf{1}_{r\neq 0}\Bigg(\frac2{r^2}
         +\frac{1}{4\pi^2}\intcl \left(\frac{\d w_1}{ w_1+1}+\frac{\d w_1}{ w_1-1}\right)\intcl \big(\frac{\d w_2}{ w_2+w_1}-\frac{\d w_2}{ w_2-w_1}\big) \times
         \\ & \qqq\qqq\qqq\qqq \left( \F_\kappa\left[ \int_{\R^2} \d\y h_n(\y-\x) \left( |\w\cdot\y| - |\kappa|\right)^+ \right](r) - \frac2{r^2} \right)
       \\
        &\qqq+\frac1{4\pi^2} \intcl \left(\frac{\d w_2}{ w_2+1}+\frac{\d w_2}{ w_2-1}\right)\intcl \left(\frac{\d w_1}{ w_1+w_2}-\frac{\d w_1}{ w_1-w_2}\right) \times
        \\ & \qqq\qqq\qqq\qqq \left(\F_\kappa\left[ \int_{\R^2} \d\y h_n(\y-\x) \left( |\w\cdot\y| - |\kappa|\right)^+ \right](r) - \frac2{r^2} \right) \Bigg)\Bigg](k),
    \eeql
    $\d k$-almost everywhere for each $\x\in\R^2$,   where $h_n$ is the sequence of Dirac approximation functions given in \eqref{e:gtod-separable}.
\end{proposition}
\proof  For each $\y\in\R^2$, the payoff function $ k \mapsto C_2(\y,k)$ has compact support and is thus in $L^1(\R)$, while $ r \mapsto \F_k\big[ C_2(\y,k)\big](r)$ is in $L^1(\R)$ by \eqref{e:c} and is continuous in $\y\in\R^2 $ for each $r\in\R^*$.  Therefore, by Proposition \ref{prop:strong-always}, we have for each $\x\in\R^2$,
    \beql{eq:strong-always-DC}
        C_2(\x,k) & = \Fi_r\left[\lim_{n\to\infty} \mathbf{1}_{r\neq 0} \inner{N_{\d \w}^{C_2}, \intRd \d\y \theh_n(\y-\x) \F_\kappa\left[ \big(|\w \cdot \y | - |\kappa |\big)^+\right](r) }_{\w}\right](k), 
    \eeql   
$\d k$-almost everywhere. Let $\psi^n_e(\w) =\intRd \d\y \theh_n(\y-\x) \F_\kappa\left[ \big(|\w \cdot \y | - |\kappa |\big)^+\right](r) $ and
$\phi^n_e=\psi^n_e - \lim_{\boldsymbol\infty}\psi^n_e$.  By Proposition \ref{ex:1} and 
Lemma \ref{l:dc2} with $h(\y)\equiv h_n(\y-\x)$ that is multiplicatively separable,
\[
    \inner{N^{C_2}, \psi^n_e} = \inner{T^{C_2}, \phi^n_e}=\inner{T^2, \phi^n_e}- \phi^n_e(\boldsymbol{0}).
\]
Substituting \eqref{e:t2p} with $h(\y) = h_n(\y-\x)$ and $\frac2{r^2}\cos(r \w\cdot\y) = \frac2{r^2}\big(1-(1-\cos(r \w\cdot\y))\big) = \frac2{r^2}-\F_\kappa\left[ \left( |\w\cdot\y| - |\kappa|\right)^+ \right](r)$ into the right-hand side, then plugging the resulting expression for $\inner{N^{C_2}, \psi^n_e}$ into \eqref{eq:strong-always-DC}, yields
\begin{align*}
     C_2(\x,k) =& \Fi_r\Bigg[\lim_{n\to\infty}\mathbf{1}_{r\neq 0}\Bigg( \frac2{r^2}\int_{\R^2} \d\y h_n(\y-\x)
     \\
         &- \frac{1}{4\pi^2}\intcl \left(\frac{\d w_1}{ w_1+1}+\frac{\d w_1}{ w_1-1}\right)\intcl \big(\frac{\d w_2}{ w_2+w_1}-\frac{\d w_2}{ w_2-w_1}\big)\times\\
         &\qqq\qqq \int_{\R^2} \d\y h_n(\y-\x)\left(\frac2{r^2} - \F_\kappa\left[ \left( |\w\cdot\y| - |\kappa|\right)^+ \right](r) \right)
        \\
        &- \frac1{4\pi^2} \intcl \left(\frac{\d w_2}{ w_2+1}+\frac{\d w_2}{ w_2-1}\right)\intcl \left(\frac{\d w_1}{ w_1+w_2}-\frac{\d w_1}{ w_1-w_2}\right)\times\\
         &\qqq\qqq\int_{\R^2} \d\y h_n(\y-\x) \left(\frac2{r^2} - \F_\kappa\left[ \left( |\w\cdot\y| - |\kappa|\right)^+ \right](r) \right)
        \Bigg)\Bigg](k).
\end{align*}  
Substituting $\int_{\R^2} \d\y h_n(\y-\x) = 1$, and taking Fourier transforms with respect to $\kappa$ out of the $\d\y$ integrals yields the required result.\ \finproof\\

\section{Benefits of Unrestricted Neural Network Spanning Versus Other Discrete
Spanning Strategies} \label{sec:benefits}
\def\theNN{G}\def\theNN{\widetilde{F}}

The delicate nature of the explicit distribution derived in the previous section for the dispersion call 
shows that the solution formula \eqref{e:Tp1} for $N^F$ can be nontrivial and requires case-by-case analysis.  
When the target payoff $F$ is not smooth, even if an analytical representation can be derived,  it is likely to be numerically intractable in high dimension (see Remark \ref{r:dc-sol-combin}), and other numerical methods must be developed.  Since the hedging error \eqref{eq:repli} corresponds to the prediction error of a one-hidden-layer feedforward neural network, training such a network may constitute an efficient alternative for discrete spanning.

In this empirical section the strike variable $k$ of the target payoff $F(\x,k)$ is viewed as a fixed parameter, and we simply write $F(\x)$ for ease of notation.
For two positive integers $d$ and $\then$, let $\nn_{d,\then }$ denote the family of functions that take a vector $\x \in \R^d$ as input and return a value in $\R$ through the sequential mapping
\begin{align}\label{eq:net_unres}
\R^d \ni \x \stackrel{\theNN}{\longmapsto } \alpha + \boldsymbol \mu \cdot \x + \sum_{i=1}^{\then} \nu_i \left(\eta_i(\vec w^{(i)}\cdot \vec x - k_i)\right)^+   \in \R \,
    ,
\end{align}
where $\boldsymbol\mu, \vec w^{(i)} = [w^{(i)}_1,\cdots w^{(i)}_d]^\top $ are vector versions of the quantities introduced in \eqref{eq:repli}, while the sign of each parameter $\eta_i$ determines whether a basket call or put is used.  The sign of each parameter $\nu_i$ determines whether a long or short position is taken in basket option $i$ with strike $k_i>0$, in quantity $ \norm{ \nu_i \eta_i} $.

\begin{remark} When spanning with basket calls only, i.e.~$\eta_i>0$, approximating $F(\vec x)$ by $\theNN(\x) \in \nn_{d,\then }$ given by \eqref{eq:net_unres} is equivalent to discretizing \eqref{eq:cont-repli} for fixed $k>0$ (in other words, $\nu$ in \eqref{eq:cont-repli} would be allowed to depend on $k$). This can be seen by rewriting \eqref{eq:net_unres} as
\beql{e:changek}
          \theNN(\x) = \alpha + \boldsymbol \mu \cdot \x + \sum_{i=1}^{\then} \underbrace{\nu_i \eta_i\frac{k_i}{k_{\phantom i}}}_{=:\nu_i'}\Big(\underbrace{\frac{k}{k_i}\vec w^{(i)}}_{=:\w'^{(i)}}\cdot \vec x - k\Big)^+,
\eeql
which is a discretization of \eqref{eq:cont-repli} with explicit affine terms to the right-hand side.
\end{remark}

 The family of functions \eqref{eq:net_unres} corresponds to an unrestricted one-hidden-layer residual neural network with ReLU activation function architecture characterized by the set of trainable parameters
\[ 
    \theta = (\vec W, \boldsymbol \mu,  \vec k,\alpha,  \boldsymbol \nu,  \boldsymbol \eta ),
\]
where $ \vec W=  [\vec w^{(1)}, \cdots, \vec w^{(\then)}] ^\top\in\R^{\then \times d}$ stores the $w_j^{(i)}$ in matrix form, while $\vec k \in (\R_+^*)^{\then} ,  \boldsymbol \nu \in \R^{\then}, $ and $\boldsymbol \eta\in \R^{\then}$ store the $k_i, \nu_i$ and $\eta_i$ in vector form. 
We seek an approximation
\begin{align}
    \widehat{F} \in \argmin_{\theNN \in \nn_{d,\then }} \eh \left[(F(\x) - \theNN(\x))^2\right],
    \label{eq:loss}
\end{align}
where $\widehat{\ex}[.]= \frac{[.]_1+\cdots+[.]_\them}\them$ denotes the sample mean over $\them$ observations drawn from values of $\vec x$ that may be deterministically or randomly sampled.
Additional methodological details can be found in the paper's github. 

The NN spanning error $F-\hat F$ can classically be decomposed into three components \citep*[see][Sections 4.2 and 5.1]{bach2024learning}: a bias or approximation error term reflecting the distance between $F$ and $\nn_{d,n}$, an estimation or statistical error when the sample size $m$ is too low, and a numerical optimization error due to local minima in  \eqref{eq:loss}.  The last component is notoriously difficult to analyze.  As a consequence there is no theoretical a priori error control on the global error $F-\hat F$; this error can only be assessed empirically. 
When found too high, static hedging based on our NN spanning approach should be combined with delta-hedging of the residual payoff mismatch.

Below is a summary of our main empirical results, which we compared to three other spanning strategies (see Algorithm \ref{alg:es} for implementation details, involving an activation function $\psi$): 
\begin{itemize}[leftmargin=*,topsep=0pt,itemsep=1pt,partopsep=0pt,parsep=0pt,listparindent=0pt]
    \item Spanning with single-asset payoffs: This ``marginal'' approach is attractive to practitioners because single-asset vanilla options are more liquid and can often be traded on exchanges.
    \item Spanning with predetermined basket payoffs, i.e.\ with fixed basket weights $w_1^{(i)}, \cdots, w_d^{(i)}$ and strikes $k_i$:  This level of control may be beneficial to practitioners in order to define a tractable universe of spanning instruments.  Another benefit of this approach is that it can be solved using classical linear regression techniques.
    \item Spanning with long-only basket payoffs, i.e.\ the components of the  basket weights $\vec w^{(i)}$ of each basket payoff $1 \leq i \leq \then $ are  positive. 
\end{itemize}

We use the Adam stochastic gradient optimizer of \citet*{KingmaB14} provided by the PyTorch package in Python to train our spanning networks. We divide the dataset into 10 batches and train over 1,000 epochs for a total of 10,000 gradient steps. The learning rate is initially set at 0.01 and decreases by a factor of 0.8 every 300 epochs. A 0.1\% regularization defined as the squared Euclidean norm of all network parameters is integrated to the loss function. 

{\def\phi{\theta}
\def\theNN{\widetilde{F}^\psi}

\begin{algorithm}[!t]
\small
\LinesNumbered
\SetAlgoLined
\SetKwInOut{AlgName}{name}
\SetKwInOut{Input}{input}\SetKwInOut{Output}{output}
\Input{$\{(\vec x_1, F(\vec x_m)), \dots, (\vec x_m, F(\vec x_m))\}$, a partition $B$
of $  \{1 \dots m \} $ into subsets $batch$, a number of basket payoffs $n$, a number of epochs $E\in\N^{*}$, an $learning\_ rate  >0$ and a weight decay $\zeta$ . Adam optimizer default setting: $\beta_1 = 0.9, \beta_2 = 0.999.$}
\Output{Trained parameters  $\hat{\phi}$ of the spanning network.} 
Define the spanning network architecture $\theNN_\phi \in \nn^{\psi}_{d,n}$ with activation function $\psi$,
$$
\R^d \ni \x \stackrel{\theNN}{\longmapsto } \alpha + \boldsymbol \mu \cdot \x + \sum_{i=1}^{\then} \nu_i \left(\eta_i\big(\psi(\vec w^{(i)})\cdot \vec x - k_i\big)\right)^+   \in \R 
$$

\uIf{unrestricted basket payoffs cf.\ \eqref{eq:net_unres}}{
$$\psi( \vec w^{(i)} ) =  \vec w^{(i)} $$
}\uElseIf{single-asset payoffs}{
\[
\psi( \vec w^{(i)} ) = \left(\omega(\vec w^{(i)} )_1 \mathbf{1}_{\omega(\vec w^{(i)} )_1 = \Norm{\omega(\vec w^{(i)} )}_\infty},\dots, \omega(\vec w^{(i)} )_d \mathbf{1}_{\omega(\vec w^{(i)} )_d = \Norm{\omega(\vec w^{(i)})}_\infty}\right),
\]
with
$$\omega(\vec w^{(i)} ) = \left(\frac{\exp( w^{(i)}_1)}{\sum_{j=1}^d \exp( w^{(i)}_j) },\cdots, \frac{\exp( w^{(i)}_d)}{\sum_{i=j}^d \exp( w^{(i)}_j)} \right)$$
}\uElseIf{predetermined basket payoffs}{
$\psi( \vec w^{(i)} ) =  \vec w^{(i)} $ with $\vec w^{(i)}$ frozen in the SGD steps
}\ElseIf{long-only basket payoffs}{
$$
\psi( \vec w^{(i)} ) =   \left(\lvert w^{(i)}_1\rvert,\ldots,\lvert w^{(i)}_d\rvert \right)
$$
}

Define the loss function $\text{MSE}(\phi, batch) =\widehat{\ex}_{batch}\left[(F(\x) -\theNN_\phi(\x))^2\right] + \zeta\| \theta\|_2$

Initialize the network parameters $\hat{\phi}$

 \For{$\text{epoch} = 1, \dots, E$}{
  \For{$\text{batch} \in B$}{
$\hat{\phi}\leftarrow \text{AdamStep}(learning\_rate, \beta_1, \beta_2, \text{MSE}(\hat{\phi}, batch) )$\\
(batched version of Algorithm 1 in \cite{KingmaB14})
        }
 $learning\_rate = 0.8\times learning\_rate$ if $epoch 
  \mod  300 = 0$
 }
\caption{Spanning networks trained by Adam optimizer}
\label{alg:es}
\end{algorithm}
} 

\FloatBarrier
\subsection{Spanning Metrics}
\FloatBarrier
We use two metrics to assess the distance between the target payoff $F$ and a predictor $\widetilde{F}\in\nn_{d,\then }$: mean squared error $\mathrm{MSE} = \hex\big[(F(\vec x) - \widetilde{F}(\vec x))^2\big]$ for the loss function, and  mean absolute error $\mathrm{MAE}  = \hex\big| F(\vec x) - \widetilde{F}(\vec x) \big|$ for reporting.  The choice of MAE for reporting is motivated by its ease of financial interpretation as average absolute dollar mismatch between the target payoff and the spanning portfolio.
We considered using MAE loss for training, which is known to be more robust against outliers, but ultimately retained MSE loss as we sample asset prices in compact sets which eliminates outsize values, and MSE is more stable for gradient calculation.  Figure 3 of the implementation details in the paper's github shows that MAE loss produces similar results.
    
Figure \ref{fig:cp_all} shows contour plots reporting the target payoff $F$, its approximation $\hat F \in\nn_{2,40 }$ and the pointwise absolute spanning error $| F(x_1,x_2)-\hat F(x_1,x_2) |$ for all call target payoffs in Table \ref{tab:paydes} (similar results for puts can be found on github).
 From visual inspection the predicted surface provided by the unrestricted neural network spanning strategy (NN) fits the target payoff reasonably well, except in areas where the target payoff is nondifferentiable. 
\begin{figure}[!h]
     \centering
          \begin{subfigure}[b]{1.\textwidth}
         \centering
         \includegraphics[width = 1\textwidth]{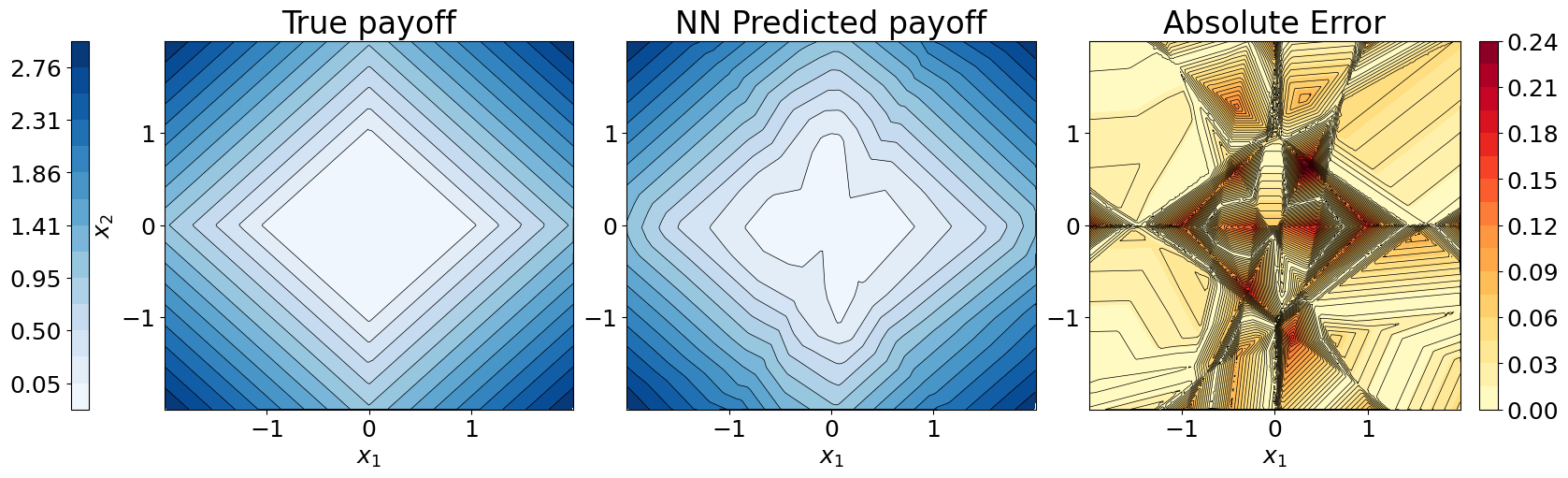}
         \caption{Dispersion call $F(x_1,x_2) = \big(|x_1| + |x_2|-1\big)^+$. MAE = 0.035 (3.5 cents per dollar of notional).}
     \end{subfigure}\vfill
     \begin{subfigure}[b]{1.\textwidth}
         \centering
         \includegraphics[width = 1\textwidth]{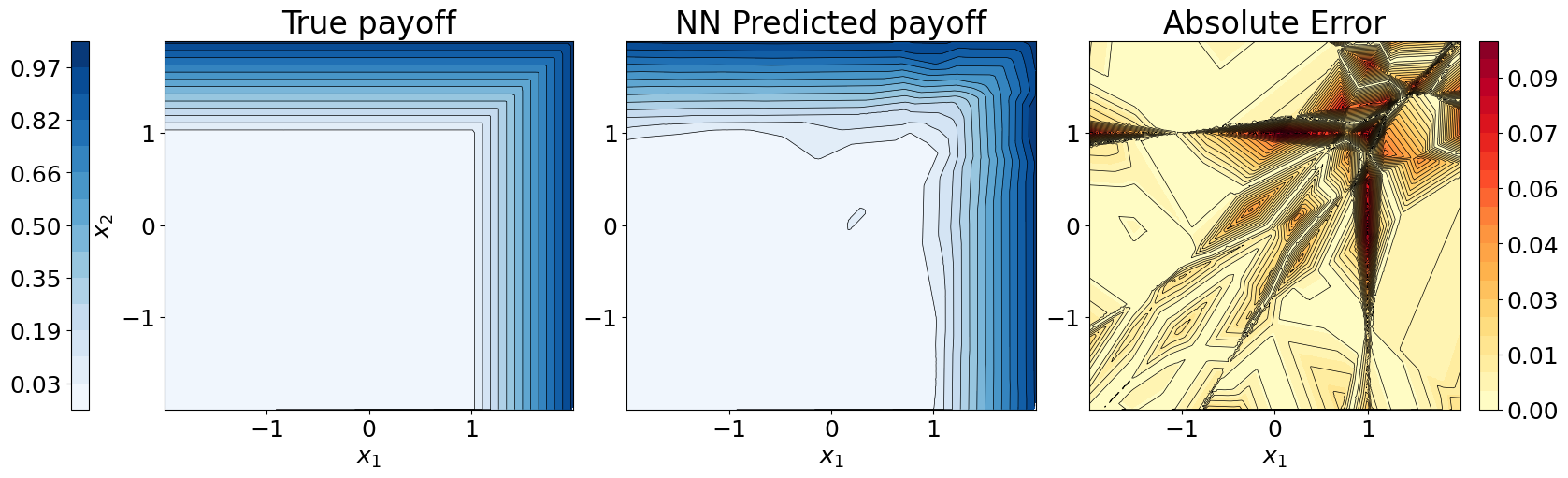}
         \caption{Best-of call $F(x_1,x_2) = \big(\max(x_1,x_2)-1\big)^+$. MAE = 0.011 (1.1 cents per dollar of notional). }
     \end{subfigure}
     \vfill
     \begin{subfigure}[b]{1.\textwidth}
         \centering
         \includegraphics[width = 1\textwidth]{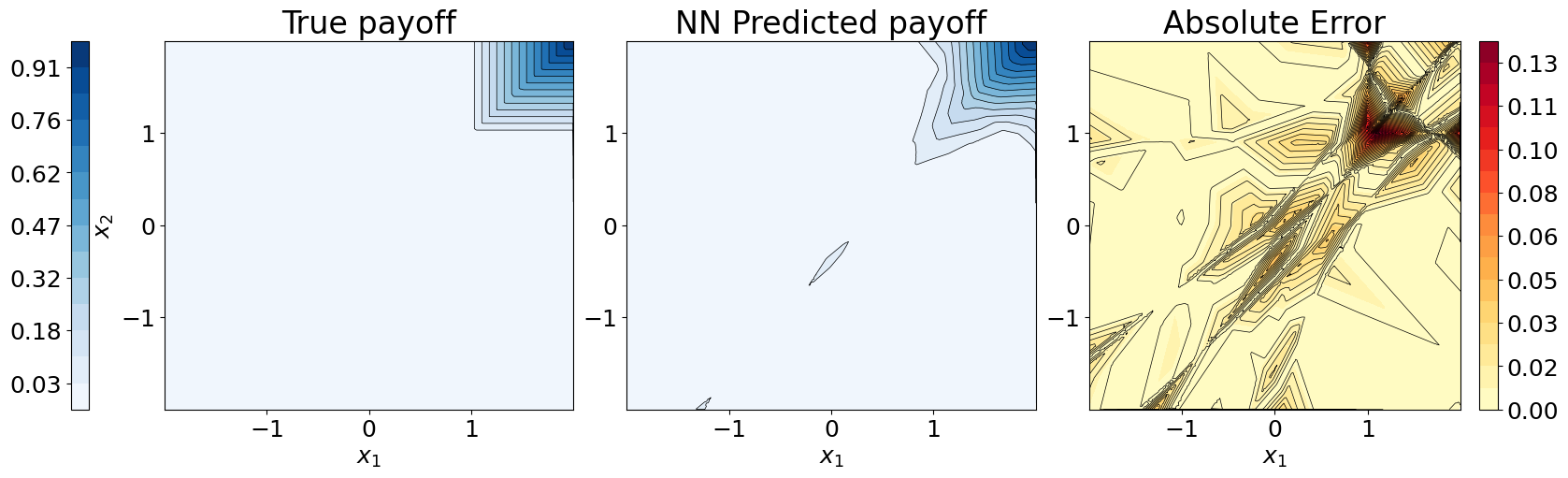}
         \caption{Worst of call $F(x_1,x_2) = \big(\min(x_1,x_2)-1\big)^+$. MAE = 0.008 (0.8 cents per dollar of notional).}
              \end{subfigure}
    \end{figure}
\begin{figure}[t]
\ContinuedFloat
     \begin{subfigure}[b]{1.\textwidth}
         \centering
         \includegraphics[width = 1\textwidth]{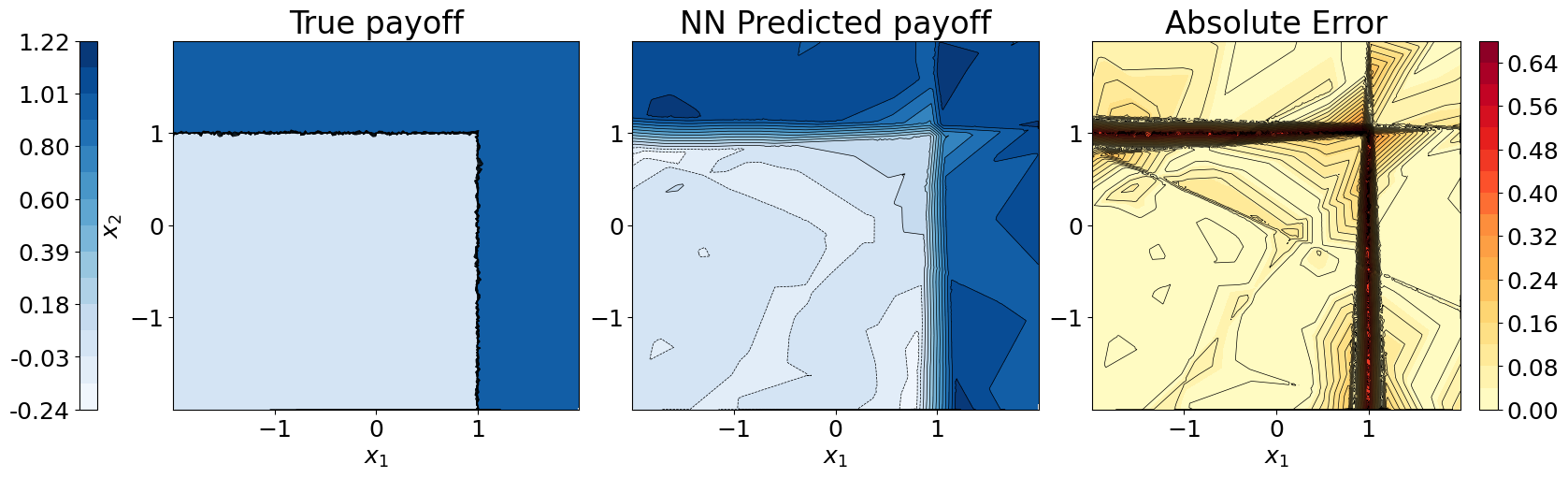}
         \caption{Best of binary call $F(x_1,x_2) = \mathbf{1}_{\max(x_1,x_2)>1}$. MAE = 0.064 (6.4 cents per dollar of notional).}
              \end{subfigure}
     \vfill
      \begin{subfigure}[b]{1.\textwidth}
         \centering
         \includegraphics[width = 1\textwidth]{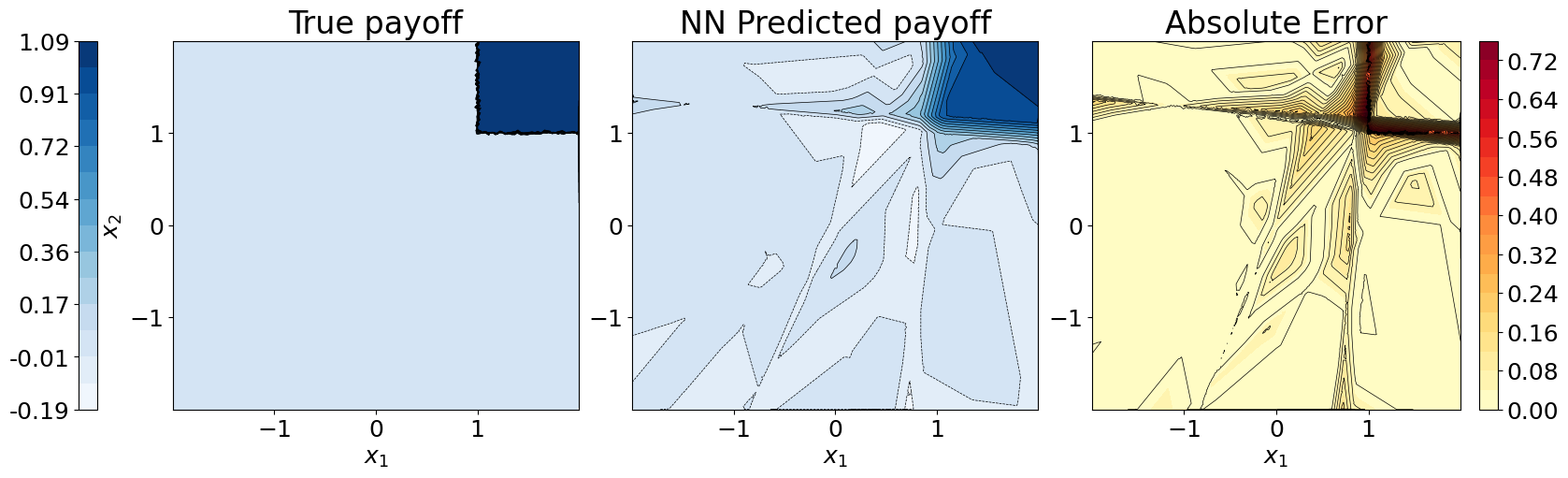}
         \caption{Worst of binary call $F(x_1,x_2) = \mathbf{1}_{\min(x_1,x_2)>1}$. MAE = 0.036 (3.6 cents per dollar of notional).}
               \end{subfigure}

    \caption{Contour plots and MAEs for one call target payoff of each kind in Table \ref{tab:paydes}:  target payoff (\textit{left}), NN prediction (\textit{center}) and absolute spanning error (\textit{right}).}
    \label{fig:cp_all}
\end{figure}
    
\FloatBarrier
\subsection{Limitations of Spanning with Single-Asset Payoffs}\label{singleAsset}
\begin{figure}[t!]
    \centering
    \includegraphics[scale=0.6]{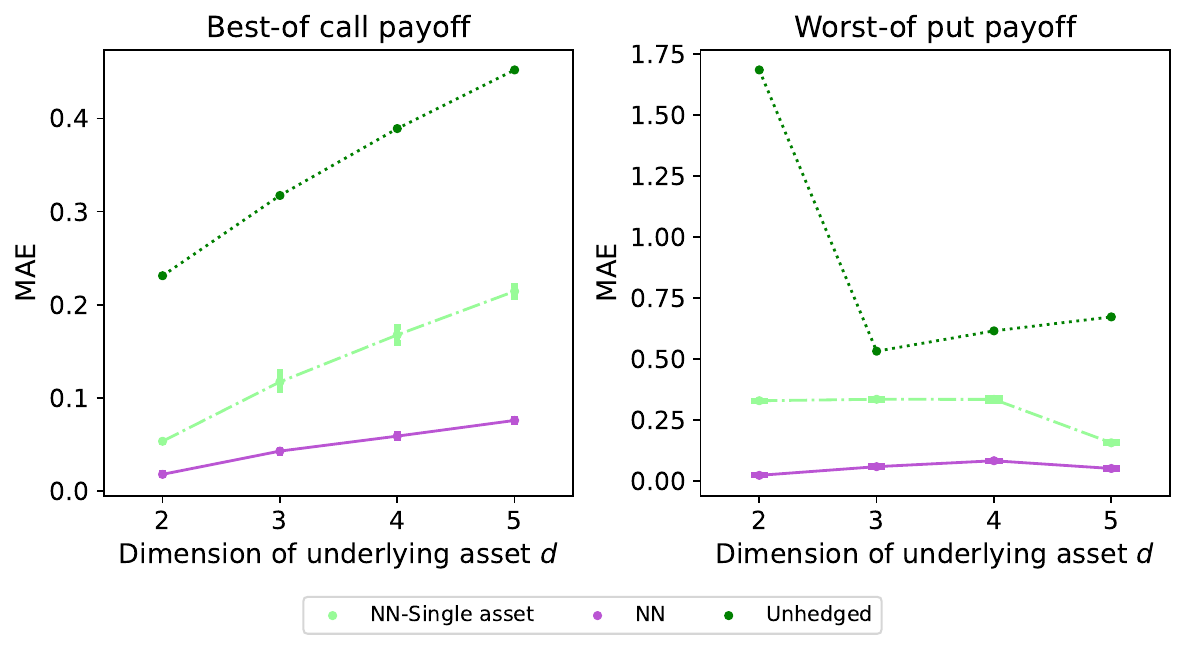}
    \caption{MAE comparison of the single-asset and unrestricted NN spanning strategies for $d=2,\dots,5$ assets: best-of call (\textit{left}) and worst-of put (\textit{right}).  Error bars are 95\% confidence intervals.}
    \label{fig:boxNNvsMono}
\end{figure}

\FloatBarrier
When spanning the target payoff $F(\vec x)$ with single-asset vanilla payoffs only, equations
\eqref{eq:net_unres}-\eqref{eq:loss}
become
\bel
    \min_{\alpha,\boldsymbol \mu,\boldsymbol \nu, \boldsymbol \eta,
    \vec k, \vec E} \widehat{\ex}\!\left[\left( F(\vec x) - \alpha - \boldsymbol \mu\cdot\vec x - \sum_{i=1}^{\then}   \nu_{i} \left(\eta_i(\vec{e}_i \cdot \vec x - k_{i})\right)^+\right)^2\right],
\eel
where $\vec{e}_i\in\{0,1\}^d$ is a ``one-hot'' vector with all coefficients equal to 0, except the coefficient corresponding to the selected asset which is equal to 1 (the index of which remains free for optimization), and $\vec E = [\vec e_1, \cdots, \vec e_{\then}]^\top \in\{0,1\}^{ \then \times d}$ is the corresponding matrix. We solve this optimization problem numerically by Adam stochastic gradient descent \citep{KingmaB14} with a restricted neural network architecture, for the best-of call and worst-of put payoffs on 2 to 5 underlying assets (see Table \ref{tab:paydes}, page \pageref{tab:paydes}).  Single-asset spanning does provide some risk reduction compared to the unhedged case for which all parameters are zero and MAE is the payoff average absolute value. However, Figure \ref{fig:boxNNvsMono} shows that the spanning error of this strategy is substantially higher than that of our core unrestricted NN approach \eqref{eq:loss}.  Mathematically, this is hardly surprising given that single-asset option spanning attempts to reproduce a ``joint distribution'' (the best-of or worst-of payoff) with ``marginal distributions'' only (the single-asset payoffs). 
\FloatBarrier
\subsection{Limitations of Spanning with Predetermined Basket Payoffs }\label{sec:limit_LS}

When spanning the target payoff $F(\vec x)$ with predetermined basket option payoffs, the strikes $k_i$, call/put selectors $\eta_i$,  and basket weights $\vec w^{(i)}$ of each basket payoff $1 \leq i \leq \then $ are fixed and the spanning problem takes the simpler form
\begin{align}
    \min_{\alpha,\boldsymbol \mu,\boldsymbol \nu} \widehat{\ex}\left[\left(F(\vec x) - \alpha - \boldsymbol \mu\cdot\vec x - \sum_{i=1}^{\then} \nu_i \left(\eta_i(\vec w^{(i)}\cdot \vec x - k_i)\right)^+\right)^2\right],
    \label{eq: opti_linear}
\end{align} 
which is a classic linear least-squares regression problem.  Due to parameter redundancy we choose unit strikes $k_i = 1$, and we also set selectors $\eta_i=1$. The explicit solution to \eqref{eq: opti_linear} is therefore given by the regression coefficients $\widehat{\boldsymbol \beta} = \left[{\boldsymbol{\mu} \atop \boldsymbol{\nu}}\right] \in \R^{d+\then }$ and constant $\hat{\alpha} \in \R$ with
\begin{align*}
        \widehat{\boldsymbol \beta} =  \widehat{\mathrm{Var}} \left(\vec z\right)^{-1} \widehat{\mathrm{Cov}} \left(\vec z, F(\vec x)\right), \qqq\hat{\alpha} &= \widehat{\ex}\left[F(\vec x)\right] - \widehat{\boldsymbol \beta}^\top \widehat{\ex}\left[\vec z \right],
\end{align*}
where $\vec z=\left[x_1,\ldots,x_d,(\vec w^{(1)} \cdot \vec x -1)^+,\ldots,(\vec w^{(\then)} \cdot \vec x -1)^+\right]^\top \in \R^{d+\then }$ is the vector of explanatory variables (underlying assets and basket payoffs), $\widehat{\mathrm{Cov}}$ is the sample covariance operator, and $\widehat{\mathrm{Var}} \left(\vec z\right)$ is the sample covariance matrix which must be nonsingular.  In practice, we found this calculation to be numerically unstable due to conditioning issues: basket payoffs tend to overlap, making the columns of $\vec z$ loosely dependent, particularly for large $\then$.  This issue is further compounded when the sampling of $\vec x$ (and thus $\vec z$) is sparse. To circumvent this practical difficulty, we used singular value decomposition (SVD) \citep[Theorems 2.5.2 and 5.5.1]{golub2013matrix}. 
\begin{figure}[!tbp]
\begin{subfigure}{\textwidth}
    \centering
    \includegraphics[scale=0.4]{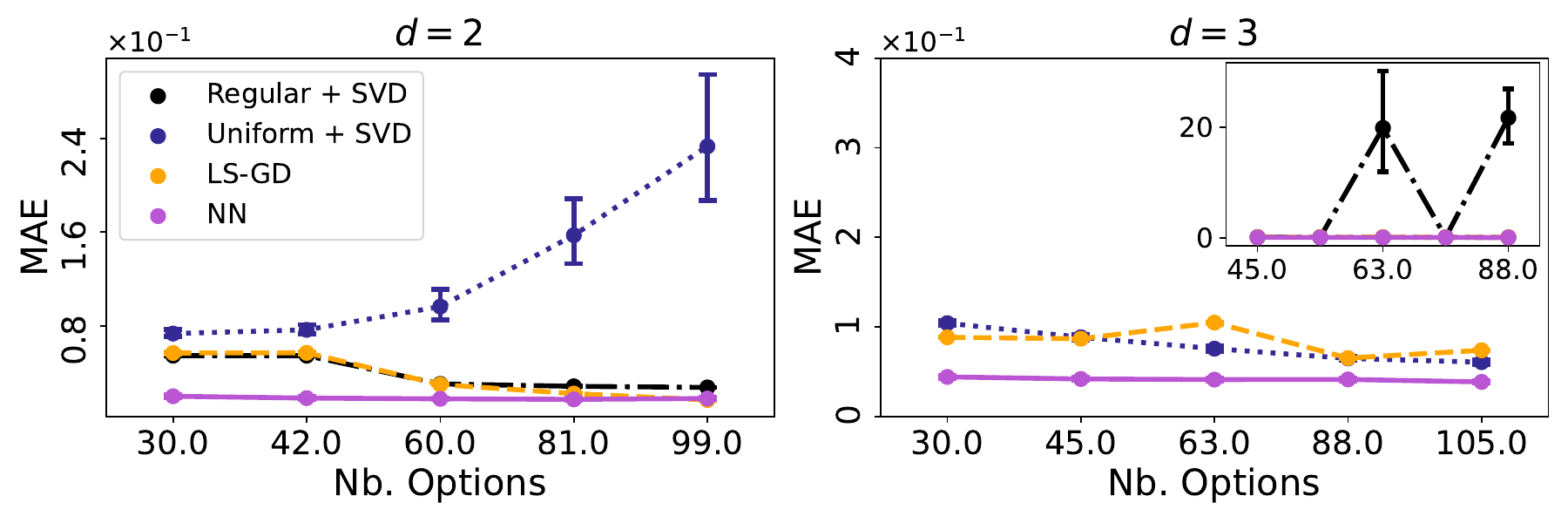}
\end{subfigure}
\begin{subfigure}{\textwidth}
    \centering
    \includegraphics[scale=0.4]{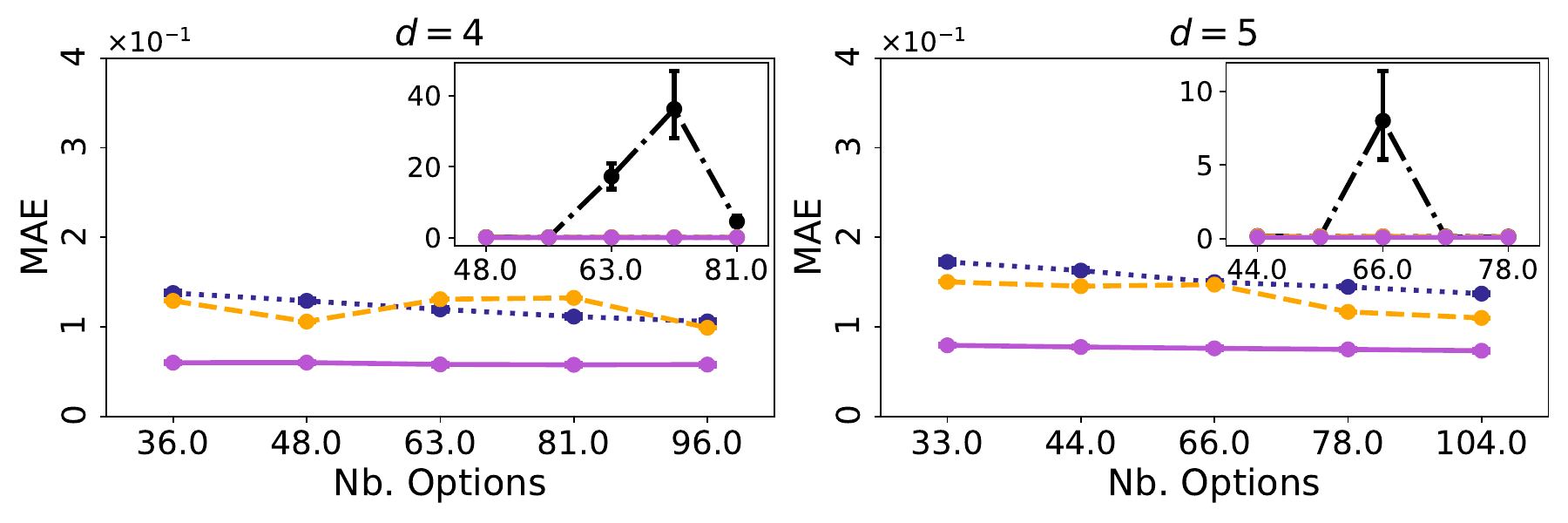}
\end{subfigure}
\caption{MAE with 95\% confidence intervals for the best-of call on $d=2$ to 5 underlying assets for methods (a) to (d), as a function of the number of basket options $\then$.  Large errors obtained with (a) are reported in inside panels for $d=3,4,5$ for readability.}
\label{fig:LS_BOC}
\end{figure}

Figure \ref{fig:LS_BOC} reports the spanning error for the best-of call on 2 to 5 underlying assets, together with 95\% confidence intervals over 30 different runs
that were obtained using the following methods:  
\begin{enumerate}[label=(\alph*),leftmargin=*,topsep=0pt,itemsep=1pt,partopsep=0pt,parsep=0pt,listparindent=0pt]
    \item SVD with regular grid sampling of basket weights $\vec w$ (regular+SVD);
    \item SVD with i.i.d.\  uniform random sampling of $\vec w$ (uniform+SVD);
    \item stochastic gradient descent with regular grid sampling of $\vec w$ (LS-GD);
    \item unrestricted neural network approach with free $\vec w$ initialized with i.i.d.\  uniform random sampling (NN).
\end{enumerate}
Each run is a new training routine with the same fixed weights $\vec w$ for methods (a) and (c), and with new random weights $\vec w$ for methods (b) and (d). Each SGD training (NN learning) takes a few seconds, while SVD regressions take less than a second. We can see that all fixed-weights methods (a) to (c) resulted in substantial spanning errors compared to (d) unrestricted NN, except perhaps in dimension $d = 2$ where the error magnitude is smaller. Remarkably enough, increasing the number $\then$ of basket payoffs does not materially reduce the spanning error for any method (a) to (d). This suggests that only a limited number of basket payoffs may be needed to obtain a satisfactory hedge of the best-of call, in contrast to the combinatorial issue identified in Remark \ref{r:dc-sol-combin}.  Finally, SVD methods (a) and (b) often perform poorly: when sampling $\vec w$ along a regular grid, the spanning error (a) even ``exploded'' in dimensions $d=3,4,5$ due to ill-conditioning of the design matrix generated by observations of $\vec z$, as shown in inner panels.

It is worth noting that, from a neural network architecture perspective, the spanning problem \eqref{eq: opti_linear} corresponds to a particular category of extreme learning machines or ELMs \citep*{huang2004extreme,huang2006extreme}.  ELMs are known to be universal approximators, but they typically require a very large number $\then$ of hidden units to achieve satisfactory performance.  In the context of payoff spanning, this suggests that only a large number $\then$ of predetermined basket payoffs in methods (a) to (c) would be able to approach the performance of the unrestricted NN method (d), which would be impractical for hedge execution.

\subsection{Summary of Restricted and Unrestricted Spanning Results}\label{subsec:dim5}
\FloatBarrier
 Figure \ref{fig:boxall5_20_50} reports the performance of four spanning strategies in dimension $d = 5, 20$ and $50$ using 50 different network initializations: (i) Adam training of single asset portfolio (NN-Single asset) as explained in Section \ref{singleAsset}, (ii) least-squares approach with gradient descent (LS-GD) as explained in Section \ref{sec:limit_LS}, (iii) Adam training of long only basket portfolio (NN-Long only), and (iv) unrestricted neural network with Adam training (NN). 
 We can see that the spanning portfolios suggested by the neural network give satisfying MAE results in most cases. Unrestricted NN outperforms all other strategies in terms of MAE and standard deviation: 
 MAE increases with the underlying asset dimension $d$ but remains fairly low. The low standard deviation figures also signals that unrestricted NN is the most stable among the four strategies. 
 \begin{figure}[!b]
\begin{subfigure}{\textwidth}
    \centering
    \hspace*{-1.3cm}\includegraphics[scale=0.55]{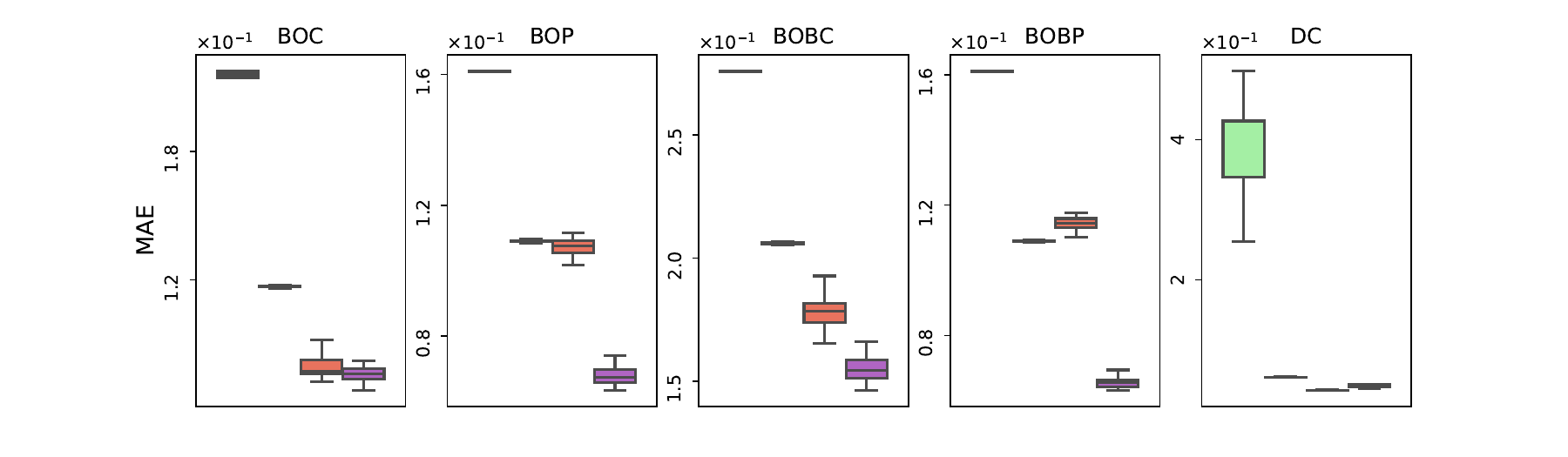}\vspace*{-0.3cm}
    \centering
    \hspace*{-1.3cm}\includegraphics[scale=0.55]{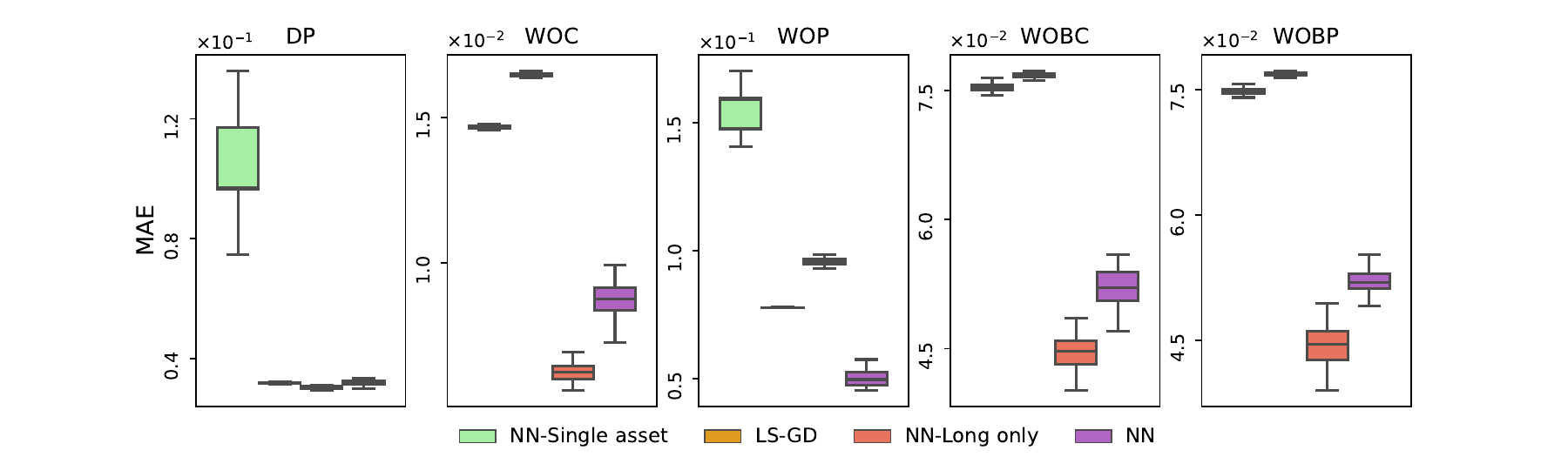}
    \caption{$d=5$ and $\then=78$.}\label{fig:boxall5}
\end{subfigure}
\end{figure}
\begin{figure}[!t]
\ContinuedFloat
\begin{subfigure}{\textwidth}
    \centering
    \hspace*{-1.3cm}\includegraphics[scale=0.55]{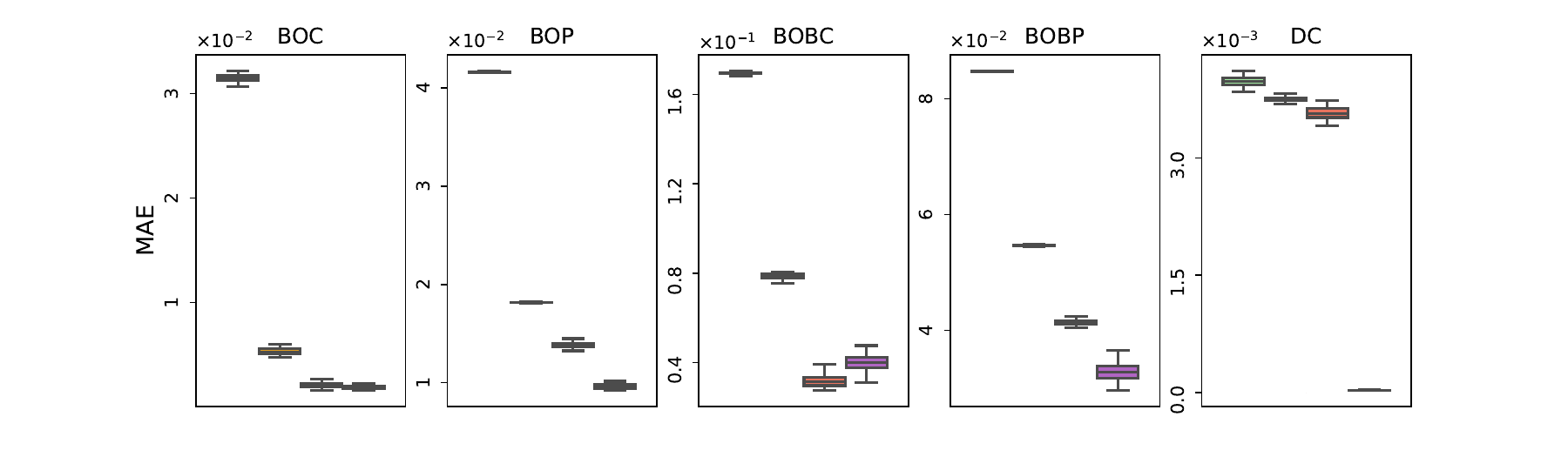}\vspace*{-0.3cm}
    \centering
    \hspace*{-1.3cm}\includegraphics[scale=0.55]{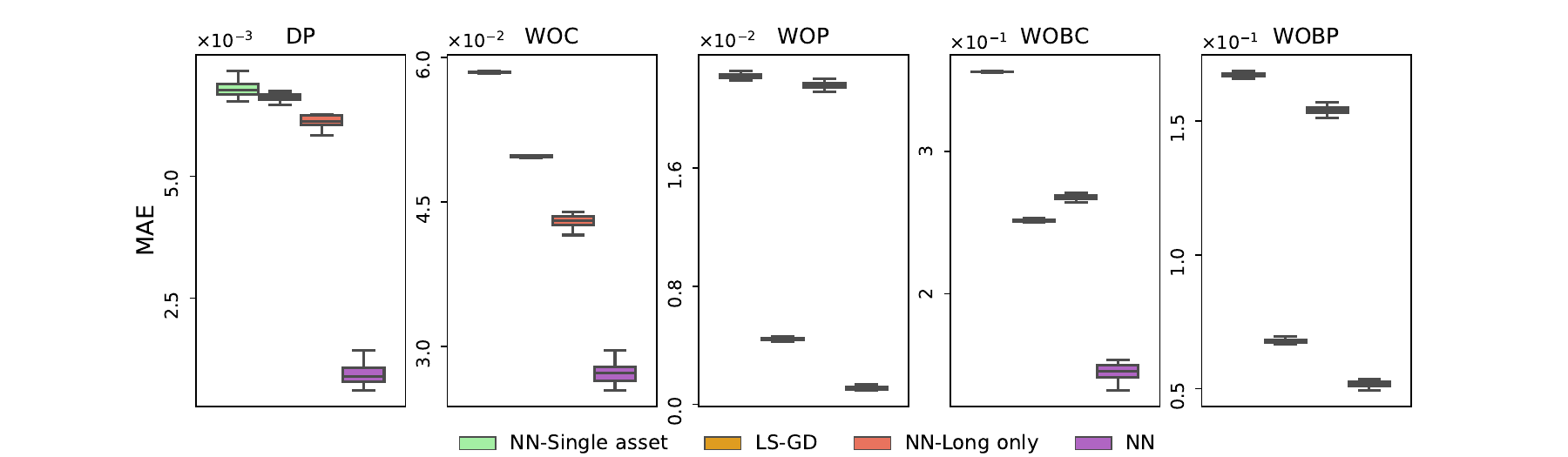}
    \caption{$d=20$ and $\then=410$.}\label{fig:boxall20}
\end{subfigure}
\begin{subfigure}{\textwidth}
    \centering
    \hspace*{-1.3cm}\includegraphics[scale=0.55]{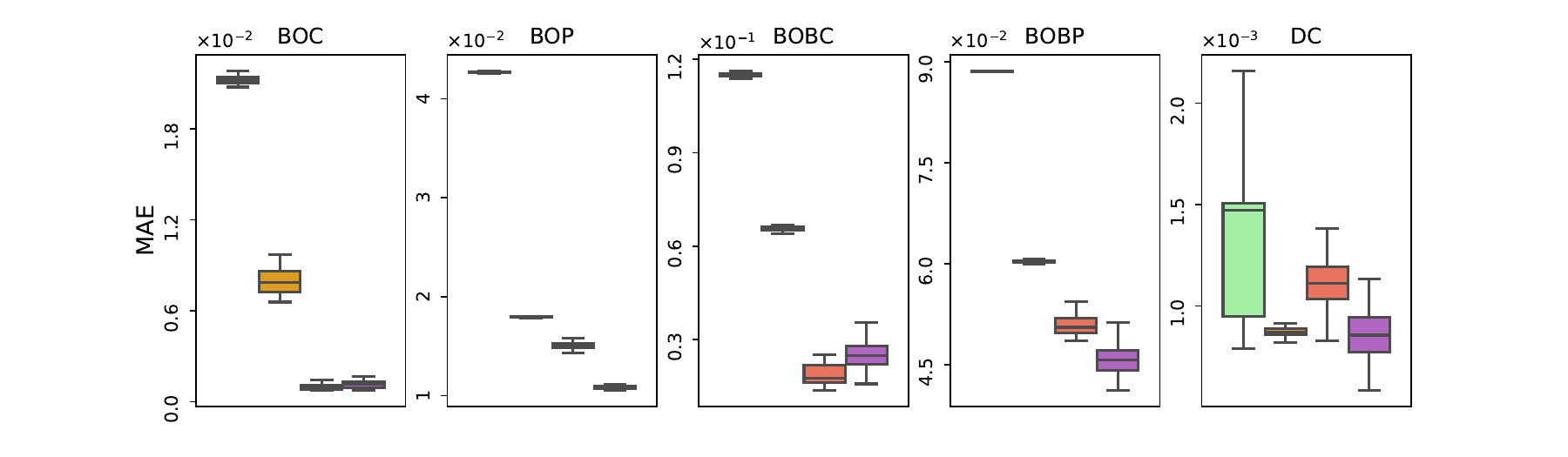}\vspace*{-0.3cm}
    \centering
    \hspace*{-1.3cm}\includegraphics[scale=0.55]{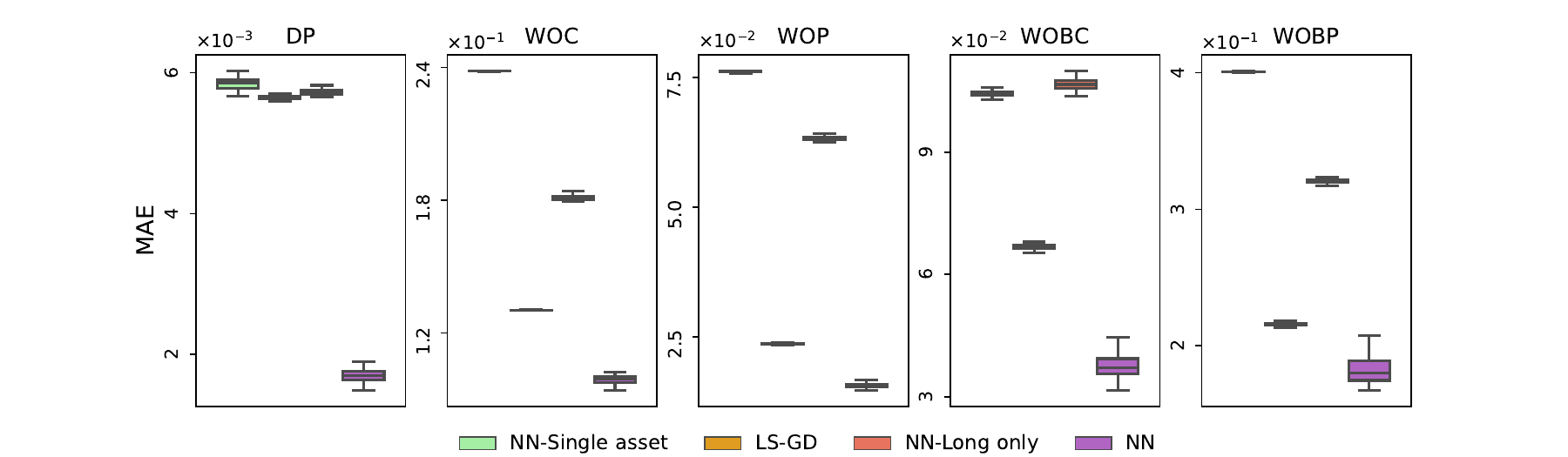}
    \caption{$d=50$ and $\then=808$.}\label{fig:boxall50}
\end{subfigure}
\caption{Average MAE and 95\% error bars over 50 runs by spanning strategy and target payoff in dimension $d=5,20$ and $50$.}\label{fig:boxall5_20_50}
\end{figure}
 \FloatBarrier
 \subsection{Stability Issues}
In linear regression models, the loss function is convex and thus easier to optimize.  In contrast, it is nonconvex in neural network parameterizations such as \eqref{eq:net_unres}-\eqref{eq:loss}: optimization is more difficult, algorithms such as Adam method typically converge to different local minima for different initializations, yet loss values usually remain small \citep*{choromanska2015loss}.  We observed this phenomenon in our study: the optimal neural network parameters vary with each training, but we obtained persistently small and stable errors.  Financially, this means that spanning performance remains strong but the particular static hedge identified based on a particular training set does not have a stable interpretative meaning with respect to the target option.
 \begin{figure}[!b]
\centering
\begin{subfigure}{\textwidth}
    \centering
    \includegraphics[width = 0.5\textwidth]{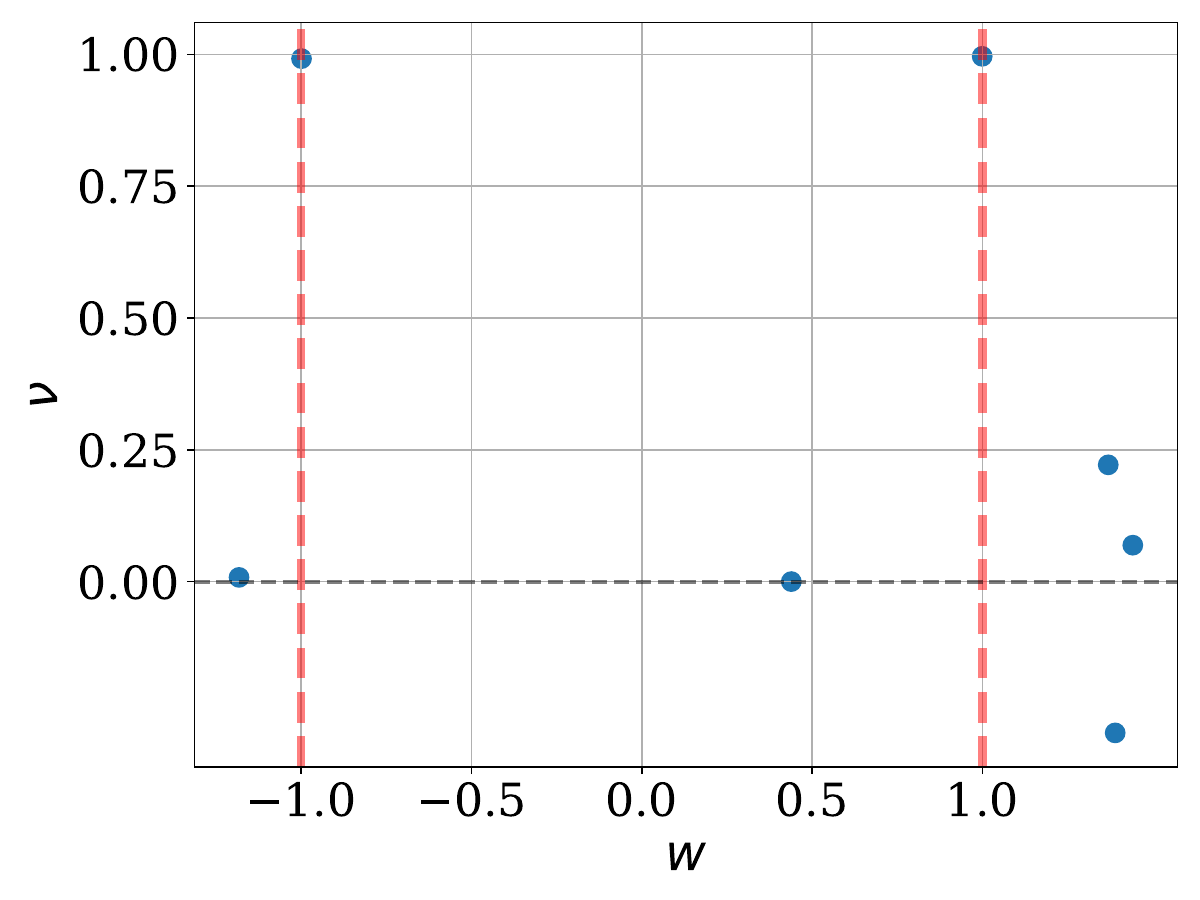}%
    \includegraphics[width = 0.5\textwidth]{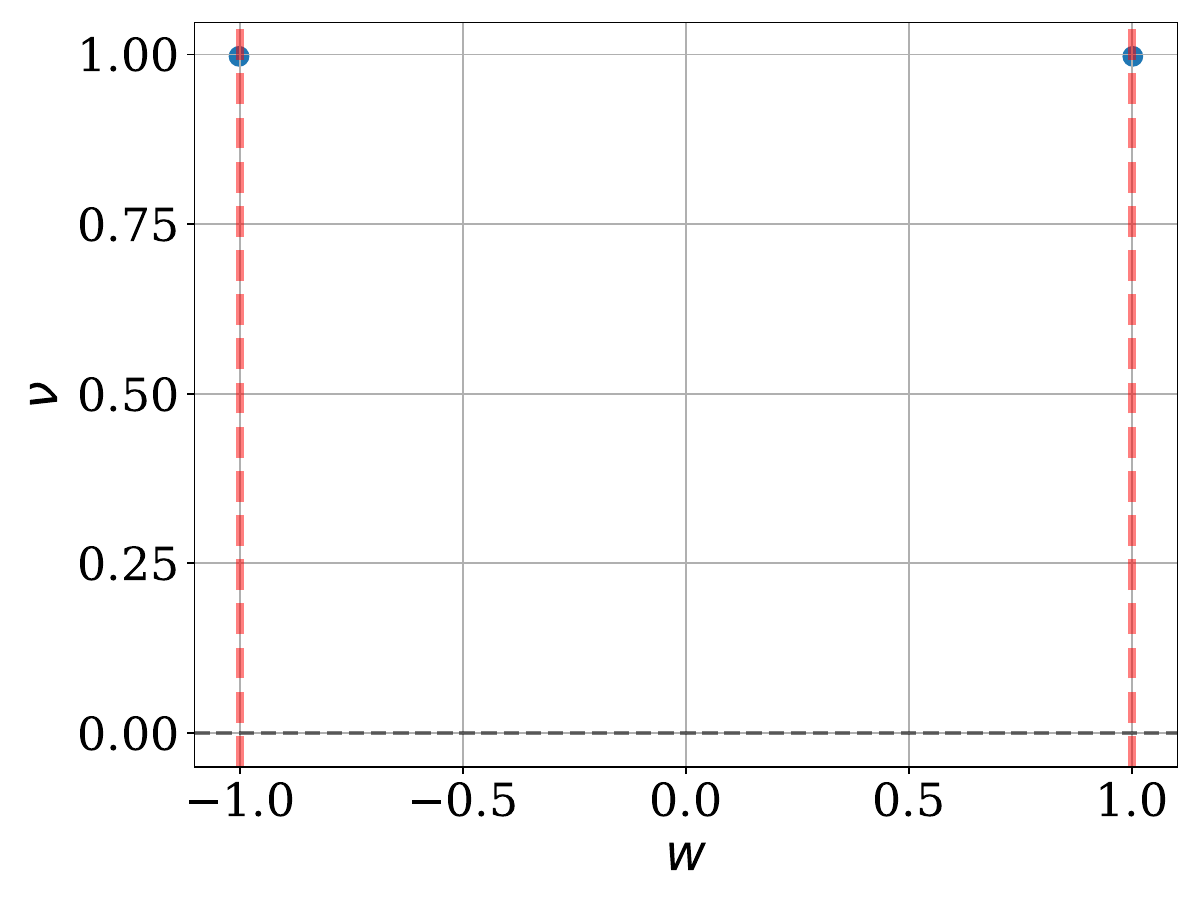}
\end{subfigure}
\caption{Scatter plots of the optimal basket call quantity $\nu_i$ against the optimal basket weight $w'^{(i)}$ (see~\eqref{e:changek}) predicted by \eqref{eq:net_unres}-\eqref{eq:loss} for the single-asset dispersion call payoff $F(x) = (|x|-1)^+$ when $\then=10$ and the training asset price $x$ is sampled in $[-2,2]$. (\textit{Left}) without regularization; (\textit{Right}) with regularization.}\label{fig:theo1}
\end{figure}
\begin{figure}[!t]
\centering
    \includegraphics[width = 0.5\textwidth]{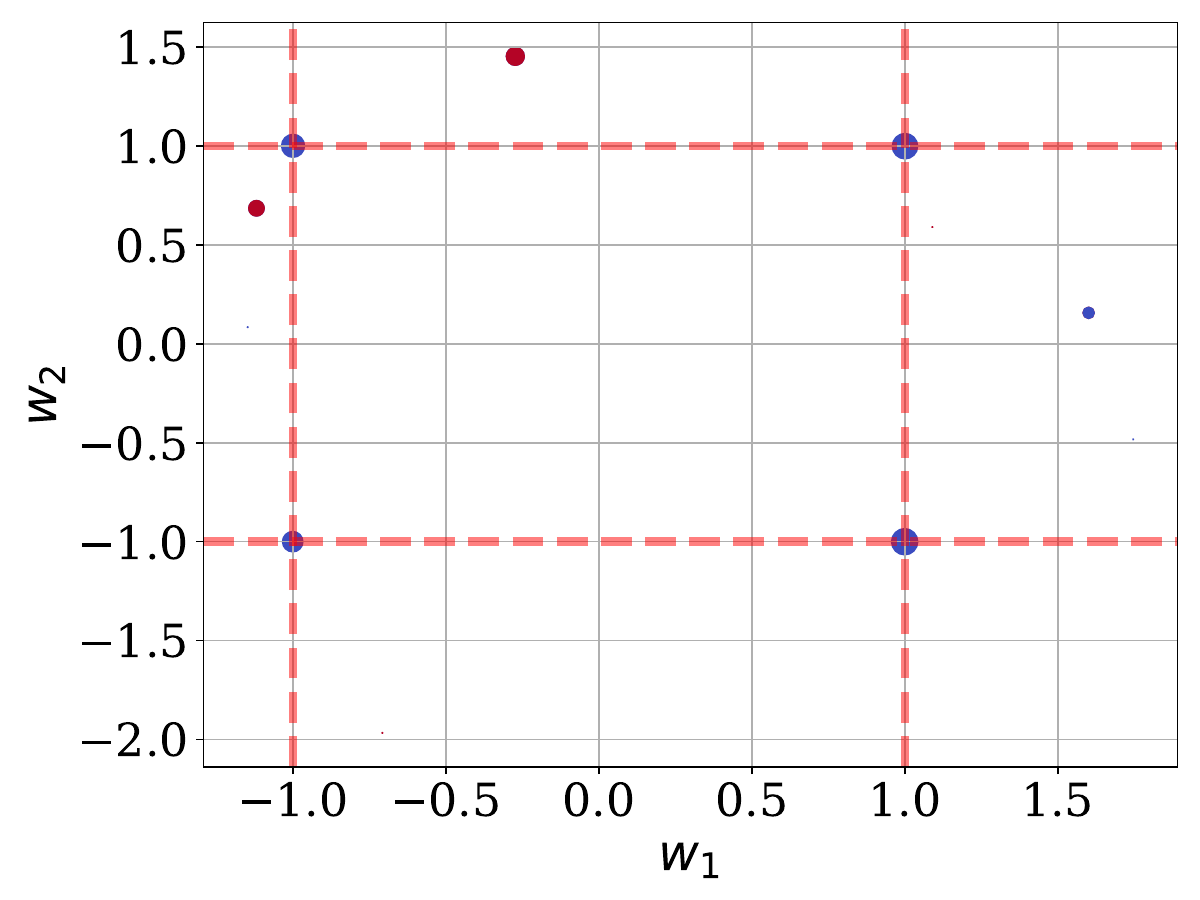}%
    \includegraphics[width = 0.5\textwidth]{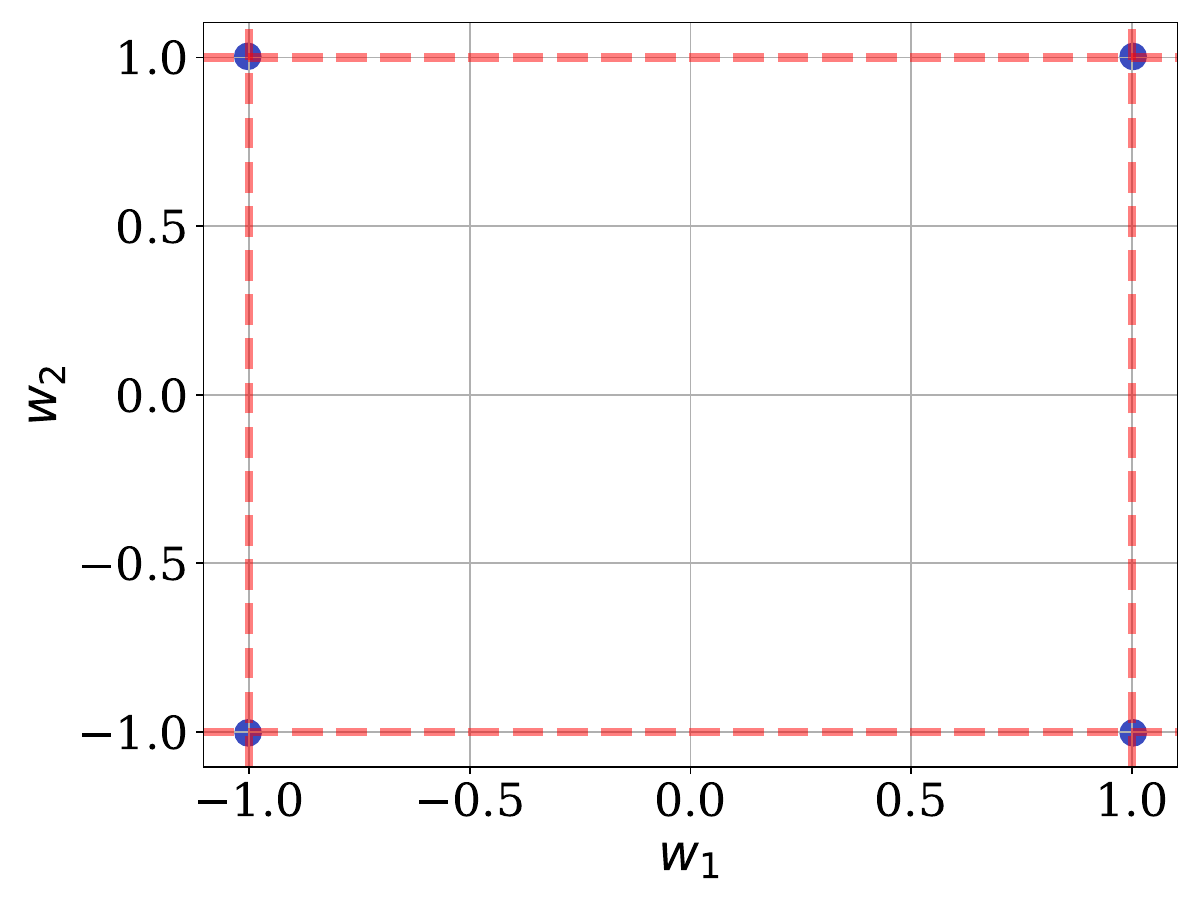}
\caption{Scatter plots of the 
optimal basket call weights $(w'^{(i)}_1,w'^{(i)}_2)$ (see.~\eqref{e:changek}) predicted by
\eqref{eq:net_unres}-\eqref{eq:loss} for the two-asset dispersion call with payoff $F(x_1,x_2) = (|x_1|+|x_2|-1)^+$ when $\then=50$. The training asset prices $(x_1,x_2)$ are sampled in $[-1,1]^2$. The blue and red points represent respectively long ($\nu_i>0$) and short ($\nu_i<0$) positions while point sizes reflect absolute quantities $|\nu_i|$. (\textit{Left}) without regularization; (\textit{Right}) with regularization.} 
\label{fig:theo2}
\end{figure}

Figures \ref{fig:theo1} and \ref{fig:theo2} show how 
the optimizer solution for the dispersion call approaches the theoretical solution that we derived in Proposition \ref{ex:1}
in dimensions $d=1,2$.  Our results are indeed remarkably consistent with the accumulation of basket call quantities $\nu_i$ predicted by theory.  In dimension $d=1$ our numerical results perfectly match the exact solution
\eqref{EQ:PROP31} (see also Remark \ref{ex:expldisp})
after we process the learning with an $\ell^2$ regularization (Figure \ref{fig:theo1}, right panel).  In dimension $d=2$ our numerical results after regularization are also consistent with the accumulation of basket calls around discontinuity points $w_1 = \pm1$ and $w_2 = \pm1$ predicted by the theoretical formulas \eqref{N2} and \eqref{e:dc2} 
(Figure \ref{fig:theo2}, right panel).

\FloatBarrier
\section{Conclusion and Perspectives}\label{sec:conclusion}

Identifying static hedges for exotic option payoffs is important for both theory and practice.  Theorem \ref{t:abs-homogeneous} expands on existing continuum replication theory of European option payoffs with vanilla options to formulate a general and rigorous solution of the continuum spanning problem.  As an application, Proposition \ref{ex:1} derives the continuum solution replicating the industry $\ell^1$ dispersion call.  In addition to the derivation of explicit solutions  for other absolutely homogeneous payoffs such as best-of and worst-of options, other formulations of the continuum spanning problem are open for future research: for example, letting $\nu$ in \eqref{eq:repli-strong} depend on $k$, or integrating the spanning portfolio over $k$, may relax the absolute homogeneity restriction to yield solutions for non-homogeneous payoffs.

Leveraging the parallel between vanilla basket calls and ReLU functions, we examined how neural networks can be used to numerically solve the corresponding discrete spanning problem and identify finite static hedges, in comparison to other restricted spanning strategies and optimization schemes such as least-squares SVD.  Our empirical study suggests that our unrestricted NN approach yields superior results in terms of static hedging error for any dimension 2 to 50.  This approach may be of practical interest for the exotic derivatives industry, particularly if combined with delta-hedging of the residual payoff mismatch, which could be investigated in future research.

\appendix

\section{Cauchy Integral}
\label{a:cauchy}

 For a function $\varphi\in L^1(\R)$ and a constant $c\in\R$, the singular integral
\begin{align*}
    \int_{-\infty}^{+\infty} \frac{\varphi(w)}{w-c} \d w = \lim_{\alpha \rightarrow 0+} \int_{-\infty}^{c-\alpha} \frac{\varphi(w)}{w-c} \d w + \lim_{\beta \rightarrow 0+} \int_{c+\beta}^{+\infty} \frac{\varphi(w)}{w-c} \d w
\end{align*}
may not exist when $\alpha$ and $\beta$ tend to $0\!+$ independently . However, the integral exists restricted to the diagonal $\alpha=\beta$ \citep[Section 1.5]{estrada2000singular}.
\begin{definition} \label{d:cpvint}
 The diagonal limit
\beql{eq:def_cpv}
    &\lim_{\epsilon \rightarrow 0+}\left(\int_{-\infty}^{c-\epsilon} \frac{\varphi(w)}{w-c} \d w +  \int_{c+\epsilon}^{+\infty} \frac{\varphi(w)}{w-c} \d w \right)=\lim_{\epsilon \rightarrow 0+}\int_{\epsilon<|w-c|} \frac{\varphi(w)}{w-c} \d w \in \R,
\eeql
denoted by $\dashint_{-\infty}^{+\infty}\frac{d w}{w-c} \varphi(w)$,
is called the Cauchy principal value 
integral of $\frac{\varphi(w)}{w-c} $ against $\d w$.
\end{definition}

\begin{definition}\label{d:ppv}
{\rm \textbf{(i)}} The Hilbert transform (up to a multiplicative factor $\pi$, see\ \citet[Section 3.1 page 83]{king_2009}) \beql{e:ht}& \intc \frac{\d w}{w-c} \varphi(w)\sp\ \varphi\in\S(\R),
\eeql
is a distribution on $\S(\R)$ \citep[Proposition 6 page 136]{reed1980functional}, 
which we denote by $ \frac{\d w}{w  - c } $ in abstract distribution notation.
{\rm\hfill\break \textbf{(ii)}} The linear form
\begin{align*}
 \intcc_{\R^2} \Big(\frac{\d w_1 }{ w_1-c_1 } \otimes  \frac{  \d w_2}{ w_2-c_2 }\Big)   \varphi(w_1,w_2):= \intc \frac{\d w_1}{w_1-c_1}     \Big(\intc \frac{\d w_2}{w_1-c_2}  \varphi(w_1,w_2) \Big)\sp \varphi   \in\S(\R^2)
 \end{align*}  
is a well defined distribution on  $\S(\R^2)$, known as the tensor (or direct) product of the distributions $\frac{\d w_1}{w_1-c_1}$ and $\frac{\d w_2}{w_2-c_2}$
on $\S(\R^2)$, which is commutative \citep[page 185]{kanwal}.
{\rm\hfill\break \textbf{(iii)}} The previous construction can be iterated to define the distribution $\prod_{j\in1..q} \frac{\d w_k}{w_j - c_j}$ for any $q\ge 1$ and $(c_j)\in\R^q$ as
\beql{e:ppv}
& 
\intcc_{\R^q} \Big(\bigotimes_{j\in 1\,..\,q} \frac{\d w_j}{w_j-c_j} \Big)
\varphi(\w)
\sp\ \varphi\in\S(\R^q).
\eeql 
We may also define, in line with the distributivity principle of commutative products, 
\beql{e:ppvprod}
& 
 \intcc_{\R^q} \Big(\bigotimes_{j\in 1\,..\,q}\big( \frac{\d w_j}{w_j-c_j} +\frac{\d w_j}{w_j+c_j} \big) \Big)
\varphi(\w) := \sum_{(\varepsilon_j )\in\{-1, 1\}^{q}}
\intcc_{\R^q} \Big(\bigotimes_{j\in 1\,..\,q} \frac{\d w_j}{w_j+\varepsilon_j c_j} \Big)
\varphi(\w),
\eeql 
as a distribution on  $\S(\R^q)$, abstractly denoted $\bigotimes_{j\in 1\,..\,q}\Big(   \frac {\d w_j}{w_j-c_j} +\frac {\d w_j}{w_j+c_j}  \Big)$.
\end{definition}

\begin{remark}
In view of the above definitions and properties, we may commute Cauchy integrals to write
\beql{e:aggregpvintc}
&\intcc_{\R^{q}}\Big(\bigotimes_{j\, \in\,  1\,..\,q}   \big(\frac{\d w_{j}}{ w_{j}- c_{j} } + \frac{\d w_{j}}{ w_{j} +c_{j} }\big)\Big) \varphi ( \w) =\\
&\qqq \intcc_{\R^{q-1}} \Big( \bigotimes_{j\, \in\, 2 \,..\,q } \big(\frac{\d w_{j}}{w_{j} - c_{j}} + \frac{\d w_{j}}{w_{j} + c_{j}}\big)\Big)   \intcc_\R  \big(\frac{\d w_{1}}{w_{1}-c_1} + \frac{\d w_{1}}{ w_{1}+ c_1} \big) \varphi ( \w).
\\
\eeql
This is in line with the definition of the tensor product of the distribution $(\frac{\d w_{1}}{w_{1}-c_1} + \frac{\d w_{1}}{ w_{1}+c_1} )$ with $\bigotimes_{j\, \in\, 2 \,..\,q } \left(\frac{\d w_{j}}{w_{j} - c_{j}} + \frac{\d w_{j}}{w_{j} + c_{j}}\right)$, equal to $\bigotimes_{j\, \in\,  1\,..\,q}  \left(\frac{\d w_{j}}{w_{j} - c_{j}} + \frac{\d w_{j}}{w_{j} + c_{j}}\right)$ \citep[page 185]{kanwal}; in bracket notation,
{\def\imath{\jmath}
\bel
 &\Bigg\langle\bigotimes_{j\, \in\,  1\,..\,q}  \left(\frac{\d w_{j}}{w_{j} - c_{j}} + \frac{\d w_{j}}{w_{j} + c_{j}}\right), \varphi ( \w) \Bigg\rangle_{w_{\imath}; \imath \,\in\, 1\,..\, q}=\\
&\qqq \Bigg \langle   \bigotimes_{j\, \in\, 2 \,..\,q } \left(\frac{\d w_{j}}{w_{j} - c_{j}} + \frac{\d w_{j}}{w_{j} + c_{j}}\right), \left\langle  \frac{\d w_{1}}{w_{1}-c_1} + \frac{\d w_{1}}{ w_{1}+c_1} 
,  \varphi ( \w) \right\rangle_{w_{1}}\Bigg\rangle_{w_{\imath}; \imath \,\in\, 2\,..\, q}.
\eel}\end{remark}

The following technical lemmas establish properties related to Cauchy integrals that will be used in Section \ref{s:disp}.

\begin{lemma}\label{l:pvcom} For any functions $\varphi_j \in \S(\R) $, $j =1,\dots,q$, there exists a constant $C > 0$ such that
\[
 \left| \int_{\epsilon<|w_2-w_1|}  \frac{\d w_2 }{w_2-w_1}  \Big( \dots \int_{\epsilon<|w_q-w_1|} \frac{\d w_q }{w_q-w_1}  \prod_{j = 1}^q \varphi_j(w_j)  \Big)  
\right| \leq C |\varphi_{1}(w_1)| \sp   w_1 \in\R , 0<\epsilon < 1,
\]
where the constant $C$ may depend on $\varphi$ but not on $w_1$ or $\epsilon$.
\end{lemma}
\proof Factoring $\varphi_1(w_1)$ out and separating integrals,
\beql{e:pvp}
&\int_{\epsilon<|w_2-w_1|}  \frac{\d w_2 }{w_2-w_1} \Big( \dots \int_{\epsilon<|w_q-w_1|} \frac{\d w_q }{w_q-w_1} \prod_{j = 1}^q \varphi_j(w_j) \Big)\\&\qqq
= \varphi_1(w_1) \prod_{j = 2}^q\int_{\epsilon<|w_j-w_1|} \d w_j \frac{\varphi_j(w_j)}{w_j-w_1} .  
\eeql
For $j = 2,\dots, q$,
\bel{}
 &\int_{\epsilon<|w_j-w_1|}  \frac{\d w_j}{w_j-w_1}\varphi_j(w_j) = \int_{\epsilon<|w_j-w_1|<1}  \frac{\d w_j}{w_j-w_1}\varphi_j(w_j)  + \int_{|w_j-w_1|\geq1} \frac{\d w_j}{w_j-w_1}\varphi_j(w_j)\\
 &\qqq=\int_{\epsilon<|w_j-w_1|<1}  \frac{\d w_j}{w_j-w_1} \big(\varphi_j(w_j) - \varphi_j(w_1)\big)  + \int_{|w_j-w_1|\geq1}  \frac{\d w_j}{w_j-w_1}\varphi_j(w_j),
\eel
where we used $ \varphi_j(w_1)\int_{\epsilon<|w_j-w_1|<1}  \frac{\d w_j}{w_j-w_1} = 0 $ in the last step (as integral of an odd function over a symmetric domain). Since $\varphi_j \in \S(\R)$, $\varphi'_j$ is bounded and by the mean value inequality, $\norm{ \varphi_j(w_j) - \varphi_j(w_1)} \leq |w_j - w_1|   \sup |\varphi_j'|$, whence 
\beql{e:pvvv1}
\left|\int_{\epsilon<|w_j-w_1|<1}  \frac{\d w_j}{w_j-w_1}\big(\varphi_j(w_j) - \varphi_j(w_1)\big)\right| \leq 2\sup\norm{\varphi_j'}.
\eeql
Turning our attention to the second integral,
\beql{e:pvvv2}
\Bigg|\int_{|w_j-w_1|\geq1}  \frac{\d w_j}{w_j-w_1}\varphi_j(w_j)\Bigg| \leq \int_{\R} \d w_j \big| \varphi_j(w_j)\big|=\Norm{\varphi_j}_{L^1(\R)}<+\infty, 
\eeql
since $\varphi_j\in \S(\R)$. Combining \eqref{e:pvvv1} and \eqref{e:pvvv2} yields
$
\left|\int_{\epsilon<|w_j-w_1|}  \frac{\d w_j}{w_j-w_1}\varphi_j(w_j) \right|< +\infty\sp j =2,\dots,q.
$
In view of \eqref{e:pvp} the lemma is thus proven.\ \finproof

\begin{lemma}\label{prop:pvcom} For any $c\in \R$ 
and functions $\varphi_j \in \S(\R), j=1,\dots,q$,
\beql{e:sch}  \dashint_{-\infty}^\infty \frac{\d w_1}{w_1 - c} \, \Big(\varphi_1(w_1)\prod_{j=2}^q\dashint_{-\infty}^\infty \frac{\d w_j}{w_j - w_1}\varphi_j(w_j)\Big)   
\eeql   
is well defined.
\end{lemma}
\proof
By Definitions \ref{d:cpvint} and \ref{d:ppv},   
\[
 \varphi_1(w_1)  \prod_{j=2}^q\dashint_{-\infty}^\infty \frac{\d w_j}{w_j - w_1} \varphi_j(w_j)
    =
    \varphi_1(w_1) \prod_{j=2}^q\lim_{\epsilon\to 0+}\int_{\epsilon<|w_j-w_1|} \frac{ \d w_j}{w_j-w_1} \varphi_j(w_j),
\]
which in view of Lemma \ref{l:pvcom} is Lebesgue-integrable with respect to $w_1$. Hence \eqref{e:sch} is well defined in accordance with \eqref{eq:def_cpv}.\ \finproof

\begin{lemma}\label{prop:intpv} For any $\varphi_1,\varphi_2 \in \S(\R) $,  
\beql{e:intpv}
    \int_\R \d w_1 \intcc_\R \frac{\d w_2}{w_1-w_2}\varphi_1(w_1)\varphi_2(w_2) = \int_\R \d w_2 \intcc_\R \frac{\d w_1}{w_1-w_2}\varphi_1(w_1)\varphi_2(w_2).
\eeql   
\end{lemma}
\proof
By Lemma \ref{l:pvcom}, both sides of \eqref{e:intpv} are well defined.  By Definition \ref{d:cpvint}, dominated convergence, Fubini's theorem, and dominated convergence again,
\bel
 &\int_\R \d w_1 \intcc_\R \frac{\d w_2}{w_1-w_2} \varphi_1(w_1)\varphi_2(w_2)=  \int_\R \d w_1 \lim_{\epsilon \to 0+} \int_\R \d w_2\frac{\varphi_1(w_1)\varphi_2(w_2)\ind_{\epsilon<|w_1-w_2|}}{w_1-w_2}\\
 =&   \lim_{\epsilon \to 0+} \int_\R \d w_1 \int_\R 
 \frac{\d w_2 \varphi_1(w_1)\varphi_2(w_2)\ind_{\epsilon<|w_1-w_2|}}{w_1-w_2}=   \lim_{\epsilon \to 0+} \int_\R \d w_2\int_\R  
\frac{\d w_1\varphi_1(w_1)\varphi_2(w_2)\ind_{\epsilon<|w_1-w_2|}}{w_1-w_2} \\
 =&\int_\R\d w_2 \lim_{\epsilon \to 0+}  \int_\R  \frac{\d w_1}{w_1-w_2} \varphi_1(w_1)\varphi_2(w_2)\ind_{\epsilon<|w_1-w_2|}= \int_\R \d w_2 \intcc_\R \frac{\d w_1}{w_1-w_2}\varphi_1(w_1)\varphi_2(w_2).\ \finproof
\eel

\section{Fourier Transform}\label{a:fourier}

\def\s{\vec s}
\begin{definition}
 For $f\in L^1 (\R^q) $, the Fourier and inverse Fourier transforms $\F f$ and $\F^{-1} f$ of $f$ are the following functions on $\R^q$ \citep[Section 2.6 and 15.6]{king_2009}:
\beql{e:ffm}
     \F f (\z) = \int_{\R^q} f(\vec s) e^{-\thermi  \z \cdot \vec s} \d \vec s
     \sp\   \F^{-1} f (\s) = \frac{1}{(2\pi)^q} \int_{\R^q} f (\z) e^{\thermi  \s\cdot \z} \d \z  .   
\eeql
\end{definition}
\noindent
When $f$ depends on several variables, we write e.g.\ $\F_{x}[f(x,y)](t)$ to indicate that the transform is taken with respect to specific variables only.

  \begin{example}\label{l:c}
For $c\in\R$, we compute
\bel
    & \F_k\left[ \big(|c| - |k |\big)^+\right](r)
    = \intR \d k e^{-\thermi rk} \big(|c| - |k |\big)^+
    = \int_{-|c|}^{|c|} \d k e^{-\thermi rk} (|c| - |k |\big)=
    \\
     & 
   \qqq \int_0^{|c|} \d k \left(e^{-\thermi rk} + e^{\thermi rk}\right) (|c| - k\big)
    = \frac{1}{r^2}(2-2\cos(r c)) = \frac{1}{r^2}(2-e^{\thermi  r c}-e^{-\thermi  r c}).\eel
Since $\int_{\R} \frac{\d r}{r^2} \big(2-2\cos(r c)\big)   = 2\pi |c|$, we have
\beql{e:c}
    & \F_k\left[ \big(|c| - |k |\big)^+\right](r) 
    = \frac{2-2\cos(r c)}{r^2}\in L^1 (\R) .
\eeql
\end{example} 

\begin{remark}\label{rem:fourl1}
   The Fourier transform is injective on $L^1(\R^q)$ \cite[Corollary 8.5]{knapp2007basic}.
   If $f \in L^1(\R^q)$, then $g=\F f $ is bounded, uniformly continuous and vanishes at infinity \cite[Proposition 8.1 and Theorem 8.3]{knapp2007basic}. 
If, in addition, $g\in L^1 (\R^q)$, then  
$\F^{-1} g 
     = f 
$
\citep*[Theorem 8.4]{knapp2007basic}.
\end{remark}
 
\begin{remark}\label{rem:fourcont}
    $\F $ and $\Fi$ are continuous automorphisms of $\S(\R^q)$ \cite[Theorem 2 page 143]{kanwal}, hence~$ \varphi= \F \Fi \varphi = \Fi \F \varphi$, for $\varphi \in \S(\R^q)$.
\end{remark}

\begin{definition} For a distribution $T$ on $\S(\R^q)$, its Fourier and inverse Fourier transforms $\F T$ and $\Fi T$ are the distributions on $ \S (\R^q) $ given as
\citep[Theorem 3 page 147]{kanwal}, \cite[Section 10.4 and 10.10]{king_2009}
\begin{align}\label{eq:def_F_dist} \inner{(\F T)_{\d\s}, \varphi(\s)}_\s = \inner{T_{\d\z},\F \varphi(\z)}_{\z}\sp 
        \inner{(\Fi T)_{\d\z}, \varphi(\z)}_{\z} = \inner{T_{\d\s},\Fi \varphi(\s)}_\s
        \sp \varphi \in\S(\R^q).
\end{align} 
\end{definition}

\begin{example} \label{ex:fourier_dist}
Below are two well-known examples of distributional Fourier transforms \citep[page 489]{king_2009}, \citep[page 415]{kammler2008}: $$ \F_x [\sgn x\, \d x]_{\d\lambda} = \frac2i \frac{\d\lambda}{\lambda} \sp\   \F_x [\cos x\, \d x](\d\lambda) = \delta_1(\d\lambda) +  \delta_{-1}(\d\lambda),$$ i.e., for $\varphi\in\S(\R),$ 
\begin{align}
    &  \inner{\F_x [\sgn x\, \d x]_{\d\lambda},\varphi(\lambda)}_{\lambda} 
     = \int_\R \d x \sgn x \int_\R\d \lambda e^{-\thermi  \lambda x} \varphi(\lambda)
    =  \frac2i\intc\frac{\d\lambda}{\lambda}  \varphi(\lambda) ,
    \label{e:sgn}\\
    &     \inner{\F_x[{\cos x\,\d x}](\d \lambda) ,\varphi(\lambda)}_{\lambda}
     = \int_{\R} \d x\cos x \int_\R\d \lambda e^{-\thermi  \lambda x} \varphi(\lambda) 
    = \pi\varphi(1)+ \pi\varphi(-1).\label{e:sincos}  \end{align}
\end{example}

\section{Proof of \eqref{e:TNbis}}\label{s:disp}

\begin{lemma}\label{l:dsxk}The measure $\d \x \delta_{\norml{\x}}(\d k) + \d \x \delta_{-\norml{\x}}(\d k) $ 
is equal to
    \beql{e:dssk0}
    & \d k   (\ind_{|x_d|<|k|}\d x_d)   (\ind_{|x_{d-1}|<|k|-|x_d|} \d x_{d-1})   \dots   (\ind_{|x_{2}|<|k|-\norml{\x_{>2}}}\d x_{2}) 
\\&\qqq\big(\delta_{|k|-\norml{\x_{>1}}}(\d x_1) + \delta_{-|k|+\norml{\x_{>1}}}(\d x_1) \big)=:    \delta_\theS(\d(\x,k))   .
    \eeql
\end{lemma}
\proof Let $f: \R^{d+1} \to \mathbb{C}$
be any $\delta_D$-\thermi  ntegrable function.
By sifting property of the Dirac mass distributions
$\delta_{\norml{\x}}(\d k)$ and  $\delta_{-\norml{\x}}(\d k)$,
\beql{e:dssk1}
    \int_{\R^{d+1}} \big(\d \x \delta_{\norml{\x}}(\d k) + \d \x \delta_{-\norml{\x}}\big)f(\x,k) = \int_{\R^{d}} \d \x f(\x,\norml{\x}) + \int_{\R^{d}} \d \x f(\x,-\norml{\x}).
\eeql
Splitting the first right-hand side integral above,
\beql{e:dssk2}
    \int_{\R^{d}} \d \x f(\x,\norml{\x}) 
    =\, & 
    \int_{\R^{d-1}}\d \x_{>1}\int_0^{\infty}\d x_1 f(\x,\norml{\x_{>1}}+x_1)
    \\
    & + \int_{\R^{d-1}}\d \x_{>1}\int^0_{-\infty} \d x_1 f(\x,\norml{\x_{>1}}-x_1).
\eeql
By change of variables $(x_1,\dots,x_d)\mapsto(k =  \norml{\x_{>1}} + x_1, x_2,\dots,x_{d}) $ in the first right-hand side integral above, and  change of variables $(x_1,\dots,x_d)\mapsto(k =   \norml{\x_{>1}} - x_1, x_2,\dots,x_{d}) $ in the second integral, where both Jacobians are 1, and adapting integral regions,
\beql{e:dssk3}
&\int_{\R^{d}} \d \x f(\x,\norml{\x}) = \int_0^{\infty} \d k \int_{-k}^{k}\d x_d\dots\int_{-k+\norml{\x_{>2}}}^{k-\norml{\x_{>2}}}\d x_{2}f(k-\norml{\x_{>1}}, \x_{>1},k) \\
&\qqq +   \int_0^{\infty} \d k \int_{-k}^{k}\d x_d\dots\int_{-k+\norml{\x_{>2}}}^{k-\norml{\x_{>2}}}\d x_{2}f(\norml{\x_{>1}}-k,\x_{>1},k).
\eeql
Following similar steps, the second right-hand side integral in \eqref{e:dssk1} may be rewritten as
\beql{e:dssk4}
&\int_{\R^{d}} \d \x f(\x,-\norml{\x}) = \int^0_{-\infty} \d k \int_{k}^{-k}\d x_d\dots\int_{k+\norml{\x_{>2}}}^{-k-\norml{\x_{>2}}}\d x_{2}f(-\norml{\x_{>1}}-k,\x_{>1},k) \\
&\qqq +  \int^0_{-\infty} \d k \int_{k}^{-k}\d x_d\dots\int_{k+\norml{\x_{>2}}}^{-k-\norml{\x_{>2}}}\d x_{2} f(k+\norml{\x_{>1}},\x_{>1},k).
\eeql
Combining \eqref{e:dssk3} and \eqref{e:dssk4} and piecing $\d k$ integrals together,
\bel&\int_{\R^{d+1}} \big(\d \x \delta_{\norml{\x}}(\d k) + \d \x \delta_{-\norml{\x}}\big)f(\x,k) \\
= & \int_{-\infty}^\infty \d k \int_{-|k|}^{|k|}\d x_d\dots\int_{-|k|+\norml{\x_{>2}}}^{|k|-\norml{\x_{>2}}}\d x_{2}f(|k|-\norml{\x_{>1}},\x_{>1},k) \\
&\qqq+\int_{-\infty}^\infty \d k \int_{-|k|}^{|k|}\d x_d\dots\int_{-|k|+\norml{\x_{>2}}}^{|k|-\norml{\x_{>2}}}\d x_{2}f(\norml{\x_{>1}}-|k|,\x_{>1},k) \\
=& \intR \d k \int_{-|k|}^{|k|}\d x_d\dots\int_{-|k|+\norml{\x_{>2}}}^{|k|-\norml{\x_{>2}}}\d x_{2} \intR \d x_1 \delta_{|k|-\norml{\x_{>1}}}(\d x_1) f(x_1,\x_{>1},k)\\
&\qqq+\int_\R \d k \int_{-|k|}^{|k|}\d x_d\dots\int_{-|k|+\norml{\x_{>2}}}^{|k|-\norml{\x_{>2}}}\d x_{2} \intR \d x_1  \delta_{-|k|+\norml{\x_{>1}}}(\d x_1) f(x_1,\x_{>1},k)\\
=& \int_{\R^{d+1}} \delta_D(\d(\x,k))f(\x,k),
\eel
where we used the sifting property again in the second equality.\ \finproof 

By sifting property of the Dirac mass distribution, 
the left-hand side in \eqref{e:TNbis} satisfies for any $\phie\in S_e$
\begin{align*}  &     \intRd \d\x  e^{-\thermi  \norml{\x}} \Fi \phi_e (\x)+     \intRd \d\x  e^{ \thermi  \norml{\x}} \Fi \phi_e (\x) \nonumber\\
&\qqq  =   \intRd \d\x \int_{\R} \delta_{\norml{\x}}(\d k ) e^{-\thermi k} \Fi \phi_e (\x) +  \intRd \d\x \int_{\R} \delta_{-\norml{\x}}(\d k )e^{-\thermi k} \Fi \phi_e (\x) \\&\qqq= \int_{\R^{d+1}} \delta_\theS(\d(\vec x, k)) e^{-\thermi k} \Fi \phi_e (\x) ,
\end{align*}
where we used Lemma \ref{l:dsxk} in the last step. 
Therefore, showing \eqref{e:TNbis} reduces to proving that
{\def\varphie{\phi_e}\beql{e:TNlast}
        &\frac{1}{2}\int_{\R^{d+1}} \delta_\theS(\d(\vec x, k)) e^{-\thermi k} \Fi \varphie (\x) 
         =    \inner{T^d_{\d \w}, \varphie (\w)}_{\w},
         \eeql
}
where $T^d$ is defined by \eqref{e:tdoe}. 
This is done in the remaining subsections by an induction procedure in line with the combinatorial structure of $T^d$ highlighted in Remark \ref{r:dc-sol-combin}.

\subsection{Dimension $ d=1$}

By definition of $ \delta_\theS$ in \eqref{e:dssk0},
{\def\varphie{\phi_e}\begin{align*}
    &\int_{\R^{2}} \delta_\theS(\d(  x, k)) e^{-\thermi k} \Fi \varphie ( x) 
     =  \int_{\R} \d k e^{-\thermi k}  \int_{\R}\big(\delta_{k}(\d x) + \delta_{-k}(\d x) \big)  \Fi \varphie(x)  \\
    &\qqq =\F_k[ \Fi \varphie(k) + \Fi \varphie(-k)](1) = \varphie(1) + \varphie(-1)\\
    &\qqq =\inner{\delta_1(\d w)+ \delta_{-1}(\d w),\varphie(w)}_{w} = 2\inner{T^1_{\d w}, \varphie (w)}_{w},
\end{align*}}
which is \eqref{e:TNlast} for $d=1$.

\subsection{Dimension $d=2$} 

Lemma \ref{lemma:fouri} below is necessary to initialize the proof by induction of Lemma \ref{l:dcc2} in general dimension $d\ge 2$.  We also include the proof of \eqref{e:TNlast} for $d=2$ in this section for ease of reading.

{\def\varphie{\varphi}
\def\Psie{\Psi}

\begin{lemma}\label{lemma:fouri}
    For $d=2$ and $\varphie  \in \S(\R^2)$, let
    \[
        \Psi_\pm (k) := \inner{ (\ind_{|x_2|<|k|} \d x_2) 
                \delta_{\pm|k|\mp|x_2|}(\d x_1), \F^{-1}\varphie(\x)}_{x_1,x_2}.       
    \]
    If $\varphie =\phi_e\in S_e$    , then
    \beql{eq:pp}
      & \Psi_+(k) + \Psi_-(k) = \ \frac{1}{2\pi^2} \int_{\R}\d w_1\sin(w_1|k|) \intcc_\R \Big(\frac{\d w_2}{w_2 + w_1}- \frac{\d w_2}{w_2 - w_1}\Big)\phi_e (w_1,w_2) 
        \\
        &\qqq + \frac{1}{2\pi^2}\int_{\R}\d w_2\sin(w_2|k|) \intcc_\R  \Big(\frac{\d w_1}{w_1 + w_2}- \frac{\d w_1}{w_1 - w_2}\Big)\phi_e(w_1,w_2) . 
    \eeql
\end{lemma}

\proof For $\varphie  \in \S(\R^2) $, we have
\begin{align*}
    \Psi_+(k) &=  \inner{ (\ind_{|x_2|<|k|} \d x_2),\inner{\delta_{|k|-|x_2|}(\d x_1), \F^{-1}\varphie (\x)}_{x_1}}_{x_2}
    \\
    &= \intR \d x_2  \ind_{|x_2|<|k|}   \F^{-1}\varphie (|k|-|x_2|,x_2)\\
&=\frac{1}{(2\pi)^2}\int_{|x_2|<|k|} \d x_2 \int_{\R^2}\d w_2 \d w_1 e^{\thermi  w_1(|k| - |x_2|)  +\thermi   w_2x_2 } \varphie (w_1,w_2).
\end{align*}
Splitting the $\d x_2$ integral at the origin, then by dominated convergence
\beql{e:dm1}
    \Psi_+(k) &= \frac{1}{(2\pi)^2}\int_{0}^{|k|} \d x_2 \int_{\R^2} \d w_2 \d w_1 e^{\thermi x_2(w_2-w_1)} e^{\thermi  w_1|k|}\varphie (w_1,w_2)\\
& \qq\qq+ \frac{1}{(2\pi)^2}\int^{0}_{-|k|} \d x_2 \int_{\R^2} \d w_2 \d w_1 e^{\thermi x_2(w_2+w_1)} e^{\thermi  w_1|k|}\varphie (w_1,w_2)\\
&= \frac{1}{(2\pi)^2}\int_{0}^{|k|} \d x_2  \Big(\lim_{\epsilon \to 0+} \int_{\R^2\setminus \{|w_2-w_1|\leq \epsilon\}}  \d w_2 \d w_1 e^{\thermi x_2(w_2-w_1)} e^{\thermi  w_1|k|}\varphie (w_1,w_2)\Big)\\
& \qq+ \frac{1}{(2\pi)^2}\int^{0}_{-|k|} \d x_2 \Big(\lim_{\epsilon \to 0+} \int_{\R^2\setminus \{|w_2-w_1|\leq \epsilon\}}  \d w_2 \d w_1 e^{\thermi x_2(w_2+w_1)} e^{\thermi  w_1|k|}\varphie (w_1,w_2)\Big).
\eeql
Since $\varphie  \in \S (\R^2)$, for any $\epsilon \geq 0$, there exists $B>0$ such that
\bel
&\Bigg| \int_{\R^2\setminus \{|w_2-w_1|\leq \epsilon\}}  \d w_2 \d w_1 \varphie (w_1,w_2) e^{\thermi x_2(w_2\pm w_1)} e^{\thermi  w_1|k|} \Bigg|\\ 
&\qqq\leq \int_{\R^2\setminus \{|w_2-w_1|\leq \epsilon\}} \d w_2 \d w_1 \Big|\varphie (w_1,w_2) e^{\thermi x_2(w_2\pm w_1)} e^{\thermi  w_1|k|}\Big| \\
&\qqq\leq \int_{\R^2\setminus \{|w_2-w_1|\leq \epsilon\}} \d w_2 \d w_1 \big|\varphie (w_1,w_2)\big|< B <\infty,
\eel
and the integral $\int_0^{\pm|k|} B \d x_2$ is finite for any $|k|<\infty$.  Hence, by dominated convergence and Fubini's theorem,
 \bel
    \Psi_+(k) &= \frac{1}{(2\pi)^2} \lim_{\epsilon \to 0+} \int_{\R^2\setminus \{|w_2-w_1|\leq \epsilon\}}  \d w_2 \d w_1 \varphie (w_1,w_2)  \int_{0}^{|k|} \d x_2 e^{\thermi x_2(w_2-w_1)} e^{\thermi  w_1|k|} \\
& \qq+ \frac{1}{(2\pi)^2} \lim_{\epsilon \to 0+} \int_{\R^2\setminus \{|w_2+w_1|\leq \epsilon\}}  \d w_2 \d w_1 \varphie (w_1,w_2)  \int^{0}_{-|k|} \d x_2 e^{\thermi x_2(w_2+w_1)} e^{\thermi  w_1|k|}.
\eel
Solving integrals with respect to $x_2$,
\bel
\Psi_+(k)&= \frac{1}{(2\pi)^2}\lim_{\epsilon \to 0+} \int_{\R^2\setminus \{|w_2-w_1|\leq \epsilon\}}  \d w_2 \d w_1 \varphie (w_1,w_2)   \frac{-\thermi  e^{\thermi  |k|w_2}+\thermi  e^{\thermi  w_1|k|}}{w_2-w_1} \\
& \qq+ \frac{1}{(2\pi)^2} \lim_{\epsilon \to 0+} \int_{\R^2\setminus \{|w_2+w_1|\leq \epsilon\}}  \d w_2 \d w_1 \varphie (w_1,w_2)  \frac{\thermi  e^{-\thermi   |k|w_2}-\thermi  e^{\thermi  w_1|k|}}{w_2+w_1} .
\eel
Similar steps would show that
\begin{align*}
    \Psi_-(k) 
    =&\ \frac{1}{(2\pi)^2}\lim_{\epsilon \to 0+} \int_{\R^2\setminus \{|w_2-w_1|\leq \epsilon\}}  \d w_2 \d w_1 \varphie (w_1,w_2)   \frac{\thermi  e^{-\thermi   |k|w_2}-\thermi  e^{-\thermi   w_1|k|}}{w_2-w_1}
    \\
    & + \frac{1}{(2\pi)^2}\lim_{\epsilon \to 0+} \int_{\R^2\setminus \{|w_2+w_1|\leq \epsilon\}}  \d w_2 \d w_1 \varphie (w_1,w_2)  \frac{-\thermi  e^{\thermi  |k|w_2}+\thermi  e^{-\thermi   w_1|k|}}{w_2+w_1}.
\end{align*}
Plugging Euler's sine formula into the above expressions and combining,
\begin{align}
    & \Psi_+(k) + \Psi_-(k) 
    \nonumber\\= &
    \frac{1}{2\pi^2} \lim_{\epsilon \to 0+} \int_\R \d w_2 \sin(w_2) \Big(\int_{\epsilon<|w_1+w_2|} \frac{\d w_1}{w_1+w_2}\varphie(w_1,w_2) - \int_{\epsilon<|w_1-w_2|}  \frac{\d w_1}{w_1-w_2}\varphie(w_1,w_2) \Big)
    \nonumber\\
     +& \frac{1}{2\pi^2}  \lim_{\epsilon \to 0+} \int_\R \d w_1 \sin(w_1) \Big(\int_{\epsilon<|w_2+w_1|}  \frac{\d w_2}{w_2+w_1} \varphie(w_1,w_2) - \int_{\epsilon<|w_2-w_1|}  \frac{\d w_2}{w_2-w_1}\varphie(w_1,w_2) \Big).    \label{e:Pk}
\end{align}
For $\varphie(w_1,w_2) = \varphi_1(w_2)\varphi_2(w_1)$ with $\varphi_1,\varphi_2\in \S(\R)$, in view of Lemma \ref{l:pvcom}, we have by dominated convergence  
\bel
&\lim_{\epsilon \to 0+} \int_\R \d w_2 \sin(w_2) \int_{\epsilon<|w_1+w_2|} \frac{\d w_1}{w_1+w_2}\varphie(w_1,w_2) \\
=& \int_\R \d w_2 \sin(w_2) \lim_{\epsilon \to 0+} \int_{\epsilon<|w_1+w_2|} \frac{\d w_1}{w_1+w_2}\varphie(w_1,w_2) =  \int_\R \d w_2 \sin(w_2) \intcc_{\R}\frac{\d w_1}{w_1+w_2}\varphie(w_1,w_2).
\eel
Following similar steps for the remaining terms in \eqref{e:Pk}, we recover \eqref{eq:pp} for $\varphie(w_1,w_2) = \varphi_1(w_2)\varphi_2(w_1)$ with $\varphi_1,\varphi_2\in \S(\R)$, from which the result for $\varphie=\phi_e \in S_e$ follows by linearity.\ \finproof

We are now in a position to prove \eqref{e:TNlast} for $d=2$. By definition of $ \delta_\theS$ in \eqref{e:dssk0},
\begin{align*}
    \delta_\theS(\d(\vec x, k))
    = \d k ( \ind_{|x_2|<|k|} \d x_2) (\delta_{|k|-|x_2|} + \delta_{-|k|+|x_2|})(\d x_1).
\end{align*}
Therefore, by Lemma \ref{lemma:fouri},
\beql{e:dd3}
\int_{\R^{3}} \delta_{\theS}&(\d(\vec x, k)) e^{-\thermi k} \Fi \phie (\x) =
\int_{\R} \d k e^{-\thermi k} (\Psi_+(k) + \Psi_-(k))  = 
  \\
    & \frac{1}{2\pi^2} \int_{\R} \d k e^{-\thermi k} \int_\R \d w_1 \sin(w_1|k|) \intcc_\R  \Big(\frac{\d w_2}{w_2 + w_1}- \frac{\d w_2}{w_2 - w_1}\Big)\phie(w_1,w_2)
    \\
    +& \frac{1}{2\pi^2}\int_{\R} \d k e^{-\thermi k} \int_\R\d w_2 \sin(w_2|k|) \intcc_\R  \Big(\frac{\d w_1}{w_1 + w_2}- \frac{\d w_1}{w_1 - w_2}\Big)\phie(w_1,w_2).
\eeql 
Let $f(w_1) := \intcc_\R \Big(\frac{\d w_2}{w_2 + w_1}- \frac{\d w_2}{w_2 - w_1}\Big)\phie(w_1,w_2)$. Substituting Euler's sine formula, then splitting the integrand while applying the change of variable $k \mapsto -k$ to the second resulting integral, the first term in \eqref{e:dd3} may be rewritten as
\begin{align*}
 & \frac{\thermi}{4\pi^2} \int_\R \d k e^{-\thermi k} \sgn(k)\int_\R \d w_1 (e^{-\thermi   w_1 k}- e^{\thermi  w_1 k}) f(w_1)
 \\
  =\ & \frac{\thermi}{4\pi^2} \left(\int_{-\infty}^\infty \d k  \sgn(k)  \int_\R \d w_1 e^{-\thermi   k(w_1+1) } f(w_1) + \int_{-\infty}^\infty \d k  \sgn(k)  \int_\R \d w_1 e^{-\thermi   k(w_1-1) } f(w_1)\right)\\
 =\ & \frac{1}{2\pi^2} \intcc_{\R}  \Big(\frac{\d w_1}{w_1+1}+\frac{\d w_1}{w_1-1}\Big)f(w_1),
\end{align*}
where we substituted \eqref{e:sgn} in the last step. In addition,
\beql{e:dd5}
    & \frac{1}{2\pi^2} \intcc_{\R}  \Big(\frac{\d w_1}{w_1+1}+\frac{\d w_1}{w_1-1}\Big)f(w_1)=\\ 
     &\qqq\frac{1}{2\pi^2} \intcc_{\R}  \Big(\frac{\d w_1}{w_1+1}+\frac{\d w_1}{w_1-1}\Big)\intcc_\R  \Big(\frac{\d w_2}{w_2 + w_1}- \frac{\d w_2}{w_2 - w_1}\Big)\phie(w_1,w_2).
\eeql
Following similar steps for the remaining terms to the right-hand side of \eqref{e:dd3}, we may conclude
in view of the definition \eqref{N2} of $T_2$
that
\begin{equation*}     
        \frac{1}{2}\int_{\R^{3}} \delta_\theS(\d(\vec x, k)) e^{-\thermi k} \Fi \phi_e (\x) =   \inner{T^2_{\d \w},\phi_e (\w)}_{\w},
    \end{equation*}
 which is \eqref{e:TNlast} for $d=2$.
\subsection{General Dimension $d\geq 2$\label{a:disp-general-dim}}

{\def\jmath{i}
For $ \phi_e\in S_e, k\in\R$, let $\Psi_d(k) \equiv \Psi_d(\x_{>d},k) $ result from the following finite recurrence: 

\beql{e:phidef}
     \Psi_1(\x_{>1},k) & = \inner{(\delta_{|k|-\norml{\x_{>1}}} + \delta_{-|k|+\norml{\x_{>1}}})(\d x_1) ,\F^{-1}\phie(x_1,\x_{>1})}_{x_1}, \x_{>1}\in\R^{d-1},
        \\
    \Psi_l(\x_{>l},k) & = \inner{ \ind_{|x_l|<|k|-\norml{\x_{>l}}} \d x_l, \Psi_{l-1}(x_l, \x_{>l}, k)}_{x_l}
    \sp \x_{>l}\in\R^{d-l}\sp l = 2, \dots, d.
\eeql
\begin{lemma}\label{l:dcc2} For any $\phi_e \in S_e, \thel  = 2 \dots, d$,  and  $(\x,k)\in \R^{d+1}$ such that $|\x_{>\thel}|\leq |k|$, 
\beql{e:recu}
    \Psi_{\thel}(\x_{>\thel},k)=& \frac2{(2\pi)^{\thel}} \sum_{j = 1}^\thel \int_\R \d w_j \,\sc_{\thel }\big(w_j (|k|-\norml{\x_{>\thel}})\big) \times
        \\
& \intcc_{\R^{\thel-1 }}\Big(\bigotimes_{\jmath\, \in\, 1\,..\,\thel\, \setminus\,  \{j\}}\big(\frac{\d w_{\jmath}}{w_{\jmath} + w_{j}} - \frac{\d w_{\jmath}}{w_{\jmath} - w_{j}}\big) \Big)\,\Phi_{\thel } (\w_{\le l},\x_{>l})
  ,\\
\eeql
where
$ \displaystyle\Phi_{\thel } (\w_{\le l},\x_{>l}) 
:=\F^{-1}_{\w_{>\thel}}[\phie(\w)](w_1,\dots, w_l ,\x_{>\thel}) \in S_e$, and $\sc_{\thel }(\cdot)  =  (-1)^{\frac{\thel -1}{2}}  \cos(\cdot)$ if $\thel$ is odd, $(-1)^{\frac{\thel-2 }{2}} \sin(\cdot)$ if $\thel$ is even.
\end{lemma}
\proof We proceed by 
induction on $l\ge 2$. 
Note that 
$\displaystyle\Phi_{2 } \in S_e $.
For any fixed $\x\in\R^d$,
an application of Lemma  \ref{lemma:fouri} to $(w_1,w_2)\stackrel{\phi_e}{\longmapsto} \Phi_{2}(w_1,w_2,\x_{>2}) $ and to the strike $\kappa :=|k|-\norml{\x_{>2}}\ge 0 $ yields
\beql{e:rec}
    &\Psi_{2}(\x_{>2},k) = \frac{1}{2\pi^2}\sum_{j = 1}^2 \int_\R \d w_j \,\sin(w_j  \kappa 
        )  \times\\
    &\qqq\qqq\intcc_\R \Big(\bigotimes_{\jmath\, \in\, 1\,..\,2\, \setminus\,  \{j\}} \big(\frac{\d w_{\jmath}}{w_{\jmath} + w_{j}} - \frac{\d w_{\jmath}}{w_{\jmath} - w_{j}}\big)\Big) \Phi_{2}(w_1,w_2,\x_{>2}),
\eeql
 which is \eqref{e:recu} for $\thel =2$}.
It remains to prove that if \eqref{e:recu} is satisfied for some index  $\thel  = 2, \dots, d-1$, then it also holds at index $\thel +1$, i.e.
\beql{e:recup}
&\Psi_{\thel+1}(\x_{>\thel+1},k) = \frac2{(2\pi)^{\thel +1}}\sum_{j =1}^{\thel+1} \int_\R \d w_j\,\sc_{\thel+1}(w_j (|k|-\norml{\x_{>\thel+1}})) \times\\
&\qqq\qqq\intcc_{\R^{\thel }} \Big(  \bigotimes_{\jmath\, \in\, 1\,..\,(\thel+1)\, \setminus\,  \{j\}} \big(\frac{\d w_{\jmath}}{w_{\jmath} + w_{j}} - \frac{\d w_{\jmath}}{w_{\jmath} - w_{j}}\big) \Big) \Phi_{\thel+1 } (\w_{\le l+1},\x_{>l+1}).
\eeql
Assuming \eqref{e:recu} for some $\thel \ge 2$,
let us thus prove \eqref{e:recup}, first in the case where $\thel $ is even. Then $\sc_{\thel }(\cdot) = (-1)^{\frac{\thel -2}{2}}\sin(\cdot)$ and
\beql{e:lll1bis}
    &\Psi_{\thel+1}(\x_{>\thel+1},k) = \inner{ \ind_{|x_{\thel+1}|<|k|-\norml{\x_{>\thel+1}}} \d x_{\thel+1}, \Psi_{\thel}(\x_{>\thel},k)}_{x_{\thel+1}} = 
     \frac{(-1)^{\frac{\thel-2}{2}}2}{(2\pi)^{\thel }}\times\\
    &\qqq \Bigg[\int^{|k|-\norml{\x_{>\thel+1}}}_0\d x_{\thel+1} \sum_{j = 1}^\thel\int_\R \d w_j\, \sin(w_j (|k|-\norml{\x_{>\thel+1}}-x_{\thel+1})) \times \\ 
    &\qqq\qqq\intcc_{\R^{\thel-1 }} \Big(  \bigotimes_{\jmath\, \in\, 1\,..\,\thel\, \setminus\,  \{j\}} \big(\frac{\d w_{\jmath}}{w_{\jmath} + w_{j}} - \frac{\d w_{\jmath}}{w_{\jmath} - w_{j}}\big) \Big) \Phi_{\thel } (\w_{\le l},\x_{>l})\\
    &\qqq+  \int_{-|k|+\norml{\x_{>\thel+1}}}^0\d x_{\thel+1} \sum_{j = 1}^\thel\int_\R \d w_j\, \sin(w_j (|k|-\norml{\x_{>\thel+1}}+x_{\thel+1})) \times \\ 
    &\qqq\qqq\intcc_{\R^{\thel-1 }} \Big(  \bigotimes_{\jmath\, \in\, 1\,..\,\thel\, \setminus\,  \{j\}} \big(\frac{\d w_{\jmath}}{w_{\jmath} + w_{j}} - \frac{\d w_{\jmath}}{w_{\jmath} - w_{j}}\big) \Big) \Phi_{\thel } (\w_{\le l},\x_{>l})\Bigg].
\eeql
Substituting Euler's sine formula into the above together with
\bel
    \Phi_{\thel } (\w_{\le l},\x_{>l}) &= \frac{1}{2\pi} \intR \d {w_{\thel+1}}e^{-\thermi  x_{\thel+1}w_{\thel+1}}\Phi_{\thel+1 } (\w_{\le l+1},\x_{>l+1}) \\
    &=\frac{1}{2\pi} \lim_{\epsilon\to 0+}\int_{\epsilon<|w_{\thel+1} \pm w_j|} \d {w_{\thel+1}}e^{-\thermi  x_{\thel+1}w_{\thel+1}}\Phi_{\thel+1 } (\w_{\le l+1},\x_{>l+1}),
\eel
which stems from slicing the Fourier transform $\Phi_{\thel } (\w_{\le l},\x_{>l})$ along $x_{\thel+1}$, then using dominated convergence, the first term inside the square bracket in \eqref{e:lll1bis} may be rewritten as \beql{e:lll1c}
&\frac{\thermi}{4\pi}\int^{|k|-\norml{\x_{>\thel+1}}}_0\d x_{\thel+1} \sum_{j=1}^l \int_{\R}\d w_j \intcc_{\R^{\thel-1 }} \Big(  \bigotimes_{\jmath\, \in\, 1\,..\,\thel\, \setminus\,  \{j\}} \big(\frac{\d w_{\jmath}}{w_{\jmath} + w_{j}} - \frac{\d w_{\jmath}}{w_{\jmath} - w_{j}}\big) \Big) \times \\ 
& \qqq \lim_{\epsilon\to 0+}\int_{\epsilon<|w_{\thel+1}- w_j|} \d {w_{\thel+1}}e^{-\thermi  x_{\thel+1}w_{\thel+1}}e^{-\thermi  w_j (|k|-\norml{\x_{>\thel+1}}-x_{\thel+1})}\Phi_{\thel+1 } (\w_{\le l+1},\x_{>l+1})\\
-&\frac{\thermi}{4\pi}\int^{|k|-\norml{\x_{>\thel+1}}}_0\d x_{\thel+1} \sum_{j=1}^l \int_{\R}\d w_j  \intcc_{\R^{\thel-1 }} \Big(  \bigotimes_{\jmath\, \in\, 1\,..\,\thel\, \setminus\,  \{j\}} \big(\frac{\d w_{\jmath}}{w_{\jmath} + w_{j}} - \frac{\d w_{\jmath}}{w_{\jmath} - w_{j}}\big) \Big) \times\\
& \qqq \lim_{\epsilon\to 0+}\int_{\epsilon<|w_{\thel+1}+w_j|} \d {w_{\thel+1}}e^{-\thermi  x_{\thel+1}w_{\thel+1}}e^{\thermi w_j (|k|-\norml{\x_{>\thel+1}}-x_{\thel+1})} \Phi_{\thel+1 } (\w_{\le l+1},\x_{>l+1}).\eeql
By a combination of Definition \ref{d:cpvint}, dominated convergence and Fubini's theorem similarly to the proof of Lemma \ref{lemma:fouri},
we may bring the outer $\d x_{\thel+1}$-integrals above inside all others. As such, the first term in \eqref{e:lll1c} is equal to
\begin{align*}
&\frac{\thermi}{4\pi} \sum_{j = 1}^\thel\int_\R \d w_j\, \intcc_{\R^{\thel-1 }} \Big(  \bigotimes_{\jmath\, \in\, 1\,..\,\thel\, \setminus\,  \{j\}} \big(\frac{\d w_{\jmath}}{w_{\jmath} + w_{j}} - \frac{\d w_{\jmath}}{w_{\jmath} - w_{j}}\big) \Big) \Bigg(\lim_{\epsilon\to 0+}\int_{\epsilon<|w_{\thel+1}+w_j|} \d {w_{\thel+1}}  \times\\
&\qqq \int^{|k|-\norml{\x_{>\thel+1}}}_0\d x_{\thel+1} e^{-\thermi  x_{\thel+1}(w_{\thel+1}-w_j)-\thermi  w_j (|k|-\norml{\x_{>\thel+1}})}   \Phi_{\thel+1 } (\w_{\le l+1},\x_{>l+1})\Bigg)\\
=&\frac{1}{4\pi}  \sum_{j = 1}^\thel\int_\R \d w_j\, \intcc_{\R^{\thel-1 }} \Big(  \bigotimes_{\jmath\, \in\, 1\,..\,\thel\, \setminus\,  \{j\}} \big(\frac{\d w_{\jmath}}{w_{\jmath} + w_{j}} - \frac{\d w_{\jmath}}{w_{\jmath} - w_{j}}\big) \Big)\Bigg( \lim_{\epsilon\to 0+}\int_{\epsilon<|w_{\thel+1}- w_j|}\d w_{\thel+1} \times\\
&\qqq \frac{e^{\thermi  w_{\thel+1}(|k|-\norml{\x_{>\thel+1}})}-e^{\thermi   w_{j}(|k|-\norml{\x_{>\thel+1}})}}{w_j-w_{\thel+1}}\Phi_{\thel+1 } (\w_{\le l+1},\x_{>l+1})\Bigg),
\end{align*}
where we solved the $\d x_{\thel+1}$ integral in the last step.  Following the same steps for the second term in \eqref{e:lll1c}, and then the second term within square brackets in \eqref{e:lll1bis}, we obtain
\bel
&\Psi_{\thel+1}(\x_{>\thel+1},k) =\frac{(-1)^{\frac{\thel -2}{2}}}{(2\pi)^{\thel +1}} \sum_{j = 1}^l \int_\R \d w_j \intcc_{\R^{\thel-1 }} \Big(  \bigotimes_{\jmath\, \in\, 1\,..\,\thel\, \setminus\,  \{j\}} \big(\frac{\d w_{\jmath}}{w_{\jmath} + w_{j}} - \frac{\d w_{\jmath}}{w_{\jmath} - w_{j}}\big) \Big) \times \\
&\Bigg( -\lim_{\epsilon\to 0+}\int_{\epsilon<|w_{\thel+1}-w_j|}\d w_{\thel+1}\frac{e^{\thermi  w_{\thel+1}(|k|-\norml{\x_{>\thel+1}})}-e^{\thermi   w_{j}(|k|-\norml{\x_{>\thel+1}})}}{w_{\thel+1}-w_j}\Phi_{\thel+1 } (\w_{\le l+1},\x_{>l+1})  \\
& +\lim_{\epsilon\to 0+}\int_{\epsilon<|w_{\thel+1}+w_j|} \d w_{\thel+1}\frac{e^{\thermi  w_{\thel+1}(|k|-\norml{\x_{>\thel+1}})}- e^{-\thermi   w_{j}(|k|-\norml{\x_{>\thel+1}})}}{w_{\thel+1}+w_j}\Phi_{\thel+1 } (\w_{\le l+1},\x_{>l+1})\\
& -\lim_{\epsilon\to 0+}\int_{\epsilon<|w_{\thel+1}-w_j|}\d w_{\thel+1}\frac{e^{-\thermi   w_{\thel+1}(|k|-\norml{\x_{>\thel+1}})} -e^{-\thermi   w_{j}(|k|-\norml{\x_{>\thel+1}})}}{w_{\thel+1}-w_j}\Phi_{\thel+1 } (\w_{\le l+1},\x_{>l+1})\\
&+\lim_{\epsilon\to 0+}\int_{\epsilon<|w_{\thel+1}+w_j|} \d w_{\thel+1}\frac{e^{-\thermi   w_{\thel+1}(|k|-\norml{\x_{>\thel+1}})}- e^{\thermi  w_{j}(|k|-\norml{\x_{>\thel+1}})}}{w_{\thel+1}+w_j}\Phi_{\thel+1 } (\w_{\le l+1},\x_{>l+1}) \Bigg).
\eel
Splitting and rearranging integrands, substituting Euler's cosine formula, and recognizing Cauchy principal integrals,
\bel
  & \Psi_{\thel+1}(\x_{>\thel+1},k) 
= \frac{(-1)^{\frac{\thel -2}{2}}}{(2\pi)^{\thel +1}} \sum_{j = 1}^l \int_\R \d w_j \intcc_{\R^{\thel-1 }} \Big(  \bigotimes_{\jmath\, \in\, 1\,..\,\thel\, \setminus\,  \{j\}} \big(\frac{\d w_{\jmath}}{w_{\jmath} + w_{j}} - \frac{\d w_{\jmath}}{w_{\jmath} - w_{j}}\big) \Big) \times  \nonumber\\ 
& \qqq \Bigg(\intcc_{\R}\frac{\d w_{\thel+1} }{w_{\thel+1}+w_j} \cos(w_{\thel+1}(|k|-\norml{\x_{>\thel+1}}))\Phi_{\thel+1 } (\w_{\le l+1},\x_{>l+1})\nonumber\\
&\qqq  - \intcc_{\R}\frac{\d w_{\thel+1} }{w_{\thel+1}- w_j} \cos(w_{\thel+1}(|k|-\norml{\x_{>\thel+1}}))\Phi_{\thel+1 } (\w_{\le l+1},\x_{>l+1})\nonumber\\
&\qqq-  \intcc_{\R}\frac{\d w_{\thel+1} }{w_{\thel+1}+w_j} \cos(w_{j}(|k|-\norml{\x_{>\thel+1}}))\Phi_{\thel+1 } (\w_{\le l+1},\x_{>l+1})\nonumber\\
&\qqq+ \intcc_{\R}\frac{\d w_{\thel+1} }{w_{\thel+1}- w_j}\cos(w_{j}(|k|-\norml{\x_{>\thel+1}}))\Phi_{\thel+1 } (\w_{\le l+1},\x_{>l+1})\Bigg).
\eel
By \eqref{e:ppvprod} for $q=2$, this is equal to
\beql{e:lll}
& \Psi_{\thel+1}(\x_{>\thel+1},k) 
= \frac{(-1)^{\frac{\thel -2}{2}}}{(2\pi)^{\thel +1}} \sum_{j = 1}^l \int_\R \d w_j \intcc_{\R^{\thel-1 }} \Big(  \bigotimes_{\jmath\, \in\, 1\,..\,\thel\, \setminus\,  \{j\}} \big(\frac{\d w_{\jmath}}{w_{\jmath} + w_{j}} - \frac{\d w_{\jmath}}{w_{\jmath} - w_{j}}\big) \Big) \times \\ 
& \qqq \Bigg(\intcc_{\R}\big(\frac{\d w_{\thel+1} }{w_{\thel+1}+w_j}-\frac{\d w_{\thel+1} }{w_{\thel+1}-w_j}\big) \cos(w_{\thel+1}(|k|-\norml{\x_{>\thel+1}}))\Phi_{\thel+1 } (\w_{\le l+1},\x_{>l+1}) \\
&\qqq - \intcc_{\R}\big(\frac{\d w_{\thel+1} }{w_{\thel+1}+w_j}-\frac{\d w_{\thel+1} }{w_{\thel+1}-w_j}\big) \cos(w_{j}(|k|-\norml{\x_{>\thel+1}}))\Phi_{\thel+1 } (\w_{\le l+1},\x_{>l+1})\Bigg).\eeql
By \eqref{e:aggregpvintc}, 
the first term above corresponding to $\cos(w_{\thel+1}\cdots)$ may be rewritten as 
\beql{e:lll1}
 &\frac{(-1)^{\frac{\thel -2}{2}}}{(2\pi)^{\thel +1}} \sum_{j = 1}^l \int_\R \d w_j  \intcc_\R \big(\frac{\d w_{\thel+1} }{w_{\thel+1}+w_j}-\frac{\d w_{\thel+1} }{w_{\thel+1}-w_j}\big) \times\\
 &\intcc_{\R^{\thel-1 }}\Big(\bigotimes_{\jmath\, \in\, 1\,..\,\thel\, \setminus\,  \{j\}} \big(\frac{\d w_{\jmath}}{w_{\jmath} + w_{j}  }
 - \frac{\d w_{\jmath}}{w_{\jmath} - w_{j}}\big)\Big)  \cos(w_{\thel+1}(|k|-\norml{\x_{>\thel+1}})) \Phi_{\thel+1 } (\w_{\le l+1},\x_{>l+1}) \\
 &= \frac{(-1)^{\frac{\thel -2}{2}}}{(2\pi)^{\thel +1}} \sum_{j = 1}^l \int_{\R} \d w_{\thel+1}\,\cos(w_{\thel+1}(|k|-\norml{\x_{>\thel+1}})) \times\\
 &\intcc_{\R}\big( \frac{\d w_j}{w_{j} + w_{\thel+1}} + \frac{\d w_j}{w_{j} - w_{\thel+1}} 
  \big)\intcc_{\R^{\thel-1 }}\Big(\bigotimes_{\jmath\, \in\, 1\,..\,\thel\, \setminus\,  \{j\}} \big(\frac{\d w_{\jmath}}{w_{\jmath} + w_{j}  }
 - \frac{\d w_{\jmath}}{w_{\jmath} - w_{j}}\big)\Big)  \Phi_{\thel+1 } (\w_{\le l+1},\x_{>l+1}) ,
\eeql
where the equality follows from Lemma \ref{prop:intpv} applied to the variables $ w_{\thel+1}$ and $ w_{j}$.
By reflective change of variables $(w_j,w_{\thel+1}) \mapsto (-w_j,-w_{\thel+1})$ to change minus signs into plus signs within Cauchy integrals, the second term in \eqref{e:lll} corresponding to $\cos(w_j\cdots)$ may be rewritten as 
\beql{e:lll2}
&\frac{(-1)^{\frac{\thel -2}{2}}}{(2\pi)^{\thel +1}} \sum_{j = 1}^l \int_{\R} \d w_j\,\cos(w_{j}(|k|-\norml{\x_{>\thel+1}})) \times
\\
&\qqq\intcc_{\R^{\thel }}\Big(\bigotimes_{\jmath\, \in\, 1\,..\,(\thel+1)\, \setminus\,  \{j\}} \big(\frac{\d w_{\jmath}}{w_{\jmath} + w_{j}  }
 - \frac{\d w_{\jmath}}{w_{\jmath} - w_{j}}\big)\Big)  \Phi_{\thel+1 } (\w_{\le l+1},\x_{>l+1}).
\eeql
Substituting  \eqref{e:lll1} and \eqref{e:lll2} into \eqref{e:lll} yields
\bel
&\Psi_{\thel+1}(\x_{>\thel+1},k) =  \frac{(-1)^{\frac{\thel -2}{2}}}{(2\pi)^{\thel +1}} \sum_{j = 1}^{\thel+1} \int_{\R}\d w_j\,\cos(w_{j}(|k|-\norml{\x_{>\thel+1}})) \times
\\
&\qqq\intcc_{\R^{\thel}}\Big(\bigotimes_{\jmath\, \in\, 1\,..\,(\thel+1)\, \setminus\,  \{j\}} \big(\frac{\d w_{\jmath}}{w_{\jmath} + w_{j}  }
 - \frac{\d w_{\jmath}}{w_{\jmath} - w_{j}}\big)\Big)  \Phi_{\thel+1 } (\w_{\le l+1},\x_{>l+1}),
\eel
thereby proving that \eqref{e:recup} holds for $\thel $ even. If $\thel $ is odd, then $\sc_{\thel }(\cdot) = (-1)^{\frac{\thel-1 }{2}}\cos(\cdot)$ and  following similar steps would prove \eqref{e:recup}, thereby establishing the identity \eqref{e:recu} for any $\thel  \in 2\,.. \, d$.\ \finproof 
 
We are now in position to prove \eqref{e:TNlast} for any $d\ge 2$.  By definition of $ \delta_\theS$ in \eqref{e:dssk0},
\begin{align}\label{eq:soo}
    \int_{\R^{d+1}} \delta_\theS(\d(\vec x, k)) e^{-\thermi k} \Fi \phie (\x) =  \int_\R \d ke^{-\thermi k}\Psi_d(k),
\end{align}
where $\Psi_d(k) \equiv \Psi_d(\x_{>d},k)$ as defined by the recurrence \eqref{e:phidef}. 
By Lemma \ref{l:dcc2},
\beql{eq:soo1}
    &\Psi_d(k) =  \frac{2}{(2\pi)^d} \sum_{j = 1}^{d} \int_{\R}\d w_j\, \sc_d(w_j |k|) \intcc_{\R^{d-1}}\Big(\bigotimes_{\jmath\, \in\, 1\,..\,d\, \setminus\,  \{j\}} \big(\frac{\d w_{\jmath}}{w_{\jmath} + w_{j}  }  - \frac{\d w_{\jmath}}{w_{\jmath} - w_{j}}\big)\Big)  \phie(\w). \eeql
Substituting \eqref{eq:soo1} into \eqref{eq:soo},
\beql{eq:ff11}
   &\int_{\R^{d+1}} \delta_\theS(\d(\vec x, k)) e^{-\thermi k} \Fi \phie (\x) = \frac{2}{(2\pi)^d} \sum_{j = 1}^{d} \int_\R \d k e^{-\thermi k}  \times  \\
    & \qqq\int_\R \d w_j \; \sc_d (w_j|k|)\intcc_{\R^{d-1}}\Big(\bigotimes_{\jmath\, \in\, 1\,..\,d\, \setminus\,  \{j\}} \big(\frac{\d w_{\jmath}}{w_{\jmath} + w_{j}  }  - \frac{\d w_{\jmath}}{w_{\jmath} - w_{j}}\big)\Big)  \phie(\w).
\eeql

In odd dimension $d\geq3$, substituting \eqref{e:sincos} yields \eqref{e:TNlast} as required. In even dimension, replacing $\sc_d (\cdot) = (-1)^{\frac{d-2}{2}}\sin(\cdot)$ into \eqref{eq:ff11} yields
\bel
&\int_{\R^{d+1}} \delta_\theS(\d(\vec x, k)) e^{-\thermi  k} \Fi \phie (\x) = \frac{(-1)^{\frac{d-2}{2}}2}{(2\pi)^d} \sum_{j = 1}^{d} \int_{\R} \d k e^{-\thermi k} \int_\R\d w_j \sin(w_j|k|)  \times \\
&\qqq\qqq\underbrace{\intcc_{\R^{d-1}}\Big(\bigotimes_{\jmath\, \in\, 1\,..\,d\, \setminus\,  \{j\}} \big(\frac{\d w_{\jmath}}{w_{\jmath} + w_{j}  }  - \frac{\d w_{\jmath}}{w_{\jmath} - w_{j}}\big)\Big)  \phie(\w)}_{=: f_j(w_j)},
\eel
by which we recognize the same integrals as  in  \eqref{e:dd3}.
Substituting Euler's sine formula, then splitting the integrand while applying the change of variable $k \mapsto -k$ to the second resulting integral,
\bel
 &\int_{\R^{d+1}} \delta_\theS(\d(\vec x, k)) e^{-\thermi k} \Fi \phie (\x) \\
 =\ &  \frac{\thermi(-1)^{\frac{d-2}{2}}}{(2\pi)^d} \sum_{j = 1}^{d}  \int_\R \d k e^{-\thermi k} \sgn(k)\int_\R \d w_j (e^{-\thermi   w_j k}- e^{\thermi  w_j k}) f_j(w_j) 
 \\
  =\ &  \frac{	\thermi(-1)^{\frac{d-2}{2}}}{(2\pi)^d} \sum_{j = 1}^{d} \Bigg(\int_{-\infty}^\infty \d k  \sgn(k)  \int_\R \d w_j e^{-\thermi   k(w_j+1) } f_j(w_j) +\\
  &\qqq\qqq\qqq\qqq \int_{-\infty}^\infty \d k  \sgn(k)  \int_\R \d w_j e^{-\thermi   k(w_j-1) } f_j(w_j)\Bigg) \\
 =\ & \frac{(-1)^{\frac{d-2}{2}}2}{(2\pi)^d} \sum_{j = 1}^{d} 
 \intcc_\R  \Big(\frac{\d w_j}{w_j+1}+\frac{\d w_j}{w_j-1}\Big) f_j(w_j),
\eel
where we substituted \eqref{e:sgn} in the last step. Substituting $f_j(w_j)$ into the above equation, we recover 
\begin{align*}
    & \int_{\R^{d+1}} \delta_\theS(\d(\vec x, k)) e^{-\thermi k} \Fi \phie (\x) =      \frac{(-1)^{\frac{d-2}{2}}2}{(2\pi)^d}\times  \\
    &\qqq \sum_{j = 1}^{d} \intcc_{\R}\big( \frac{\d w_j}{w_{j} - 1} +
  \frac{\d w_j}{w_{j} + 1}\big)\intcc_{\R^{d-1}}\Big(\bigotimes_{\jmath\, \in\, 1\,..\,d\, \setminus\,  \{j\}} \big(\frac{\d w_{\jmath}}{w_{j} + w_{\jmath}}
 -\frac{\d w_{\jmath}}{w_{\jmath} - w_{j}}\big)\Big)  \phie(\w) ,
\end{align*}
which is \eqref{e:TNlast}.
}

\section{Correspondence Between Carr-Madan Spanning and Basket Call Spanning in Dimension $d = 1$}\label{a:carr-madan}

 \begin{proposition}\label{p:carr-madan}
    In dimension $d=1$ and for $k>0$, the Carr-Madan spanning formula \eqref{eq:carrmadan} of a twice differentiable  
    payoff $F(x)\equiv F(x,k)$ such that $ F(\lambda x,\lambda k)= \lambda  F(x,k)$  for all $\lambda>0$, and $\partial_x F(0,k) =  \partial^2_{x^2} F(0,k) = 0$,
        can be rewritten in the form \eqref{eq:cont-repli} for $x\ge 0$, with
\beql{e:thenud}
\nu(\d w) = \frac{1}{w^3}(\partial^2_{x^2}F)\left(\frac{1}{w}, 1\right)\d w .\eeql      
 \end{proposition}

\proof By change of variable $K\mapsto w = k/K $ in \eqref{eq:carrmadan},
   \begin{align}
       F(x) &= \int_0^\infty  \frac{k}{w^2} \left( x - \frac{k}{w}\right)^+ (\partial^2_{x^2}F)\left(\frac{k}{w}, k\right)\d w = \int_0^\infty  \frac{k}{w^3} \left( wx - {k}\right)^+ (\partial^2_{x^2}F)\left(\frac{k}{w}, k\right)\d w \nonumber\\
       &= \int_{-\infty}^\infty  \frac{k}{w^3} \left( wx - {k}\right)^+ (\partial^2_{x^2}F)\left(\frac{k}{w}, k\right)\d w\sp  x\in \R_+ , \label{eq:cm-to-cont}
   \end{align}
    where we used in the last step that $(wx - k)^+ = 0 $ 
    for $w< 0<\frac{k}{x}$. Differentiating $F(\lambda x, \lambda k) =  \lambda F(x,k)$ twice with respect to $x$ yields $ \partial^2_{x^2}F(\lambda x,\lambda k) =  \lambda^{-1} \partial^2_{x^2}F(x, k) $, whence
    \beql{e:ahd}\frac{k}{w^3}(\partial^2_{x^2}F)\left(\frac{k}{w}, k\right) = \frac{1}{w^3} (\partial^2_{x^2}F)\left(\frac{1}{w}, 1\right)\sp k>0.
    \eeql
    Substituting the above into \eqref{eq:cm-to-cont}, we recover \eqref{eq:cont-repli} for $ x \geq 0$  and $\nu(\d w)$ given by \eqref{e:thenud}.\ \finproof
 
\begin{example} 
Consider the one-dimensional smooth payoff
$F(x, k) =G_1(x,k) = \mathbf1_{x\neq0} \sqrt{x^2} e^{-{\frac{k^2}{x^2} }} , k>0, x\in \R_+$ (see Propositions \ref{e:Mdkk}-\ref{ex:Mex2}).  We have
    \begin{align*}
        \partial_{x} G_1(x,k) = \mathbf{1}_{x\neq 0}
        \frac{e^{-\frac{k^2}{x^2}}(2k^2+x^2)}{x\sqrt{x^2}}\sp \partial^2_{x^2} G_1(x,k) = \mathbf{1}_{x\neq 0}\frac{2k^2e^{-\frac{k^2}{x^2}}(2k^2-x^2)}{x^4\sqrt{x^2}}, 
    \end{align*}
    hence $G_1(0,k) = \partial_x G_1(0,k) = 0$. The Carr-Madan spanning formula for $G_1$ then reads
   \bel
   G_1(x,k) = \int_{0}^\infty (x-K)^+ \frac{2k^2e^{-\frac{k^2}{K^2}}(2k^2-K^2)}{K^4\sqrt{K^2}}\d K.
   \eel
Equivalently, Proposition \ref{p:carr-madan} yields
    \begin{align*}
    G_1(x,k)     = \int_{-\infty}^{\infty}(wx - k )^+ 2e^{-{w^2}}(2w^2-1) \d w
    \sp\  k>0, x \in \R_+,
\end{align*}
which is a representation of the form \eqref{eq:cont-repli} (with $(wx - k )^+=0$ for $w<0$).\ \finenv
\end{example}

\bibliographystyle{chicago}\bibliography{mybib}
\end{document}